\documentclass[11pt]{article}
\usepackage{longtable}
\usepackage{graphicx}
\newcommand{\mysection}[1]{\section{#1}
   \hspace{0.8cm}\setcounter{equation}{0}}
\renewcommand{\theequation}{\arabic{section}.\arabic{equation}}
\newcommand{\myappendix}{\appendix
   \renewcommand{\theequation}{\Alph{section}.\arabic{equation}}
   \noindent {\Large \bf Appendices}}

\newcounter{mytab}
\newcommand{\mycap}[3]{\refstepcounter{mytab}
	\label{#1} \\ \vspace{#2} Table \Roman{mytab}: #3}
\newcommand{\mycapw}[3]{\refstepcounter{mytab}
	\label{#1} \\ \vspace{#2} \parbox{6in}{Table \Roman{mytab}: #3}}

\newlength{\dummysp}
\newcommand{\spc}{\hbox{\hspace{\dummysp}}}
\newcommand{\beq}{\begin{equation}}
\newcommand{\eeq}{\end{equation}}
\newcommand{\diag}{\mathop{{\hbox{diag} \, }}\nolimits}
\newcommand{\tr}{\mathop{{\hbox{tr} \, }}\nolimits}
\newcommand{\tra}[1]{\mathop{{\hbox{tr}_{#1} \, }}\nolimits}
\newcommand{\mtxt}[1]{\mathop{\hbox{{\small #1}}}\nolimits}
\newcommand{\ttxt}[1]{\mathop{\hbox{{\tiny #1}}}\nolimits}
\newcommand{\stxt}[1]{\mathop{\hbox{{\scriptsize #1}}}\nolimits}
\newcommand{\bbar}[1]{{\overline{#1}}}
\newcommand{\half}{{1 \over 2}}
\newcommand{\third}{{1 \over 3}}
\newcommand{\twthird}{{2 \over 3}}
\newcommand{\beqa}{\begin{eqnarray}}
\newcommand{\eeqa}{\end{eqnarray}}
\newcommand{\nnn}{ \nonumber \\ }
\newcommand{\mod}{{\; \mtxt{mod} \; }}
\newcommand{\p}{{\partial}}
\newcommand{\W}{{\cal W}}
\newcommand{\Zbf}{{{\bf Z}}}
\newcommand{\Rbf}{{{\bf R}}}
\newcommand{\Db}{{\bar {\cal D}}}

\newcommand{\s}{{\sigma}}
\newcommand{\vev}[1]{{\langle #1 \rangle}}
\newcommand{\bigvev}[1]{{\left\langle #1 \right\rangle}}
\newcommand{\ord}[1]{{{\cal O}(10^{#1})}}
\newcommand{\ordnt}[1]{{{\cal O}(#1)}}
\newcommand{\gappeq}{\mathrel{\rlap {\raise.5ex\hbox{$>$}}
{\lower.5ex\hbox{$\sim$}}}}
\newcommand{\lappeq}{\mathrel{\rlap{\raise.5ex\hbox{$<$}}
{\lower.5ex\hbox{$\sim$}}}}
\newcommand{\myref}[1]{(\ref{#1})}
\newcommand{\eelat}{{\Lambda_{E_8 \times E_8}}}
\newcommand{\elat}{{\Lambda_{E_8}}}
\newcommand{\ux}{$U(1)_X$}
\newcommand{\eetee}{$E_8 \times E_8$}
\newcommand{\gsmc}{$G_{SM} \times G_C$}
\newcommand{\LamH}{\Lambda_H}
\newcommand{\LamHsq}{\Lambda_H^2}
\newcommand{\LamSB}{\Lambda_{\ttxt{SUSY}}}
\newcommand{\MSB}{M_{\ttxt{SUSY}}}
\newcommand{\LamX}{\Lambda_X}
\newcommand{\LamC}{\Lambda_C}
\newcommand{\LamU}{\Lambda_U}
\newcommand{\LamTh}{{\Lambda_{SU(3)^3}}}
\newcommand{\uone}{$U(1)$}
\newcommand{\ket}[1]{{ | #1 \rangle }}
\newcommand{\bra}[1]{{ \langle #1 | }}
\newcommand{\nasm}{$SU(3)_C \times SU(2)_L$}
\newcommand{\icoup}[1]{{\alpha_{#1}^{-1}}}
\newcommand{\dbtw}{{\delta b_2}}
\newcommand{\dbth}{{\delta b_3}}
\newcommand{\dbY}{{\delta b_Y}}
\newcommand{\dbYp}{{\delta b_Y'}}
\newcommand{\dkY}{{\delta k_Y}}
\newcommand{\dkYm}{{\delta k_Y^{\stxt{min}}}}
\newcommand{\bfe}[1]{\vspace{5pt} {\bf #1 \hspace{1pt}}}
\newcommand{\emset}{$\{ V, a_1, a_3, a_5 \}$}
\newcommand{\GNA}{{G_{\stxt{NA}}}}
\newcommand{\GUO}{{G_{\stxt{UO}}}}
\newcommand{\GSM}{{G_{\stxt{SM}}}}
\newcommand{\GGUT}{{G_{\stxt{GUT}}}}
\newcommand{\bGS}{{b_{\stxt{GS}}}}
\newcommand{\ipr}[2]{{\langle #1 | #2 \rangle }}
\newcommand{\tK}{{\tilde K}}
\newcommand{\tbeta}[1]{{b_{#1}^{\stxt{tot}}}}
\newcommand{\GSME}{{SU(3)_C \times SU(2)_L \times U(1)_Y}}
\newcommand{\ben}{\begin{enumerate}}
\newcommand{\een}{\end{enumerate}}

\newcommand{\swZ}{\sin^2 \theta_W (m_Z)}
\newcommand{\bsa}{${\rm BSL}_{\rm A}$}
\newtheorem{defn}{Definition}
\def\[{\left [}
\def\]{\right ]}
\def\({\left (}
\def\){\right )}
\begin{document}

\begin{titlepage} 

\baselineskip=14pt

\hfill    LBNL-48864

\hfill    UCB-PTH-01/36

\hfill    hep-th/0108244

\hfill    Aug.~31, 2001

\begin{center}

\vspace{50pt}

{ \bf \Large Spectra in Standard-like ${\bf Z_3}$ Orbifold Models}

\end{center}

\vspace{5pt}

\begin{center}
{\sl Joel Giedt${}^*$}

\end{center}

\vspace{5pt}

\begin{center}

{\it Department of Physics, University of California, \\
and Theoretical Physics Group, 50A-5101, \\
Lawrence Berkeley National Laboratory, Berkeley,
CA 94720 USA.}\footnote{This work was supported in part by the
Director, Office of Science, Office of High Energy and Nuclear
Physics, Division of High Energy Physics of the U.S. Department of
Energy under Contract DE-AC03-76SF00098 and in part by the National
Science Foundation under grant PHY-95-14797.}

\end{center}

\vspace{5pt}

\begin{center}

{\bf Abstract}

\end{center}

\vspace{5pt}

General features of the spectra of matter states in
all 175 models found in a previous work by the author
are discussed.  Only twenty patterns of representations
are found to occur.  Accomodation of the Minimal
Supersymmetric Standard Model (MSSM) spectrum is
addressed.  States beyond those
contained in the MSSM and nonstandard hypercharge
normalization are shown to be generic, though some
models do allow for the usual hypercharge normalization
found in $SU(5)$ embeddings of the Standard Model
gauge group.  The minimum value of the hypercharge
normalization consistent with accomodation of the MSSM
is determined for each model.  In some cases,
the normalization can be smaller than that
corresponding to an $SU(5)$ embedding of the
Standard Model gauge group, similar to
what has been found in free fermionic models.
Bizzare hypercharges typically occur for
exotic states, allowing for matter which
does not occur in the decomposition of $SU(5)$
representations---a result which has been
noted many times before in four-dimensional
string models.  Only one of the twenty patterns of
representations, comprising seven of the 175 models,
is found to be without an anomalous $U(1)$.
The sizes of nonvanishing vacuum expectation
values induced by the anomalous $U(1)$ are studied.
It is found that large radius
moduli stabilization may lead to the breakdown
of $\sigma$-model perturbativity.
Various quantities of interest in effective supergravity
model building are tabulated for the set of 175 models.
In particular, it is found that string moduli
masses appear to be generically quite near the
gravitino mass.
String scale gauge coupling unification is shown
to be possible, albeit contrived, in an example model.
The intermediate scales of exotic particles are estimated
and the degree of fine-tuning is studied.

\vfill

\begin{tabbing}

{}~~~~~~~~~\= blah  \kill
\> ${}^*$ E-Mail: {\tt JTGiedt@lbl.gov}

\end{tabbing}

\end{titlepage}

\renewcommand{\thepage}{\roman{page}}
\setcounter{page}{2}
\mbox{ }

\vskip 1in

\begin{center}
{\bf Disclaimer}
\end{center}

\vskip .2in

\begin{scriptsize}
\begin{quotation}
This document was prepared as an account of work sponsored by the United
States Government. Neither the United States Government nor any agency
thereof, nor The Regents of the University of California, nor any of their
employees, makes any warranty, express or implied, or assumes any legal
liability or responsibility for the accuracy, completeness, or usefulness
of any information, apparatus, product, or process disclosed, or represents
that its use would not infringe privately owned rights. Reference herein
to any specific commercial products process, or service by its trade name,
trademark, manufacturer, or otherwise, does not necessarily constitute or
imply its endorsement, recommendation, or favoring by the United States
Government or any agency thereof, or The Regents of the University of
California. The views and opinions of authors expressed herein do not
necessarily state or reflect those of the United States Government or any
agency thereof of The Regents of the University of California and shall
not be used for advertising or product endorsement purposes.
\end{quotation}
\end{scriptsize}

\vskip 2in

\begin{center}
\begin{small}
{\it Lawrence Berkeley Laboratory is an equal opportunity employer.}
\end{small}
\end{center}

\newpage
\renewcommand{\thepage}{\arabic{page}}
\setcounter{page}{1}
\def\thefootnote{\arabic{footnote}}
\setcounter{footnote}{0}

\baselineskip=14pt

\mysection{Introduction}
Since its introduction, the {\it heterotic string} \cite{GHMR85}
has offered the possibility that it may provide
a unifying description of all fundamental
interactions.  However, the theory as originally
formulated has a ten-dimensional space-time.
To construct a four-dimensional theory, one
typically associates six of the spatial dimensions
of the original theory with a very small compact
space.  One route to ``compactifying'' the six
extra dimensions, which has been the subject
of intense research for several years now,
is to take the six-dimensional space
to be an {\it orbifold} \cite{DHVW85,DHVW86}.

Four-dimensional heterotic string theories
obtained by orbifold compactification
take two broad paths to the treatment of internal string
degrees of freedom not associated with four-dimensional
space-time.  On the one hand, these degrees of freedom
are associated with two-dimensional {\it free fermionic}
fields \cite{ABK87};
on the other, some are associated with
two-dimensional {\it bosonic} fields propagating
in a constant background.

Remarkable progress in the construction of
realistic four-dimensional {\it free fermionic}
heterotic string models \cite{AEHN88}
has been made in the last several years:
a high standard has been established
recently by Cleaver, Faraggi, Nanopoulos
and Walker in their construction 
and analysis \cite{CFN99} of a Minimal Superstring
Standard Model based on the free
fermionic model of Ref.~\cite{FNY90}.
The Minimal Superstring Standard Model
has only the matter content of the Minimal
Supersymmetric Model\footnote{For a review of the MSSM, 
see for example Refs.~\cite{mssmr}.}
(MSSM)
at scales significantly
below the string scale $\LamH \sim 10^{17}$ GeV.
Furthermore, the hypercharge normalization
(discussed in detail below) is conventional.

Similarly realistic four-dimensional {\it bosonic} heterotic
string models have not yet been engineered,
though the foundations of such an effort
were laid some time ago
\cite{DHVW85,DHVW86,INQ87,IMNQ88}.  Some of the most
promising models were of the $Z_3$ orbifold variety,
with nonvanishing {\it Wilson lines} (discussed below)
chosen such that
the matter spectrum naturally had three generations.
One such model was introduced
by Ib\'a\~nez, Kim, Nilles and Quevedo
in Ref.~\cite{IKNQ87}, which we will refer
to as the Bosonic Standard-Like-I (BSL-I) model.
The model was subsequently
studied in great detail by two
groups: Ib\'a\~nez, Nilles, Quevedo et al.~in
Refs.~\cite{FINQ88b,FIQS90}; Casas and Mu\~noz in
Refs.~\cite{CM88}.  As is often the case in
supersymmetric models, the vacuum in the BSL-I model
is not unique; different choices lead to different
low energy effective theories.  A particularly
encouraging vacuum was the one chosen
by Font, Ib\'a\~nez, Quevedo and Sierra (FIQS)
in Section~4.2 of Ref.~\cite{FIQS90}; in what
follows, we will refer to this effective
string-derived theory as the FIQS model.
Departures from realism in the FIQS model were pointed
out recently in \cite{GG00} and \cite{Gie01a}.  
In the latter article,
we suggested that a scan over three generation
constructions analogous to the BSL-I model be
conducted, in the search for a more realistic model.
Ultimately, we would like to attempt models comparable to
the free fermionic Minimal Superstring Standard Model.
Part of the purpose of this paper is
to report some of our progress toward this goal.

This article is devoted to a
model {\it dependent} study of bosonic standard-like
$Z_3$ orbifolds.  Model {\it independent}
analyses are appealing because they paint a wide swath
and highlight general predictions of a class of
theories.  Too often, however, one is left wondering
whether the limiting assumptions made in
such analyses really reflect the properties
of some class of explicit, consistent
underlying theories.  At some point it is necessary
to get dirt on oneself and investigate whether or
not the broad assumptions made in model
independent analyses are ever valid.  This is
one of the motivations for model dependent
studies such as the one contained here.
Another reason to study explicit string constructions
is that certain peculiarities are
more readily apparent under close examination.
One well-known example, which will be discussed
in detail below, is the generic presence of
exotic states with hypercharges which do not
occur in typical Grand Unified Theories\footnote{ 
For a review of non-supersymmetric GUTs see  
Refs.~\cite{gutrvw,Sla81} and for supersymmetric
extensions see Refs.~\cite{sgtrvw}.} (GUTs).

One objection to model dependent studies in
four-dimensional string theories is that the
number of possible constructions is enormous.
However, in at least one respect the enormity
is not as great as it would appear.
Already in the second of the two seminal papers by
Dixon, Harvey, Vafa and Witten, it was realized
that many ``different'' orbifold models are
in fact equivalent \cite{DHVW86}.  Casas, Mondragon and Mu\~noz 
(CMM) have shown in detail how equivalence
relations among orbifold compactifications can be
used to greatly reduce the number of {\it embeddings}
(in the present context, a set \emset\
of sixteen-dimensional vectors)
which must be studied in order to produce all
physically distinct models within a given
class of constructions \cite{CMM89}.
In particular, they applied these techniques to a special
class of bosonic standard-like heterotic string models;
for convenience, we will refer to this as the
$BSL_A$ {\it class.}  For completeness,
we give its technical definition below.
The meanings of the terms used here will
be made clear in Section~\ref{mss},
as much as is required to follow the
discussion in the remainder of this article.
For further details, the interested
reader is encouraged to consult the various reviews
\cite{orbrvw,BL99}, texts \cite{orbtxt}, and references therein.
In simpler terms, the definition given
here implies that
we follow the construction outlined in \cite{INQ87},
with three generations by the method
suggested in \cite{IKNQ87},
subject to additional restrictions
imposed by CMM (items (iii) and (iv) below).
\begin{defn}
The ${\rm {\bf BSL}}_{\bf A}$ {\bf class}
consists of all bosonic \eetee\ heterotic $Z_3$ orbifold 
models with the following properties:
\ben
\item[(i)]
symmetric treatment of left- and right-movers
and a shift embedding $V$ of the twist operator $\theta$;
\item[(ii)] two nonvanishing Wilson lines $a_1,a_3$
and one vanishing Wilson line $a_5=0$;
\item[(iii)] observable sector gauge group
\beq
G_O = SU(3) \times SU(2) \times U(1)^5 ;
\label{CMMgo}
\eeq
\item[(iv)] a quark doublet representation
$(3,2)$ in the {\it untwisted sector.}
\een
\end{defn}
CMM found that models satisfying
(i-iv) may be described (in part) by
one of just nine {\it observable sector}
embeddings; here, ``observable'' refers to the first
eight entries of each of the nonvanishing embedding vectors,
$V, a_1, a_3$; it is this which determines
properties (iii) and (iv) listed above.
In a previous article \cite{Gie01b}, we showed that
these nine observable sector embeddings are equivalent
to a smaller set of six embeddings.  To fully
specify a model, the observable sector embedding
must be completed with a {\it hidden sector} embedding---the
last eight entries of each of the nonvanishing
embedding vectors, $V, a_1, a_3$.
In Ref.~\cite{Gie01b} we enumerated all possible ways
to complete the embeddings in the hidden sector,
using equivalence relations to reduce this set to
a ``mere'' 192 models.
Surprisingly, only five hidden sector gauge groups
$G_H$ were found to be possible.
These possibilities
are shown in Table \ref{gcs}.

\begin{table}[ht!]
\begin{center}
\begin{tabular}{cc}
Case & $G_H$ \\ \hline 
1 & $SO(10) \times U(1)^3$ \\
2 & $SU(5) \times SU(2) \times U(1)^3$ \\
3 & $SU(4) \times SU(2)^2 \times U(1)^3$ \\
4 & $SU(3) \times SU(2)^2 \times U(1)^4$ \\
5 & $SU(2)^2 \times U(1)^6$ \\ \hline
\end{tabular}
\caption{Allowed hidden sector gauge groups $G_H$. \label{gcs}}
\end{center}
\end{table}

The $Z_3$ orbifold models studied here have
$N=1$ local
supersymmetry (supergravity) at the string scale.  In our
analysis, we assume that this supersymmetry
is broken dynamically via gaugino condensation
of an asymptotically free condensing group $G_C$
in the hidden sector.  That is, the vacuum expectation
value ({\it vev}) of
the gaugino bilinear $\vev{\lambda \lambda}$
acquires a nonvanishing value.  This operator has
mass dimension three; we therefore
define the dynamically generated {\it condensation
scale} $\LamC$ by
\beq
\vev{ \lambda \lambda } = \LamC^3.
\eeq
To estimate the value of $\LamC$, consider
the one loop evolution of the
running gauge coupling $g_C(\mu)$ of $G_C$:
\beq
{d g_C \over d \ln \mu} = \beta(g_C) = {b_C g_C^3 \over  16 \pi^2}.
\label{bfdef}
\eeq
The $\beta$ function coefficient $b_C$ is given by
\beq
b_C = -3 C(G_C) + \sum_R X_C(R).
\label{bcd}
\eeq
Here, $C(G_C)$ is the eigenvalue
of the quadratic Casimir operator for the adjoint
representation of the group $G_C$ while
$X_C(R)$ is the {\it Dynkin index} for the representation $R$,
given by $\tra{R}(T^a)^2 = X_C(R)$ in a Cartesian
basis for the generators $T^a$;
we adhere to a normalization where $X_C=1/2$
for an $SU(N)$ fundamental representation.
The sum runs over chiral supermultiplet
representations.
Provided $b_C$ is negative, the coupling turns strong
at low energies and the dynamical scale
$\LamC$ is generated, in analogy to
$\Lambda_{\stxt{QCD}}$.  
The running of gauge couplings
from an initial unified value $g_H \sim 1$ at a
unification scale,
which in our case is the
string scale $\LamH \sim 10^{17}$ GeV, gives
\beq
\LamC \sim \LamH \exp (8 \pi^2/ b_C g_H^2),
\label{lces}
\eeq
where we have identified $\LamC$ with the
Laundau pole of the running coupling.

Soft mass terms in the low energy effective
lagrangian split the masses of supersymmetry multiplets,
and thereby break supersymmetry;
partners to Standard Model (SM) particles are
generically heavier by the soft mass scale $\MSB$.
The soft terms arise from nonrenormalizable interactions
in the supergravity lagrangian, with masses
proportional to the gaugino condensate
$\vev{ \lambda \lambda }$,
suppressed by inverse powers of the (reduced) Planck mass,
$m_P \equiv 1 / \sqrt{8 \pi G} = 2.44 \times 10^{18}$ GeV.
On dimensional grounds, one expects that the observable
sector supersymmetry breaking scale $\MSB$ is given by
\beq
\MSB \approx \zeta \cdot \vev{ \lambda \lambda} / m_P^2
= \zeta \cdot \LamC^3 / m_P^2,
\label{sbs}
\eeq
with (naively) $\zeta \sim \ordnt{1}$.
For supersymmetry to protect the gauge hierarchy $m_Z \ll m_P$
between the electroweak scale and the fundamental scale,
one requires, say, $\MSB \lappeq 10$ TeV.  Then
\myref{sbs} with $\zeta \sim \ordnt{1}$ implies
$\LamC \lappeq 4 \times 10^{13}$ GeV.  On the other hand,
direct search limits \cite{PDG00} on charged superpartners
require, say, $\MSB \gappeq 50$ GeV, which translates
into $\LamC \gappeq 7 \times 10^{12}$ GeV.
More precise results may be obtained, for instance,
with the detailed supersymmetry breaking models
of Bin\'etruy, Gaillard and Wu (BGW) \cite{BGW} as well
as subsequent ellaborations by Gaillard and
Nelson \cite{GN00}.  These calculations confirm
the naive expectation \myref{sbs}, except that
\beq
\ord{-2} \lappeq \zeta \lappeq \ord{-1},
\label{1a}
\eeq
which tends to increase $\LamC$.  For example,
the lower bound implied by $\MSB \gappeq 50$ GeV
changes to $\LamC \gappeq 9 \times 10^{12}$ GeV
if $\zeta \approx 0.4$, near the upper end of
the range \myref{1a}.
The result is that
\beq
\ord{13} \lappeq {\LamC \over \mtxt{GeV}} \lappeq \ord{14}
\label{1b}
\eeq
is a reasonably firm estimate.

For $G_C = SU(2)$ with no matter,
one has $b_C = -6$.  Substituting
into \myref{lces}, one finds $\LamC \sim 10^{11}$ GeV.
On the other hand, \myref{lces} is a crude estimate;
studies of the BGW effective theory show that
the naive estimate \myref{lces}
can receive significant corrections
due to a variety of effects, and deviations
by an order of magnitude are certainly possible.
Thus, a more reliable bound is $\LamC \lappeq 10^{12}$ GeV.
Since $b_C > -6$ when $G_C$ charged matter
is present, the limit $\LamC \lappeq 10^{12}$ GeV is saturated by the
case with no matter.  In the models considered here, as will be seen
below, $SU(2)$ groups always have many, many matter
representations, and it is unlikely that {\it all} of them
would acquire effective mass couplings
{\it at the unification scale} $\LamH$
so that $b_C = -6$ and 
$\LamC \sim 10^{12}$ GeV could be achieved.
In any case, $10^{12}$ GeV is below the lower bound
in \myref{1b}, set by $\MSB \gappeq 50$ GeV,
the firmer of the soft scale requirements,
so having $b_C = -6$ is marginal at best.
Case 5 of Table \ref{gcs} was therefore considered
to be an unviable hidden sector gauge group.
Certainly, Cases 1 to 4 appear more
promising.
Eliminating the models with the Case 5 gauge group,
only 175 models remain.  The matter spectra of these
models are the topic of discussion for the present paper.

Quite commonly in the models considered here,
some of the \uone\ factors contained in the
gauge group $G = G_O \times G_H$ are apparently anomalous:
$\tr Q_a \not= 0$.
Redefinitions of the charge generators
allow one to isolate
this anomaly such that only one \uone\
has an apparent trace anomaly.
We denote this factor of $G$ as \ux.
The associated anomaly is canceled by the Green-Schwarz
mechanism \cite{GS84}:  tree level couplings between
the \ux\ vector multiplet and the two-form
field strength (dual to the
universal axion)
are added to the effective action
in such a way that the one loop \ux\ anomaly is
canceled \cite{UXR}; the \ux\ only appears to
be anomalous.
When the cancellation is done in
a supersymmetric fashion, a Fayet-Illiopoulos (FI)
term $\xi$ for \ux\ is induced; we have, for example,
described this effect at the effective supergravity
level in the Appendix of \cite{Gie01a}.  The result of
these considerations is an effective D-term for \ux\
of the form:
\beq
D_X = \sum_i {\p K \over \p \phi^i} {\hat q}^X_i \phi^i + \xi,
\qquad \xi = {g_H^2 \tr \hat Q_X \over 192 \pi^2} m_P^2.
\label{1.2}
\eeq
The \ux\ generator $\hat Q_X$ has a normalization consistent
with unification (discussed further below),
${\hat q}^X_i$ is the charge of
the scalar $\phi^i$ with respect to $\hat Q_X$,
$K$ is the K\"ahler potential and
and $g_H$ is the unified coupling mentioned above.
Since the scalar potential of the effective supergravity
theory at the string scale $\LamH$ contains
the term $g_H^2 D_X^2 / 2$,
some scalar fields generically shift to cancel the FI
term (i.e., $\vev{D_X}=0$ to leading order)
and get vevs of order $\sqrt{|\xi|}$.
Adopting the terminology
of \cite{Gie01a}, we will refer to these
as {\it Xiggs} fields, 
since they are associated with the breaking
of \ux\ (and typically other factors of $G$)
via the Higgs mechanism.
Generally, the
way in which the FI term may be canceled is
not unique and continuously connected vacua
result.  Pseudo-Goldstone modes, {\it D-moduli} \cite{GG00},
parameterize the flat directions;
dynamical supersymmetry
breaking and loop effects are required to select
the true vacuum and render these
scalar fields massive \cite{GG00,Gai01}.
(Moduli parameterizing flat directions of the
scalar potential are a generic feature of
supersymmetric field theories \cite{BDFS82}.
An example of D-moduli was noted previously in the study
of D-flat directions in \cite{CM88},
parameterized there by the quantity
``$\lambda$,'' which interpolated between
various vacuua.  Such moduli have also been
noted in the study of
flat directions in free fermionic string models,
for instance in Ref.~\cite{CCF01}.)
The FI term $\xi$ has mass dimension two and its
square root therefore gives the approximate scale of
\ux\ breaking, which we hereafter denote
\beq
\LamX \equiv \sqrt{|\xi|}
= { \sqrt{|\tr \hat Q_X |} \over 4 \pi \sqrt{12} }
\times g_H m_P.
\label{1c}
\eeq
In the examples below we will find by explicit calculation
of $\tr \hat Q_X$ in each of the 175 models
that $\LamX \approx \LamH \sim 0.2 \, m_P$.

In Section~\ref{mss} we discuss the determination of
the spectrum of massless states from the
underlying string theory.  We discuss in
careful detail how the gauge group $G$
is determined.
We then describe in similar detail
how one determines the irreducible representations 
({\it irreps}) and \uone\ charges of matter
states.  In Section~\ref{mds} we
make observations
on the general features of the 175
models, as determined from the spectrum
of massless states and their \uone\ charges.
We find that only 20 patterns of
irreps occur in the 175 models.  
In Section~\ref{hyc}, we delve into difficulties
associated with the electroweak hypercharge.  
We explore
the most natural definition of hypercharge:
to embed it into an $SU(5)$
gauge group which also contains 
the $SU(3) \times SU(2)$ of 
the observable gauge group $G_O$.
As a further condition, we require that
the $SU(5)$ is a subgroup of the observable
$E_8$ factor of the ``parent'' \eetee\ theory.
We find that none of the 175 models can
accomodate the full MSSM spectrum when this is
done; although adequate $SU(3) \times SU(2)$
irreps are present, the hypercharge quantum
numbers are not correct for enough of the irreps.
We will explain how the presence of states with
unusual hypercharge values corresponds to the
phenomenon of {\it charge fractionalization}
in orbifolds.
The absense of states with correct hypercharges
for the $SU(5)$ embedding leads us to the less attractive
alternative of engineering a hypercharge
which is a general linear combination of the several
$U(1)$s contained in $G$ {\it and} generators
of the Cartan subalgebras of nonabelian
factors contained in the hidden gauge group $G_H$.
We find that this {\it does} allow for
the accomodation of the MSSM spectrum.
At the same time, rather bizzare hypercharges
for extra matter are found to be generic,
as well as nonstandard hypercharge normalization.
In Section~\ref{app} we illustrate these
unconventional results with a detailed examination
of one of the 175 models.  We describe
various assignments of the MSSM to the
spectrum of 153 chiral multiplets of matter
states present in the model, and the hypercharges and
nonstandard hypercharge normalizations which
occur.  In spite of
nonstandard hypercharge normalization,
it is found that
successful unification of gauge couplings
at the string scale $\LamH \sim 10^{17}$
GeV is possible.
However, the unification scenario in this
model is rather ugly, since it requires that
exotic states with fractional electric
charges be introduced at intermediate
scales---between the
electroweak scale and the string scale.
We suggest how one might circumvent
phenomenological difficulties with
fractionally charged states having
intermediate scale masses.
In Section~\ref{con} we make concluding remarks and
suggest directions for further research.
In Appendix~A we review cancellation of the
{\it modular anomaly.}  In Appendix~B we present
our more lengthy sets of tables.

Since each model contains $3 \times \ordnt{50}$
matter irreps
and eight or nine independent
\uone\ generators, it is for obvious reasons that
we do not provide in full detail the spectra and charges
of all 175 models.
However, upon request, complete tables 
of the matter spectra and \uone\ charges
are available from the author.

\mysection{Determination of Spectra}
\label{mss}
Several textbooks discussing heterotic
orbifolds are available \cite{orbtxt}.  In addition,
many reviews have been written over the 
years~\cite{orbrvw},
including the recent (and widely
available) review
by Bailin and Love \cite{BL99}.
Rather than repeat lengthy discussions
given elsewhere, we have chosen to avoid many
details of the underlying string theory and present
a somewhat heuristic description.  Our intent is to
provide just enough information to allow
one to determine the spectrum of gauge and
matter states below the string scale, for the class
of orbifold models considered here.
To this end, we provide a set of ``recipes''
for the spectrum determination at the close of
this section.  These are designed as a tool
for the ``string novice'' who merely wishes to
study these models from a low energy,
phenomenological point of view.

To make contact with the world of particle physics,
one is interested in the effective theory produced
by heterotic string theory at energy scales far below the
string scale $\LamH \sim 10^{17}$ GeV.  The first
step in constructing such a theory is to determine
the string states with masses much less than $\LamH$.
Secondly, one must derive the interactions between
these states and an appropriate description for these
interactions.  In the context of perturbative string
theory, there exist systematic methods for the
accomplishment of these tasks, subject to certain
technical difficulties which we will not discuss
here, since for the most part we work only at leading order in
string perturbation theory.  The perturbation series
corresponds to string
world-sheet (the two-dimensional surface
swept out by the string) diagrams of increasing complexity.
These are labeled by the {\it genus} of the diagram,
starting at genus zero, often referred to as ``tree level''
in string theory.  The next order, genus one, is
often referred to as the ``one loop level'' in string
theory, because the world-sheet diagram is a two-dimensional
torus.  Interactions are described
by scattering amplitudes between string states.  In
particular, these amplitudes can be studied in the
limit where external momenta are taken to be much
less than the string scale, often referred to as
the {\it zero-slope limit} \cite{Sch71}.  One then matches
the results onto a field theory; that
is, one constructs a local field theory lagrangian
which, when quantized, would have single particle
states with the same properties (mass, spin, charge,
etc.)~as the low-lying string
states and scattering amplitudes which match the
string scattering amplitudes at low external momenta.
Thus, one can talk about the ``particle'' states
which arise from the ``field theory limit'' of the
string.

A study of the heterotic string at tree level shows that
the string states are organized into a tower of
mass levels, with the lowest level of states
massless.  For the four-dimensional heterotic
string, subject to certain qualifications which
will not trouble us here (e.g., the large radius
limit of the extra dimensions where massive string states
can drop below $\LamH$), the only string states
with masses significantly below $\LamH$ are those which lie
at the massless level of the string.
However, genus one corrections can be significant
if, for example, an anomalous \ux\ is present.
On the effective field theory side, this
correction is represented by the FI term
which is induced from cancellation of the
\ux\ anomaly.  The tree level
spectrum of masses can be dramatically altered.
For this reason, we hereafter refer to the states which
are massless at tree level as {\it pseudo-massless.}
Many of the pseudo-massless states
have masses near $\LamH$ once the
one loop corrections are accounted for!
This is because the Xiggses acquire $\ordnt{\LamX}$
vevs; explicit calculations detailed below
show that $\LamH/1.73 \leq \LamX \leq \LamH$
in the 175 models studied here, indicating that
$\LamX$ is more or less the string scale $\LamH$.
The Xiggs vevs cause several
chiral (matter) superfields to
get effective ``vector'' superpotential couplings
\beq
W \ni {1 \over m_P^{n-1}} \vev{\phi^1 \cdots \phi^n} A A^c .
\label{ems}
\eeq
Here, $A$ and $A^c$ are conjugate with respect to the
gauge group which survives after spontaneous symmetry
breaking caused by the \ux\ FI term.
The right-hand side of \myref{ems} is an effective
supersymmetric mass term, which generally results in masses
\beq
m_{eff} \sim \ordnt{\LamX^n / m_P^{n-1}} 
\approx \ordnt{\LamH^n / m_P^{n-1}} .
\label{mef}
\eeq
With $n=1$ in \myref{mef}, the effective masses
are near the string scale.  Due to the numerous
gauge symmetries present in the models considered
here, as well as discrete symmetries
known as {\it orbifold selection rules}
(see for example \cite{HV87,FIQS90,BL99}), not all operators
of the form $A A^c$ will have couplings with $n=1$
in \myref{ems}.  Because of this, a hierarchy of mass scales
is a general prediction of models with a \ux\ factor
(all but seven of the 175 models studied here).
We return to this point in Section~\ref{app},
where we briefly discuss gauge coupling unification.

By construction, the spectrum is that of an $N=1$ four-dimensional
locally supersymmetric theory.  Furthermore,
the compact space is a six-dimensional $Z_3$ {\it orbifold}
(defined below).  Certain parts of the spectrum are well-known to
be present by virtue of these facts alone \cite{DHVW85}.
We will not discuss these states
in this section except to
note their existence:  the supergravity
multiplet, the {\it dilaton} supermultiplet and
nine chiral multiplets $T^{ij}$ whose scalar
components correspond to the {\it K\"ahler-}
or {\it T-moduli}
of the compact space.  (See for example
\cite{Gre97} for a discussion of toric
moduli.)

The remainder of the spectrum depends on the
choice of {\it embedding,} and it is this part
of the spectrum which we must calculate separately for
each of the 175 models.  The embedding-dependent spectrum
consists of massless chiral multiplets of
matter states and massless vector multiplets of gauge states.
Once the vacuum shifts to cancel the FI
term, some gauge symmetries are spontaneously broken
and chiral matter multiplets (which are linear
combinations of Xiggses) get ``eaten'' by some of the vector
multiplets to form massive vector multiplets.
Examples of the ``degree of freedom balance sheet''
may be found for example in \cite{GG00}.

\subsection{The ${\bf Z_3}$ Orbifold}
The six-dimensional $Z_3$ orbifold
may be constructed from a six-dimensional Euclidean
space $\Rbf^6$.  One defines basis vectors
$e_1, \ldots, e_6$ satisfying
\beq
e_i^2 = e_{i+1}^2 = 2 \, R_i^2, 
\qquad e_i \cdot e_{i+1} = -1 \, R_i^2,
\qquad i=1,3,5,
\label{2f}
\eeq
with a vector $x \in \Rbf^6$
having real-valued components:
\beq
x = \sum_{i=1}^6 x^i e_i ,
\qquad x^i \in \Rbf \quad \forall \;
i = 1, \ldots, 6.
\label{comp}
\eeq
Each of the three pairs $e_i,e_{i+1} \; (i=1,3,5)$
define a two-dimensional subspace which
is referred to below as the ``$i$th complex plane.''
The $i$th such pair also defines
a two-dimensional $SU(3)$ root lattice, obtained from
the set of all linear combinations of the
form $n_i e_i + n_{i+1} e_{i+1}$ with
$n_i,n_{i+1}$ both integers.
Taking together all six basis vectors $e_1,\ldots,e_6$,
we obtain the $SU(3)^3$ root lattice $\LamTh$,
formed from all linear combinations of the
basis vectors $e_1, \ldots, e_6$ with integer
coefficients:
\beq
\LamTh = \left\{ \left. \; \sum_{i=1}^6 \ell^i e_i \; \right|
\; \ell^i \in \Zbf \; \right\} .
\eeq
Note that the {\it radii} $R_i$ in \myref{2f} are
not fixed; neither are angles not appearing in
\myref{2f}, such as $e_1 \cdot e_3$.  These free
parameters determine the size and shape of the
unit cell of the lattice $\LamTh$, and are
encoded in the 
T-moduli $T^{ij}$ mentioned above.
These moduli depend on
the metric $G_{ij}=e_i \cdot e_j$ ($i,j=1,\ldots,6$)
of the six-dimensional compact
space, as well as an antisymmetric two-form
$B_{ij}$.  Of particular interest are the
{\it diagonal} T-moduli $T^i \equiv T^{ii}$.
Up to normalization conventions on the
$T^i$ and $B_{ij}$,
the diagonal T-moduli are defined by
\beq
T^i = \sqrt{\det G^{(i)}} + i B_{i,i+1}, \quad i=1,3,5.
\eeq
Here, $G^{(i)}$ is the metric of the $i$th
complex plane:
\beq
G^{(i)} = \pmatrix{ e_i \cdot e_i & e_i \cdot e_{i+1} \cr
e_{i+1} \cdot e_i & e_{i+1} \cdot e_{i+1} \cr}
= R_i^2 \pmatrix{2 & -1 \cr -1 & 2 \cr}.
\eeq

Translations in $\Rbf^6$ by elements of
$\LamTh$,
\beq
x \to x + \ell, \qquad \ell \in \LamTh,
\qquad \forall \; x \in \Rbf^6,
\eeq
form what is
referred to as the {\it lattice group.}
A rotation $\theta$ 
in $\Rbf^6$ is defined, with action
on the basis vectors:
\beq
\theta \cdot e_i = e_{i+1}, \qquad \theta \cdot e_{i+1} = - e_i - e_{i+1},
\qquad i=1,3,5.
\label{pnh}
\eeq
Typically, $\theta$ is referred to as
the orbifold {\it twist operator.}
It is easy to check that $\theta^3 =1$.
The twist operator $\theta$ generates the orbifold {\it point group,}
\beq
Z_3 = \{ 1, \theta, \theta^2 \}.
\eeq
It can be seen from \myref{pnh} that the
twist operator maps any element of $\LamTh$ into $\LamTh$.
Consequently, we can define the product group
generated by the combined action of the
point group and the lattice group.
This group is referred to as the {\it space group} $S$ and
a generic element is written $(\omega,\ell)$,
with $\omega \in Z_3$ and $\ell \in \LamTh$.
Acting on any element $x \in \Rbf^6$,
\beq
(\omega,\ell) \cdot x = \omega \cdot x + \ell
= \sum_{i=1}^6 [x^i (\omega \cdot e_i) + \ell^i e_i],
\eeq
where $\omega \cdot e_i$ can be
obtained from \myref{pnh}.
It is not hard to check the multiplication rule
\beq
(\omega,\ell) \cdot (\omega',\ell')
= (\omega \omega', \omega \ell' + \ell).
\label{ormu}
\eeq
The space group has four generators:
$(\theta,0), \; (1,e_1), \; (1,e_3)$ and $(1,e_5)$.
For example, using (\ref{pnh},\ref{ormu}) one can write
\beq
(1,e_2) = (\theta,0) \cdot (1,e_1) \cdot (\theta,0) \cdot(\theta,0).
\eeq
Certain
points $x_f \in \Rbf^6$ are fixed under the action
of space group elements with $\omega=\theta$:
\beq
(\theta,\ell) \cdot x_f = \theta \cdot x_f + \ell = x_f.
\eeq
It is not hard to solve this equation; one finds
that the {\it fixed points} are in one-to-one correspondence
with elements of $\LamTh$:
\beq
x_f(\ell) = (1-\theta)^{-1} \cdot \ell.
\label{2.3}
\eeq

To define the orbifold, denoted $\Omega = \Rbf^6 / S$,
one demands that
points $x,x' \in \Rbf^6$ be treated as equivalent
if they are related to each other under the
action of the space group $S$.
\begin{defn}
The points $x,x' \in \Rbf^6$ are {\bf equivalent}
on the orbifold $\Omega = \Rbf^6 / S$,
notated $x' \simeq x$, if and only if
there exists $(\omega,\ell) \in S$ such that
$x' = (\omega,\ell) \cdot x$.
\end{defn}
A space constructed in this way is often
referred to as a {\it quotient space,}
because we ``divide out'' by the action
of a discrete transformation group, in
this case the space group $S$.
It is worth noting that quotient
space constructions for extra dimensions
were applied in a field theory context
some years prior to the construction of
four-dimensional strings on orbifolds,
with important consequences such as chiral
fermions \cite{CGS}.

On the orbifold, most of the fixed points \myref{2.3}
are equivalent to each other.  There are only 27
inequivalent fixed points, which can
be obtained from \myref{2.3} using
\beq
\ell(n_1,n_3,n_5) = n_1 e_1 + n_3 e_3 + n_5 e_5,
\quad n_i = 0,\pm 1.
\eeq
Note the correspondence between this
parameterization of the fixed points
and the generators of the space group
which are elements of the lattice group:
$(1,e_1), \; (1,e_3)$ and $(1,e_5)$.

\subsection{Boundary Conditions}
At the classical level, the location
of the string in the six-dimensional
compact space is specified by a two parameter
map $X_{\stxt{cl}}(\s,\tau)$ which has
a component expression of the form \myref{comp}:
\beq
X_{\stxt{cl}}(\s,\tau)=
\sum_{i=1}^6 X^i_{\stxt{cl}}(\s,\tau) \, e_i.
\eeq
The parameter $\s$ labels points along the string,
with $\s \to \s + \pi$ as one goes once around
the string; $\tau$ labels proper time in
the frame of the string.  The heterotic theory
is a theory of closed strings, so
$X_{\stxt{cl}}(\s,\tau)$ and $X_{\stxt{cl}}(\s + \pi,\tau)$
should be equivalent points on the orbifold.
This requirement is extended to the quantized
theory $X_{\stxt{cl}}(\s,\tau) \to X(\s,\tau)$,
with $X(\s,\tau)$ a quantum operator.
As a consequence of Definition~2, $X(\s,\tau)$
need only be closed up to a space group element.
For the ``$(\omega,\ell)$ sector,''
\beq
X(\s + \pi,\tau) = (\omega,\ell) \cdot X(\s,\tau).
\label{2.1}
\eeq

If we apply some other space group element $(\omega',\ell')$ to
\myref{2.1}, we find
\beq
(\omega',\ell') \cdot X(\s + \pi,\tau) =
\left[ (\omega',\ell') \cdot (\omega,\ell) 
\cdot (\omega',\ell')^{-1} \right] \cdot
(\omega',\ell') \cdot X(\s,\tau).
\eeq
Because $(\omega',\ell') \cdot X(\s,\tau)$ and $X(\s,\tau)$
are equivalent on the orbifold, the boundary
condition
\beq
X(\s + \pi,\tau) =
(\omega',\ell') \cdot (\omega,\ell)
\cdot (\omega',\ell')^{-1} \cdot X(\s,\tau)
\label{2.2}
\eeq
must be treated as equivalent to \myref{2.1}.
That is, boundary conditions in the same {\it conjugacy class}
as $(\omega,\ell)$,
\beq
\left\{ \; (\omega',\ell') \cdot (\omega,\ell)
 \cdot (\omega',\ell')^{-1}
\quad \left| \quad \omega' \in Z_3, 
\; \ell' \in \LamTh \; \right. \right\},
\eeq
are equivalent because they are
related to each other under the action of
the space group \cite{DHVW86}.  There are
27 such conjugacy classes associated with sectors
twisted by $\theta$.  There exists a correspondance
between each of these conjugacy classes and
one of the 27 inequivalent fixed points of
the $Z_3$ orbifold.  Since these sectors do not mix with
each other under the action of the space group,
we regard them as 27 {\it different} twisted sectors.  

Nontrivial boundary conditions are typically
extended to internal string degrees of freedom
$\Psi(\s,\tau)$ not associated with the location of the
string in the six-dimensional compact space.
For the $(\omega,\ell)$ sector,
which has \myref{2.1}, the extension
may be written schematically as
\beq
\Psi(\s+\pi,\tau) = U[(\omega,\ell)] \cdot \Psi(\s,\tau) .
\label{2r}
\eeq
Consistency requires this extension to
be a {\it homomorphism} of the space group:
\beq
U[(\omega,\ell)] \cdot U[(\omega',\ell')]
\simeq U[(\omega,\ell)\cdot(\omega',\ell')],
\label{2p}
\eeq
where ``$\simeq$'' denotes equivalence,
the precise meaning of which depends
on the nature of $\Psi(\s,\tau)$.
As mentioned above, the space group has
four generators; it is therefore sufficient
to specify the action of $U$ for these generators,
since the homomorphism requirement then determines
$U$ for any other element of the space group.

In particular, there exist
sixteen internal bosonic degrees
of freedom $X^I(\s,\tau), \; I=1,\ldots,16$;
these are employed in the construction
of a current algebra which is the source of
gauge symmetry in the effective field theory.
In the twisted sectors, the $X^I(\s,\tau)$ are typically
assigned nontrivial boundary conditions
according to a homomorphism $U$.
As described above, we may define $U$ through
a map of the space group generators
into the internal degrees of freedom.
In the construction studied here, this consists of
a set of shifts:
\beqa
U[(\theta,0)]^I_J \, X^J(\s,\tau)
& = & X^I(\s,\tau) + \pi V^I, \nnn
U[(1,e_i)]^I_J \, X^J(\s,\tau)
& = & X^I(\s,\tau) + \pi a_i^I, \qquad \forall \; i=1,3,5.
\label{edf}
\eeqa
The vector $V$ is referred to as the {\it shift embedding}
of the space group generator $(\theta,0)$;
equivalently, $V$ embeds the twist operator $\theta$.
Likewise, the vectors $a_i$ embed the
other three space group generators $(1,e_i)$,
$i=1,3,5$ respectively.  They
are referred to as {\it Wilson lines}
because of their interpretation as background
gauge fields in the compact space.  (It is worth noting
that nontrivial
gauge field configurations in an extra-dimensional
compact space were used by Hosotani in a field theory
context to achieve gauge symmetry breaking \cite{Hos83};
the nontrivial $a_1,a_3$ in the \bsa\ models represent
a ``stringy'' version of the Hosotani mechanism,
allowing one to obtain standard-like $G$.)

Taking together the embeddings \myref{edf},
and using the space group multiplication
\beq
(1,e_1)^{n_1} \cdot (1,e_3)^{n_3} \cdot (1,e_5)^{n_5} \cdot
(\theta,0) =
(\theta,n_1 e_1 + n_3 e_3 + n_5 e_5),
\eeq
the embedding of the twisted
boundary condition \myref{2.1} for each
of the 27 twisted sectors corresponding to
$(\omega,\ell) = (\theta,n_1 e_1 + n_3 e_3 + n_5 e_5)$
is described by a sixteen-dimensional
embedding vector $E(n_1,n_3,n_5)$:
\beqa
X^I(\s+\pi,\tau) & = & U[(\theta,n_1 e_1 + n_3 e_3 + n_5 e_5)]^I_J
\, X^J(\s,\tau) \nnn
& = & X^I(\s,\tau) + \pi E^I(n_1,n_3,n_5),
\label{2q} \\
E(n_1,n_3,n_5) & = & V + n_1 a_1 + n_3 a_3 + n_5 a_5.
\label{evd}
\eeqa
Consistency conditions \cite{IMNQ88,BLT88} for \emset\ following
from the homomorphism condition \myref{2p}
have been accounted for in
the embeddings enumerated in \cite{Gie01b}.
For example, $(\theta,n_1 e_1 + n_3 e_3 + n_5 e_5)^3=(1,0)$
implies that we must have
\beq
U[(\theta,n_1 e_1 + n_3 e_3 + n_5 e_5)^3]^I_J
\, X^J(\s,\tau) = X^I(\s,\tau) + 3 \pi E^I(n_1,n_3,n_5)
\simeq X^I(\s,\tau).
\eeq
This last step is true because the $X^I(\s,\tau)$
propagate on the \eetee\ root torus where
\beq
X^I(\s,\tau) \simeq X^I(\s,\tau) + \pi L^I,
\qquad \forall \; L \in \eelat,
\eeq
and the embedding vectors are constrained
to satisfy $3 E(n_1,n_3,n_5) \in \eelat$.
The results of a detailed study of these aspects
of the underlying string theory 
\cite{IMNQ88,BLT88} have been built into
the embeddings given in \cite{Gie01b}
and the recipes given below.

As noted above, the boundary conditions
are labeled by the conjugacy classes of the
space group; it is clear that in the general
case, the extension $U$ in \myref{2r}---and
more specifically the embedding
$E(n_1,n_3,n_5)$---will be different
for each conjugacy class.
In the description of string states,
it is therefore convenient to decompose the
Hilbert space into sectors, with each sector
corresponding to a particular conjugacy class.
For the $Z_3$ orbifold, one
has an {\it untwisted} sector, 27 {\it twisted} sectors
corresponding to fixed point (conjugacy class)
labels $(n_1,n_3,n_5)$, $n_i = 0,\pm 1$,
and 27 {\it antitwisted} sectors with similar labeling.
The 27 (anti)twisted sectors are often lumped together
and regarded as a single (anti)twisted sector, since the
(anti)twist (i.e., the point group element)
is identical among them; we prefer
not to do this here.
The term ``twisted state,'' when applied to a particle,
must be understood to refer to the string state
taken to the field theory limit, since it is not
possible to go from one end of a particle to the other!
The antitwisted sectors of the $Z_3$ orbifold
merely contain the antiparticle
states of the twisted sectors, so
we need not discuss them below.

\subsection{${\bf E_8 \times E_8}$:  Progenitor}
Prior to FI gauge symmetry breaking,
the gauge group $G$ is a rank sixteen subgroup of \eetee.
The theory on the orbifold involves
``twisting'' the \eetee\ heterotic string.
Even though $G$ is a subgroup of \eetee, its
description on the string side reflects
the \eetee\ symmetry of the original theory.  That is,
$G$ is ``embedded into \eetee.''  To clarify what
is meant by this phrase, we rehearse a well-known
example.

Recall that each irrep
of a Lie group\footnote{For a review of Lie
algebras and groups see for example
Refs.~\cite{lietxt,lienoee,Sla81}.}
can be identified with
a weight diagram; points on the weight diagram
are labeled by weight vectors.  Well-known
examples are the flavor $SU(3)_F$ weight diagrams of
hadrons containing only $u,d,s$ valence quarks.
In this case, the weight vectors are two-dimensional,
$(\lambda_1,\lambda_2)$,
with entries corresponding to eigenvalues
of two basis elements $H^1,H^2$ of a Cartan
subalgebra of $SU(3)_F$.  If we work in the
limit $m_u=m_d$, $m_s \gg m_u$, then $SU(3)_F$ is
not a good symmetry, but the flavor isospin
subgroup $SU(2)_F$ is.  In a well-chosen
basis for $SU(3)_F$,
the weight diagrams of
$SU(2)_F$ are one-dimensional subdiagrams
of the $SU(3)_F$ weight diagrams.  The points
of the one-dimensional $SU(2)_F$ weight
diagrams are labeled by eigenvalues of the 
basis element $I_3$
of a Cartan subalgebra of $SU(2)_F$.  However,
we could just as well continue to label states
by the $SU(3)_F$ weight vectors; the isospin
quantum numbers would be determined by an
appropriate linear combination
\beq
I_3 = \alpha^1 H^1 + \alpha^2 H^2
\label{2.5}
\eeq 
of $SU(3)_F$ Cartan generators.  The additional
information contained in the two-dimensional
$SU(3)_F$ weight vectors, strangeness, determines
quantum numbers under a global $U(1)_S$
symmetry group which commutes with
$SU(2)_F$.  The generator of $U(1)_S$
is given by
\beq
S = s^1 H^1 + s^2 H^2.
\label{2.6}
\eeq
Consistency of this decomposition requires
that for any irrep $R$ of $SU(3)_F$,
\beq
\tra{R} (I_3 S) = 0 \quad
\Rightarrow \sum_{i,j=1}^2 \kappa^{ij} \alpha^i s^j = 0, 
\label{2s}
\eeq
where $\kappa^{ij}$ is defined by
\beq
\tra{R} (H^i H^j) = X(R) \, \kappa^{ij}.
\eeq
To summarize, the symmetry
group is $G_F = SU(2)_F \times U(1)_S$; states are
conveniently labeled by $SU(3)_F$ weight vectors, which allow
one to determine the quantum numbers with
respect to $G_F$; the weight diagrams of $SU(2)_F$ are
best recognized as subdiagrams of $SU(3)_F$ weight diagrams.
We say that $G_F$ is embedded into $SU(3)_F$.

In complete analogy, an irrep of
the gauge symmetry group $G$ of a given orbifold
model will be described by a set of basis states labeled by
weight vectors of \eetee.
The weights with respect to nonabelian
factors of $G$ as well \uone\ charges of the irrep
are determined by these \eetee\ weight
vectors, just as was the case in the $SU(3)_F$
example above.  The weights of the adjoint representation
are referred to as {\it roots.}  Massless states in
the untwisted sector correspond to a subset of the
states in the \eetee\ adjoint representation.
For this reason, we shall often have occasion to refer to the
\eetee\ root system.  For \eetee, the adjoint
representation {\it is} the fundamental representation
and higher dimensional representations are obtained
from tensor products of the adjoint representation
with itself.  These higher dimensional representations
appear at higher mass levels in the ten-dimensional
uncompactified \eetee\ heterotic string.
These representations are
relevant to the massless spectrum in the twisted
sectors of the four-dimensional theory, in a peculiar
way which will be described below.
Weight vectors add when the tensor products
are taken to form higher dimensional
representations; consequently, the weight diagrams of
the higher dimensional representations fill
out a weight lattice, spanned by the basis vectors
of the adjoint representation weight diagram.
In the case of \eetee, this is the
root lattice $\eelat$, which is described
in most modern string theory texts \cite{orbtxt};
it was also reviewed in Appendix~A of our
previous article \cite{Gie01b}.

Briefly, the root lattice for $E_8$ is given by
\beq
\elat  =  \left\{ (n_1, \ldots, n_8), \; (n_1 + \half, \ldots, n_8 + \half)
\quad \left| \quad n_1, \ldots, n_8 \in \Zbf, \; \;
\sum_{i=1}^8 n_i = 0 \mod 2 \right. \right\}
\label{eld}
\eeq
and $\eelat = \elat \oplus \elat$, the direct sum
of two copies of $\elat$.  The sixteen entries
of a root lattice vector $(n_1, \ldots, n_8; n_9, \ldots, n_{16})$
correspond to eigenvalues with respect to
a basis of the \eetee\ Cartan subalgebra,
which we write as $H^I \; (I = 1, \ldots, 16)$
and which is Cartesian:
\beq
\tra{R} (H^I H^J) = X(R) \; \delta^{IJ} ,
\label{2a}
\eeq
where the trace is taken over an
\eetee\ irrep $R$.  In particular,
the adjoint representation (A) corresponds to
the elements $\alpha \in \eelat$ with 
$\alpha^2 = 2$.  These are the
480 nonzero roots of \eetee, which
take the form $\alpha = (\beta;0)$ or
$\alpha = (0;\beta)$ with
$\beta \in \elat, \; \beta^2=2$.   
It is not hard to check
from \myref{eld} that $X(A)=60$, which
is twice the value typically used by phenomenologists.
Thus, the $H^I$ in \myref{2a} and the
eigenvalues in \myref{eld} are larger
by a factor of $\sqrt{2}$ than the
phenomenological normalization.
{\it Positive roots} are nonzero roots
which have their
first nonzero entry positive, according to an (arbitrary)
ordering system.  {\it Simple roots} are positive roots which
cannot be obtained from the sum of two positive roots.  The
number of simple roots is equal to the rank of the Lie
algebra, which for \eetee\ is sixteen.  We label the simple
roots $\alpha_1, \ldots, \alpha_{16}$.
Of particular importance is the map
of roots $\alpha_i$ into the Cartan
subalgebra defined by
\beq
H(\alpha_i) = \sum_{I=1}^{16} \alpha_i^I H^I.
\label{rtm}
\eeq
From this, one defines an inner product on
the root space:
\beq
\langle \alpha_i | \alpha_j \rangle
\equiv \tra{A} \left[ H(\alpha_i) \cdot H(\alpha_j) \right].
\label{ird}
\eeq
Using \myref{2a}, it is not hard to see that
\beq
\langle \alpha_i | \alpha_j \rangle
= X(A) \; \alpha_i \cdot \alpha_j .
\label{2b}
\eeq
It can be seen that the Dynkin index
$X(A)'$ of the basis \myref{rtm}
is related to the index of \myref{2a}
by $X(A)'=2 X(A)$.  Thus, the generators
\myref{rtm} are larger by a factor
of 2 than the phenomenological
normalization; we return to this point
in Section \ref{hyc} below.
The Cartan matrix of a Lie algebra is
defined by
\beq
A_{ij} = {2 \; \langle \alpha_i | \alpha_j \rangle
\over \langle \alpha_j | \alpha_j \rangle },
\label{dcm}
\eeq
where $i,j$ run over the simple roots.
Using \myref{2b} and $\alpha_i^2 = 2$,
it is easy to check that \myref{dcm}
is simply expressed in terms of the
sixteen-dimensional simple root vectors:
\beq
A_{ij} = {\alpha_i \cdot \alpha_j}.
\label{2c}
\eeq
In the orbifold constructions below,
a subset of the \eetee\ simple roots
survive, and by computing the submatrices
according to \myref{2c}, we can identify
the nonabelian factors in the surviving
gauge group $G$, using widely available
tables for the Cartan matrices of Lie
algebras (e.g., Ref.~\cite{lietxt}).
Finally, it is worth mentioning that
by taking all linear combinations of the sixteen
simple roots with integer-valued coefficients,
one recovers the root lattice $\eelat$.  That is,
\beq
\eelat = \left\{ \left. \; \sum_{i=1}^{16} 
m^i \alpha_i \quad \right| 
\quad m^i \in \Zbf \; \right\} .
\label{2l}
\eeq

\subsection{Recipes}
We next write down without proof recipes
for the generation of the spectrum of pseudo-massless
states.  Where possible, we have attempted to
motivate the rules in a heuristic fashion,
avoiding a detailed discussion
of the underlying string theory.  For further
details, see the reviews \cite{orbrvw,BL99},
texts \cite{orbtxt}, and references therein.

\bfe{Nonzero root gauge states.}  We write these
states as $\ket{\alpha}$ where $\alpha$ satisfies:
\beq
\alpha^2 = 2 , \qquad \alpha \in \eelat ,
\label{2d}
\eeq
\beq
\alpha \cdot a_i \in \Zbf, \quad \forall \; i=1,3,5,
\label{2e}
\eeq
\beq
\alpha \cdot V \in \Zbf .
\label{nzr}
\eeq
Eq.~\myref{2d} merely states that $\alpha$ is
an \eetee\ root.  For nontrivial \emset, 
several roots of \eetee\ will
not satisfy (\ref{2e},\ref{nzr}).
Consequently, the nonzero roots of $G$ will be a subset of the
\eetee\ roots.  The states 
$\ket{\alpha}$ are eigenstates of the
generators $H^I$ of the \eetee\ Cartan subalgebra:
\beq
H^I \ket{\alpha} = \alpha^I \ket{\alpha}, 
\qquad I=1,\ldots,16.
\label{2j}
\eeq
To determine $G$, one first (fully)
decomposes the solutions of (\ref{2d}-\ref{nzr})
into orthogonal subsets.  That is, for $a \not= b$
the subset $\{ \alpha_{a1}, \ldots, \alpha_{a n_a} \}$
is orthogonal to the subset $\{ \alpha_{b1}, \ldots, \alpha_{b n_b} \}$
provided 
\beq
\alpha_{ai} \cdot \alpha_{bj} = 0, \qquad
\forall \quad i=1,\ldots,n_a,
\quad j=1,\ldots,n_b .
\eeq
The $a$th such subset corresponds to a nonabelian 
simple subgroup $G_a$ of $G$,
and the solutions $\alpha_{a1}, \ldots, \alpha_{a n_a}$ 
belonging to this subset are the nonzero roots
of $G_a$.  One next determines which of the
$\alpha_{a1}, \ldots, \alpha_{a n_a}$ are simple
roots.  From the
simple roots one can compute the Cartan 
matrix for $G_a$ using \myref{2c}
and thereby determine the group $G_a$.

As an example, in all of the \bsa\ 
embeddings, there are precisely eight solutions to
(\ref{2d}-\ref{nzr})
which {\it do not} have all first eight
entries vanishing:
\beq
\alpha_{1,1} \, , \, \alpha_{1,2} =
(\underline{1,-1},0,0,0,0,0,0;0,\ldots,0), \qquad
\alpha_{2,1}\, , \, \ldots \, , \, \alpha_{2,6} =
(0,0,\underline{1,-1,0},0,0,0;0,\ldots,0).
\label{nz32}
\eeq
Here (and elsewhere below),
all permutations of underlined entries
should be taken.  These are the nonzero
roots of the observable sector gauge
group $G_O$, and should reproduce
\myref{CMMgo}.  The first set 
in \myref{nz32} is orthogonal to
all vectors in the second set; therefore, these
two sets correspond to different simple factors,
one with two nonzero roots and the other with six;
the two groups must be $SU(2)$ and $SU(3)$.
It is easy to check that the simple roots are
\beq
\alpha_{1,1} = (1,-1,0,0,0,0,0,0;0,\ldots,0),
\label{2m}
\eeq
\beq
\alpha_{2,1} = (0,0,1,-1,0,0,0,0;0,\ldots,0), \qquad
\alpha_{2,2} = (0,0,0,1,-1,0,0,0;0,\ldots,0).
\label{2.39}
\eeq
The simple roots \myref{2.39} give the correct Cartan
matrix for $SU(3)$, using \myref{2c}.

\bfe{Zero root gauge states.}  We write these states
in an orthonormal basis
$\ket{I}$, where $I=1, \ldots, 16$.  These correspond
to gauge states for the Cartan subalgebra of $G$,
in the Cartesian basis $H^I$ discussed above.
They of course have vanishing \eetee\ weights:
\beq
H^I \ket{J} = 0, \quad \forall \; I,J=1,\ldots,16.
\eeq
The group $G$ typically has a nonabelian part $\GNA$ which
is a product of $m$ simple factors, and a \uone\ part
$\GUO$ which is a product of $n$ {\uone}s:
\beq
G = \GNA \times \GUO, \qquad
\GNA = G_1 \times G_2 \times \cdots \times G_m ,
\qquad \GUO = U(1)_1 \times U(1)_2 \times \cdots
   \times U(1)_n .
\label{gde}
\eeq
For the 175 orbifold models under consideration,
the simple factors $G_a$ ($a=1,\ldots,m$)
are either $SU(N)$ or $SO(2N)$ groups.  Each $G_a$
has its own Cartan subalgebra with
a corresponding basis $H_a^1, \ldots, H_a^{r_a}$,
where $r_a$ is the rank of $G_a$.
Each basis element $H_a^i$ is a linear combination
of the \eetee\ Cartan basis elements $H^I$:
\beq
H_a^i = \sum_{I=1}^{16} h_a^{iI} H^I .
\label{cgc}
\eeq
This is the analogue of \myref{2.5}.
It should not be too surprising that 
corresponding linear
combinations of the \eetee\ Cartan gauge states
$\ket{I}$ are taken to obtain Cartan gauge states
of $G_a$:
\beq
\ket{a;i} = \sum_{I=1}^{16} h_a^{iI} \ket{I} .
\eeq
Similarly, the generator $Q_a$ of the factor $U(1)_a$
may be written as
\beq
Q_a = \sum_{I=1}^{16} q_a^I H^I
\label{Qdf}
\eeq
(this is the analogue of \myref{2.6})
and the corresponding gauge state
\beq
\ket{a} = \sum_{I=1}^{16} q_a^I \ket{I} .
\eeq
It is convenient to choose the states
$\ket{a}$ to be orthogonal (we discuss normalization
below):
\beq
\ipr{a}{b} = q_a \cdot q_b = 0 \quad
\mtxt{if} \quad a \not= b.
\label{u1o}
\eeq
For the Cartan states
$\ket{a;i}$, it is more convenient that their inner
product reproduce the Cartan matrix $A^a$ for the
group $G_a$:
\beq
\ipr{a;i}{b;j} = h_a^i \cdot h_b^j = \delta_{ab} A^a_{ij}.
\eeq
It is hopefully apparent from \myref{2c}
that this equation is satisfied
if we take $h_a^i$ to be the
sixteen-dimensional simple root
vectors for $G_a$: $h_a^i \equiv \alpha_{ai}$.
We therefore rewrite \myref{cgc} as
\beq
H_a^i = H(\alpha_{ai}) = \sum_{I=1}^{16} \alpha_{ai}^I H^I ,
\label{2g}
\eeq
where we use the notation of \myref{rtm};
as mentioned there, these generators are
larger by a factor of two than the
phenomenological normalization.

Naturally, we want the $\GNA$ Cartan states orthogonal to
the $\GUO$ states:
\beq
\ipr{a}{b;j} = q_a \cdot \alpha_{bj} = 0,
\qquad \forall \; a,b,j.
\label{qao}
\eeq
It can be seen from the definitions above
that this gives for any irrep $R$ of \eetee\
\beq
\tra{R} (Q_a H_b^j) = 0,
\eeq
which is the analogue of \myref{2s}.
The $q_a$ are therefore chosen to be orthogonal to
the simple roots and to each other.  With $n$
\uone\ factors, as in \myref{gde}, the choice of $q_a$
is determined only up to reparameterizations
which preserve the orthogonality conditions 
(\ref{u1o},\ref{qao}).
In practice, most choices for the \uone\
generators lead to several of them being
anomalous.  It is then useful to make
redefinitions such that only one \uone\
is anomalous.  Let
\beq
t_a = \tr Q_a, \quad t_b = \tr Q_b, \quad
s_a = q_a^2, \quad s_b = q_b^2,
\eeq
with $t_a, t_b$ both nonzero.  
Then define generators $Q_a' = \sum_I (q_a')^I H^I$
and $Q_b' = \sum_I (q_b')^I H^I$ via
\beq
q_a'  =  t_b q_a - t_a q_b , \qquad
q_b'  =  t_a s_b q_a + t_b s_a q_b .
\eeq
It is easy to see that 
$\tr Q_a' = t_b t_a - t_a t_b = 0$,
so that the anomaly is isolated to $Q_b'$.
Furthermore, orthogonality is maintained:
\beq
q_a' \cdot q_b' = t_a t_b (s_b q_a^2 - s_a q_b^2)
= t_a t_b (s_b s_a - s_a s_b) = 0.
\eeq
By repeating this process,
one can easily isolate the anomaly
to a single factor, \ux.

\bfe{Untwisted matter states.}  We denote these states
as $\ket{K;i}$, $i=1,3,5$.  Here, $K$ is a sixteen-vector,
denoting weights under the \eetee\ Cartan generators
$H^I$:
\beq
H^I \ket{K;i} = K^I \ket{K;i},
\qquad I=1,\ldots,16.
\label{2k}
\eeq
Furthermore, $K$ must satisfy
\beq
K^2 = 2 , \qquad K \in \eelat ,
\label{2h}
\eeq
\beq
K \cdot a_i \in \Zbf, \quad \forall \; i=1,3,5.
\label{2i}
\eeq
\beq
K \cdot V = \third \mod 1 ,
\label{ums}
\eeq
It can be seen from comparison 
to (\ref{2d}-\ref{nzr}) 
that the weights $K$ of untwisted matter states
differ from the weights of nonzero
root gauge states
only in the last condition,
\myref{nzr} versus \myref{ums}:  untwisted matter
states correspond to a different 
subset of the nonzero \eetee\ roots
which satisfy \myref{2e}.
(The remaining subset corresponds 
to untwisted antimatter states.)  The
multiplicity of three carried by the index $i$ 
in $\ket{K;i}$ corresponds
to a ground state degeneracy in 
the underlying theory \cite{DHVW85},
which we will not discuss here.  It is one of the nice
features of the $Z_3$ orbifold which aids in easily
obtaining three generation constructions.  However, it
also means that for fixed $K$,
the three generations $i=1,3,5$ have
identical \uone\ charges and are
in identical irreps,
as can easily be checked using
(\ref{Qdf},\ref{2g},\ref{2k}):
\beqa
H_a^j \; \ket{K;i} & = & \alpha_{aj} \cdot K \; \ket{K;i},
\label{wtd} \\
Q_a \; \ket{K;i} & = & q_a \cdot K \; \ket{K;i}.
\label{2n}
\eeqa
That is, the weight $\lambda_{aj}^K = \alpha_{aj} \cdot K$
is independent of $i$ and similarly for the charge
$q_a^K = q_a \cdot K$.

In order to determine the matter
spectrum, we need more than just
the weights \myref{wtd};
we need to be able to
group the basis states 
$\ket{K_1;i}, \ldots, \ket{K_{d(R)};i}$
which make up a given irrep 
$R$ of dimension $d(R)$.
Suppose an incoming matter state $\ket{K;i}$ 
interacts with a gauge supermultiplet
state corresponding to
a nonzero root $\alpha_{aj}$ of $G_a$.  This interaction
is described by inserting a current $J(\alpha_{aj})$,
which acts like a raising or lowering operator
with respect to some $SU(2)$ subgroup of $G_a$:
\beq
\bra{K';i} J(\alpha_{aj}) \ket{K;i}
= \ipr{K';i}{K+\alpha_{aj};i}
= \delta_{K',K+\alpha_{aj}} .
\eeq
For fixed family index $i$, vectors
$K'$ related to $K$ by the addition of
one of the nonzero roots
of $G_a$ are in the same irrep.
Collecting all vectors $K'$ related to
$K$ in this way (and satisfying
(\ref{2h}-\ref{ums})), we fill
out the vertices of a weight diagram 
of an irrep of $G_a$.  
Due to \myref{qao},
$K'$ and $K$ give the same \uone\ charges
(as they must):
\beq
q_b \cdot K' = q_b \cdot \alpha_{aj} + q_b \cdot K
= q_b \cdot K .
\eeq

\bfe{Twisted non-oscillator matter states.}  We denote these
as $\ket{\tK;n_1,n_3,n_5}$, where $n_i =0,\pm 1$
specify which of the 27 fixed points (conjugacy
classes) the state
corresponds to and $\tK$ is a sixteen-vector
giving the weights with respect to the \eetee\
Cartan generators $H^I$, similar to
Eqs.~(\ref{2j},\ref{2k}) above.  However, the $\tK$
{\it do not} correspond to points on $\eelat$.
Rather (cf. \myref{evd}),
\beq
\tK^2 = 4 / 3,
\qquad
\tK = K +  E(n_1,n_3,n_5),
\qquad 
K \in \eelat.
\label{tps}
\eeq
The condition
$\tK^2 = 4/3$ guarantees $\tK \not\in \eelat$
since all elements $L \in \eelat$ have $L^2 = 0 \mod 2$,
as can be checked by inspection of \myref{2l}.
Weights and charges under $G$ are calculated as
for the untwisted states, only now the
shifted weights $\tK$ are used.
In particular,
\beqa
Q_a \; \ket{\tK;n_1,n_3,n_5} 
& = & q_a \cdot \tK \; \ket{\tK;n_1,n_3,n_5} \nnn
& = & \left[ q_a \cdot K + q_a 
\cdot E(n_1,n_3,n_5) \right] \; \ket{\tK;n_1,n_3,n_5}.
\label{2t}
\eeqa
Thus, the twisted matter states have 
charges shifted by
\beq
\delta_a(n_1,n_3,n_5) = q_a \cdot E(n_1,n_3,n_5)
\eeq
from what would occur
in the decomposition of
\eetee\ representations onto
a subgroup with \uone\ factors.
The quantity $\delta_a(n_1,n_3,n_5)$ is
the {\it Wen-Witten defect} \cite{WW85},
a problematic contribution
which is uniform for a given twisted
sector.
It is precisely this feature
which is responsible for difficulties accomodating
the hypercharges of the MSSM 
spectrum and the generic appearance of
states with fractional electric charge,
as will be discussed below.  
Comparison
to \myref{evd} shows that with $a_5 \equiv 0$,
the embedding vector $E(n_1,n_3,n_5)$ is
independent of $n_5$.  
It follows that
states which differ only by the value of $n_5$
have identical \uone\ charges and are
in identical irreps of the gauge group $G$.  This is
how three generations in twisted
sectors are naturally generated in
the class of models considered here.
Filling out irreps of $G_b$
is accomplished by collecting all $\tK'$
which are related to $\tK$ through
$\tK' = \tK + \alpha_{bj}$, similar
to what was done for untwisted states.
Of course, the other quantum numbers
$n_1,n_3,n_5$ must match.

It was stated above that higher dimensional
irreps of \eetee\ are, in a way, relevant to
massless states in the twisted sectors.  We
are now in a position to address this comment.
In Section~\ref{app} we will discuss a model
with an embedding such that
\beq
3E(1,1,n_5) = (0,0,-1,-1,-1,5,2,2;3,1,1,0,1,0,0,0).
\eeq
It is easy to check that a solution to \myref{tps}
is obtained if
\beq
K = (0,0,0,0,0,-2,-1,-1;-1,-1,0,0,0,0,0,0).
\eeq
However, $K^2=8$, so this is not a root of \eetee,
but the weight of a higher dimensional \eetee\ irrep.
Of course, the weight of the state
$\ket{\tK;n_1,n_3,n_5}$ is $\tK$ and not $K$,
so it seems unimportant that $K^2>2$.
However, $q_a \cdot K$ in \myref{2s}
would be the ``conventional''
charge while $q_a \cdot E(1,1,n_5)$ is the
Wen-Witten defect; in this interpretation
the charge $q_a \cdot K$ which would
occur if the defect were absent is
that of the decomposition a higher dimensional 
\eetee\ irrep.
If nothing else, it creates the illusion that
some massive states of the uncompactified
\eetee\ heterotic string are shifted down
into the massless spectrum when compactified
on the six-dimensional orbifold.

Finally, we note that projections
analogous to (\ref{2i},\ref{ums}) are not
required in the twisted sectors of
a $Z_3$ orbifold \cite{IMNQ88,BLT88}.  As a result,
study of this orbifold is significantly
simpler than most other orbifold constructions,
where projections in the twisted
sectors are rather complicated.

\bfe{Twisted oscillator matter states.}  We denote these
as $\ket{\tK;n_1,n_3,n_5;i}$, where $i =1,3,5$
conveys an additional multiplicity of three,
due to different ways to excite
the vacuum in the underlying string theory
with the analogue of harmonic oscillator
raising operators; three types of
oscillators---corresponding to the three
complex planes of the six-dimensional
compact space---excite the vacuum to
generate a massless state. 
The $\tK$
are again shifted \eetee\ weights, but they
have a smaller norm (to compensate for
energy associated with the excited vacuum):
\beq
{\tK}^2 = 2 / 3,
\qquad
\tK = K +  E(n_1,n_3,n_5),
\qquad 
K \in \eelat.
\eeq
The determination of weights, irreps and
charges is identical to that for the other matter states
discussed above.

\mysection{Discussion of Spectra}
\label{mds}
Automating the matter spectrum recipes
given in the previous section, we have determined
the spectra for all 175 models.
We now make some general observations
based on the results of this analysis.
Ignoring the various $U(1)$ charges,
only 20 patterns of irreps were found 
to exist in the 175 models.
These are summarized
in Tables \ref{pt1}-\ref{pt4} (Appendix~B).
In all 175 models, twisted oscillator matter states
are singlets of $\GNA$ (cf.~\myref{gde}).
Singlets notated $(1,\dots,1)_0$ are either
untwisted matter states or twisted
non-oscillator matter states
while singlets notated $(1,\dots,1)_1$
are twisted oscillator matter states.  
Only Patterns 2.6, 4.5, 4.7 and 4.8 have no
twisted oscillator states.  In Table~\ref{unpat} (Appendix~B)
we show the irreps in the untwisted
sector for each of the twenty patterns.
Comparing to Tables~\ref{pt1}-\ref{pt4},
it can be seen that the majority of
states in any given pattern are twisted
non-oscillator states.

In Table~\ref{tb1} (Appendix~B) we have cross-referenced
the 175 embeddings enumerated in \cite{Gie01b} with
the twenty patterns given here.  We now describe
the labeling of models in Table~\ref{tb1}.  We emphasize
that the tables referenced in the
following itemized list are {\it not} the
tables contained in {\it this} article!  Rather,
table references in the following list
correspond to tables in our {\it previous} article,
Ref.~\cite{Gie01b}.
Models are labeled in the format ``$i.j$'' where:
\ben
\item[(a)]
for $i=1,2,4$ or $6$, $i$ is the CMM observable 
sector embedding according to the labeling 
of Table I of Ref.~\cite{Gie01b} and $j$ is the
hidden sector embedding label
as per the corresponding choice
of table from the set Tables III-VI of Ref.~\cite{Gie01b};
\item[(b)]
$i=8$ corresponds to the CMM observable
sector embedding 8 according to the labeling
of Table I of Ref.~\cite{Gie01b} and $j$ is the
hidden sector embedding
according to the labeling of Table VII of Ref.~\cite{Gie01b} ;
\item[(c)]
$i=10$ also corresponds to the CMM observable
sector embedding 8 according to the labeling
of Table I of Ref.~\cite{Gie01b}, but now $j$ is the
hidden sector embedding
according to the labeling of Table VIII of Ref.~\cite{Gie01b};
\item[(d)]
$i=9$ corresponds to the CMM observable
sector embedding 9 according to the labeling
of Table I of Ref.~\cite{Gie01b} and $j$ is the
hidden sector embedding
according to the labeling of Table IX of Ref.~\cite{Gie01b};
\item[(e)]
$i=11$ also corresponds to the CMM observable
sector embedding 9 according to the labeling
of Table I of Ref.~\cite{Gie01b}, but now $j$ is the
hidden sector embedding
according to the labeling of Table X of Ref.~\cite{Gie01b}.
\een
We remind the reader that CMM observable
sector embeddings 3, 5 and 7 do not appear
because they are equivalent to 1, 4 and 6
respectively, as shown in Ref.~\cite{Gie01b}.

All patterns except Pattern 1.1 have
an anomalous \ux\ factor.
We have determined the FI term for each of the
models in the other 19 patterns.
We find that all models
within a particular pattern have the
same FI term; the
corresponding values
of $\LamX$, defined in
\myref{1c} above, are displayed
in Table \ref{fir}.
As will be discussed in greater detail
in Section~\ref{gcu}, Kaplunovsky \cite{Kap88} has
estimated the string scale to be
\beq
\LamH \approx g_H \times 5.27 \times 10^{17} \mtxt{GeV}
= 0.216 \times g_H m_P.
\label{3d}
\eeq
Using the values in Table~\ref{fir},
it is easy to check that
\beq
\LamH/1.73 \leq \LamX \leq \LamH .
\label{3e}
\eeq
The effective supergravity lagrangian
describing the field theory limit of the string
is nonrenormalizable.  In principle, all superpotential
and K\"ahler potential operators
allowed by symmetries of the underlying theory
should be present.  As discussed in Appendix~A,
there exist field reparameterization invariances
in the effective theory.  These invariances
relate different classical field configurations,
or vacua.  Expansion about a particular vacuum
leads to a nonlinear $\s$ model.
For instance, this is reflected in the
presence of superpotential operators such as \myref{ems}
above, with ever increasing numbers $n$ of Xiggses.
For the nonlinear $\s$ model to be perturbative,
it must be possible to truncate the sequence of operators
at some order $n_{\stxt{max}}$ and obtain a
reasonable approximation to the full theory.
Since the relevant expansion parameter for
nonrenormalizable operators is roughly
$\LamX/m_P$, which from Table~\ref{fir}
lies in the range
\beq
g_H / 8.00 \leq \LamX / m_P \leq  g_H / 4.63 ,
\label{3c}
\eeq
the nonlinear $\s$ model
has a reasonable chance to be perturbative,
provided the unified coupling satisfies
$g_H \lappeq 1$ and the number of operators
contributing to an effective coupling
(such as the $A A^c$ coupling in \myref{ems})
is not too large.  (Generically, the number
of such operators increases with dimension.)

\begin{table}
\begin{center}
\begin{tabular}{ccccc}
Pattern & $\LamX/(g_H m_P)$ & \hspace{50pt} &
Pattern & $\LamX/(g_H m_P)$ \\ \hline 
1.2 & 0.216 & & 2.6, 3.3, 4.6 & 0.170 \\
2.1, 4.2 & 0.125 & & 3.1, 4.3 & 0.148 \\
2.2, 2.3, 4.1 & 0.138 & & 3.2, 4.4, 4.8 & 0.176 \\
2.4 & 0.186 & & 3.4 & 0.181 \\
2.5 & 0.191 & & 4.5, 4.7 & 0.157 \\
\hline
\end{tabular}
\caption{The \ux\ symmetry breaking scale
$\LamX$ for each of the irrep patterns. \label{fir}}
\end{center}
\end{table}

Given the importance of nonvanishing vevs
to the perturbative expansion of the
nonlinear $\s$ model, we next
estimate the range of Xiggs vevs.
We will assume that $g_H \approx 1$ in
\myref{3c}, as suggested by analyses of
the running gauge couplings; for
example, see Section~\ref{gcu} below.
Then from \myref{3c} we have
\beq
\LamX \sim \ord{-1} \; m_P.
\label{3f}
\eeq
Furthermore, we assume that Xiggs fields
have a nearly diagonal K\"ahler potential
at leading order in an expansion about the
vacuum:
\beq
K_{\stxt{Xiggs}} = \sum_i
\bigvev{ {\p^2 K \over \p \phi^i \p \bar \phi^i} }
|\phi^i|^2 + \cdots,
\eeq
with the terms represented by ``$\cdots$'' negligible
in comparison to the explicit terms.  This
assumption is justified by the known
form for the terms in $K$ quadratic in
matter fields for $Z_3$ orbifolds
with nonstandard embedding \cite{twk}, such
as the cases considered here. 
In the limit of vanishing
off-diagonal T-moduli (i.e., $\vev{T^{ij}}=0, \; \forall \; i \not= j$),
\beq
K_{\stxt{quad.-matter}}
= \sum_i {|\phi^i|^2 \over \prod_{j=1,3,5} (T^j + \bar T^j)^{q_j^i}} .
\label{kqm}
\eeq
Here, $q_j^i$ are the {\it modular weights} of the
matter field $\phi^i$:  untwisted states $\ket{K;i}$
have modular weights $q_j^i = \delta_j^i$, while twisted
non-oscillator states $\ket{\tK;n_1,n_3,n_5}$
have modular weights $q_j^i=2/3$ and
twisted oscillator states $\ket{\tK;n_1,n_3,n_5;i}$
have modular weights $q_j^i=2/3+ \delta^i_j$.
Moduli stabilization in the BGW model
gives $\vev{T^j}=1$ or $e^{i\pi/6}$ $\, \forall \; j$.  
Assuming the former value and applying \myref{kqm},
we find
\beq
\bigvev{ {\p^2 K \over \p \phi^i \p \bar \phi^i} }_{\stxt{BGW}} =
\left\{ \begin{array}{lc}
1/2 & \mtxt{untwisted}, \\
1/2^2 & \mtxt{twisted non-oscillator}, \\
1/2^3 & \mtxt{twisted oscillator}.
\end{array} \right.
\eeq
This ignores the possible contribution of terms
$K \ni (c/m_P^2) \; f(T) \; |\phi^i|^2 |\phi^j|^2$, with both
fields $\phi^i,\phi^j$
Xiggses and $f(T)$ a function of the T-moduli.
If we assume $\vev{\phi^{i}} \sim \vev{\phi^{j}}
\sim \LamX$, these quartic terms (which include
$i$-$j$ mixing) are suppressed by $\ordnt{\LamX^2/m_P^2}$
relative to the leading terms.  However,
we still have to estimate 
$\vev{\phi^{i}}$ and $\vev{\phi^{j}}$,
so at the end of our analysis we will
have to check whether or not it was
consistent to neglect these quartic terms.
It is also unclear what the moduli-dependent
function $f(T)$ is, and whether or not
the dimensionless coefficient $c$ is $\ordnt{1}$;
an explicit calculation of such higher order
K\"ahler potential terms from the underlying
string theory apparently remains to be accomplished.

In large radius (LR) stabilization schemes 
such as in Refs.~\cite{Cas90,ILR91},
T-moduli vevs as large as 
$13 \lappeq \vev{T^j} \lappeq 16$
are envisioned.
This greatly affects our estimates
for the Xiggs vevs, since we now have
(for the larger value of $\vev{T^j} = 16$)
\beq
\bigvev{ {\p^2 K \over \p \phi^i \p \bar \phi^i} }_{\stxt{LR}} =
\left\{ \begin{array}{lc}
1/32 & \mtxt{untwisted}, \\
1/32^2 & \mtxt{twisted non-oscillator}, \\
1/32^3 & \mtxt{twisted oscillator}.
\end{array} \right.
\eeq

Let $N$ be the number of Xiggses, $q^X$ be the
average Xiggs \ux\ charge magnitude, $K''$ be the average value
for the Xiggs metric
$\vev{ \p^2 K / \p \phi^i \p \bar \phi^i }$
and $\phi$
be the average value for $|\vev{\phi^i}|$,
where ``average'' is used loosely.
Then from (\ref{1.2},\ref{1c}) we see that
$\vev{D_X} =0$ implies
\beq
\phi \sim \left( N q^X K'' \right)^{-1/2} \LamX .
\label{3a}
\eeq
In Section~\ref{app}
we will see in an explicit
example that the (properly normalized) 
nonvanishing \ux\
charges vary between $1/\sqrt{84}\approx 0.11$
to $6/\sqrt{84}\approx 0.65$.
We take this as an indication that
$1/10 \lappeq q^X \lappeq 2/3$ is reasonable.
In a typical model
there are $3 \times \ordnt{50}$ chiral
matter multiplets.  The number $N$
which may acquire vevs to cancel the FI term
varies from one flat direction to another.
A reasonable range is $1 \lappeq N \lappeq 50$,
given the enormous number of
\gsmc\ singlets in any of the models.

If a single twisted oscillator field $\phi^i$
of charge $1/10$ dominates the FI cancellation
(i.e., $\phi^i$ is the only Xiggs
or all of the other Xiggses have much
smaller vevs so that effectively $N=1$ in \myref{3a}),
then with the BGW T-moduli stabilization
\beq
\phi \sim \sqrt{10 \times 2^3} \; \LamX \sim \ordnt{1} \; m_P,
\eeq
where we have used \myref{3f}.
Such a large vev is certainly troubling.
If the large radius
value $\vev{T^j} \approx 16$ is assumed,
the result is a hundred times worse:
\beq
\phi \sim \sqrt{10 \times 32^3} 
\; \LamX \sim \ord{2} \; m_P.
\eeq
On the other hand, if we had, say, 50 Xiggs fields $\phi^i$
with more average charges of roughly $1/2$
contributing equally to cancel the FI term,
with the typical field a twisted nonoscillator
field, and the BGW stabilization of T-moduli,
\beq
\phi \sim \sqrt{2 \times 2^2 / 50} \; \LamX
\sim \ord{-2} \; m_P.
\eeq
However, for the large radius case,
\beq
\phi \sim \sqrt{2 \times 32^2  / 50} \; \LamX
\sim \ordnt{1} \; m_P.
\eeq
This examination of
\myref{1.2} indicates that
for the BGW stabilization,
Xiggs vevs are naturally
$\ord{-1 \pm 1} \; m_P$.
At the upper end, the $\s$ model would
seem to be in trouble.  The large
radius case appears to be complete catastrophe,
however we arrange cancellation of the FI
term.  To be fair, the quadratic
terms $K \ni (c/m_P^2) \; f(T) \; |\phi^i|^2 |\phi^j|^2$
mentioned above now need to be included
in the estimation of Xiggs vevs, since
they are not of sub-leading order
in the large Xiggs vev limit.

It should be noted, however, that
the principal motivation for the large
radius assumption is to produce appreciable
string scale threshold corrections to
the running gauge couplings, such as
was studied in \cite{ILR91,BL92}; there, the aim was
to achieve gauge
coupling unification at the conventional
value of approximately $2 \times 10^{16}$ GeV.
In a $Z_3$ orbifold compactification, these
large T-moduli dependent threshold corrections coming from
heavy string states are absent \cite{Nothr}.
Nevertheless, it should be clear from
the above analysis that orbifolds which {\it do} have
the T-moduli dependent string threshold corrections
{\it and} a \ux\ factor are likely to also
suffer from a problem of too large Xiggs
vevs in the large radius limit,
because of the noncanonical K\"ahler
potential.

Moderately large, yet perturbative, vevs
such as $\phi \approx m_P/5$
would require large $n$ in
\myref{ems} to generate
significant hierarchies.  This may be a virtue:
in many cases orbifold selection rules and $G$
symmetries require that leading operators
contributing to a given effective low energy
superpotential term have significantly
higher dimension than might be guessed
from \gsmc\ alone.
For example, in the FIQS model (mentioned
in the Introduction) the leading down-type quark masses
come from dimension eleven operators.
(I.e., the effective Yukawa matrix elements
are sums of vevs of seventh degree monomials
of Xiggs fields.)

The sum in \myref{1.2} allows for some terms to
be very small if others are $\ordnt{\LamX}$;
we exploited this possibility in a recent study of 
effective quark Yukawa couplings 
induced by Xiggs vevs of
rather different scales \cite{Gie01a}.
Such hierarchies in Xiggs vevs
remain to be (dynamically) motivated from a detailed
study of an explicit scalar potential
which lifts the D-moduli flat directions
\cite{GG00} mentioned in the Introduction.
The existence of these flat directions
means that the upper bound estimates
made here for Xiggs vevs are not at all robust.  Xiggs of
opposite \ux\ charge may be ``turned on''
along a particular flat direction (as in
the FIQS model).  In that case their contributions
partially cancel each other; it is technically
possible for the Xiggs vevs to be made arbitrarily
large as a result.  Of course, this would quickly
spoil the nonlinear $\s$ model expansion.

The BSL-I model mentioned in the Introduction
belongs to Pattern~1.2
and is equivalent to one of the
models 6.1-3 listed under that pattern in Table~\ref{tb1}.
(CMM found that the BSL-I model observable
sector embedding was equivalent to CMM 7,
and in \cite{Gie01b} we showed that CMM 7
is equivalent to CMM 6.)
In \cite{GG00} it was noted that the FIQS
model suffers from a problem of light
diagonal T-moduli masses; the conclusions
made there do not depend on the choice
of (hidden $SO(10)$ preserving)
flat direction, and therefore hold
for other vacua of the BSL-I model,
such as those studied by Casas and Mu\~noz \cite{CM88}.
As will shortly be
explained, the light mass
problem is a consequence of
having $G_C=SO(10)$ charged matter fields
only in the untwisted sector.  This observation
extends to {\it all}\ models of Pattern 1.2,
as well as to the models of Pattern 1.1.
Because BGW stabilize the diagonal T-moduli
with nonperturbative effects in the hidden
sector (i.e., gaugino condensation),
they simultaneously derive an effective
(soft) mass term for these fields \cite{BGW}.
If the effective moduli masses are much
larger than the gravitino mass, the 
{\it cosmological moduli problem}
\cite{CFKRR83} can be avoided.  In the
BGW effective theory, one finds for the
diagonal T-moduli
\beq
m_T \approx 2 {|\bGS - b_C| \over |b_C|} m_{\tilde G},
\label{mtm}
\eeq
where $b_C$ is the beta function coefficient
for the condensing group $G_C$, $m_{\tilde G}$
is the gravitino mass and $\bGS$ is the {\it Green-Schwarz
counterterm coefficient,} a quantity whose origin is not
important to the present discussion, but which
is briefly explained in Appendix A.  If $\bGS/b_C \approx 10$,
then $m_T \approx 20 m_{\tilde G}$; it was
argued by BGW, and others \cite{GLM99}, that this
may be heavy enough to resolve the
cosmological moduli problem.

However,
as pointed out in Ref.~\cite{GG00},
if $G_C$ has only trivial irreps in the twisted sector,
$\bGS = b_C$.  The T-moduli are massless
to the order of the approximation made
in \myref{mtm}, and the moduli problem
reappears with a vengence.  To see how $\bGS = b_C$
occurs in Patterns 1.1 and 1.2, it is only necessary
to note a few simple facts.  In Appendix A
we use well-known results to demonstrate that,
for the class of models studied here,
the Green-Schwarz coefficient is given by
\beq
\bGS = \tbeta{a} - 2 \sum_{\rho \in \stxt{tw}} X_a(R^\rho),
\qquad \forall \; G_a \in \GNA,
\label{gsc}
\eeq
where $\tbeta{a}$ is the $\beta$ function
coefficient (given by \myref{bcd} with
$G_C \to G_a$) calculated from the entire
pseudo-massless spectrum of a given
model, and the index $\rho$ runs only over
twisted matter chiral supermultiplet irreps.
In Table~\ref{tbgs} we show $\bGS$ for each of the
twenty patterns; the value is universal to all
models in a given pattern.
From \myref{gsc} it is clear that
$\bGS = \tbeta{a}$ for $G_a$ with
only trivial irreps in the twisted sector.
This occurs for $SO(10)$ in Patterns 1.1 and 1.2,
so that one has $\bGS = \tbeta{10}$;
we also recall $G_C = SO(10)$ in these patterns;
this leads to vanishing T-moduli masses
in \myref{mtm} if $b_C = \tbeta{10}$.
One might hope to get around this
by giving some of the $SO(10)$ charged matter
$\ordnt{\LamX}$ vector mass couplings so that $b_C$, the
effective coefficient which appears in
the theory below the scale $\LamX$,
is different from $\tbeta{10}$.
Pattern 1.1 does not contain $SO(10)$ charged
matter so this is fruitless.  In Pattern 1.2,
the $SO(10)$ matter is in $16$s, which have
as their lowest dimensional invariant
$(16)^4$.  To have effective vector masses for these
states from superpotential terms would require
breaking $SO(10)$.  
We leave these issues to further research.
Another way resolve the light moduli problem
in Patterns 1.1 and 1.2 would involve
alternative inflation scenarios.  For example,
light moduli could be diluted via
the thermal inflation 
of Lyth and Stewart \cite{LS96}.
Lastly, we note that the BGW result \myref{mtm}
is obtained in an effective theory which does not
account for a \ux; until it is understood how the
BGW effective theory is modified in the presence of
a \ux\ factor \cite{Gai01}, firm conclusions about the Pattern
1.2 models cannot be drawn.  (Recall that
Pattern 1.1 has no \ux\ factor.)

\begin{table}
\begin{center}
\begin{tabular}{ccccc}
Pattern & $\bGS$ & \hspace{50pt} &
Pattern & $\bGS$ \\ \hline 
1.1 & -24 & & 1.2, 2.1 & -18 \\
2.2-5, 3.1 & -15 & & 3.2-4, 4.1-4, 4.6 & -12 \\
2.6, 4.5, 4.7-8 & -9 & & & \\ 
\hline
\end{tabular}
\caption{Green-Schwarz coefficients. \label{tbgs}}
\end{center}
\end{table}

The values for $\bGS$ are problematic for
more than just the Pattern 1.1 and 1.2 models.
For example, in the $G_C=SU(5)$ Patterns 2.2-5,
the Green-Schwarz coefficient is $\bGS = -15$ and
we can constrain $-15 \leq b_C \leq -6$.
The bound $-15$ comes from a scenario
of pure $SU(5)$; i.e., no matter.  Pattern
2.2 for instance allows for the possibility
that the vector-like
$3(5 + \bar 5)$ matter acquires mass at $\LamX$,
so that effectively there is no $SU(5)$ charged
matter in the running which dynamically generates the
condensation scale.  The bound $-6$ comes from
the ``marginal'' case of very low $\LamC$
discussed in the Introduction.  For this
range of $b_C$ we have
from \myref{mtm}
\beq
0 \leq m_T \leq 3 m_{\tilde G}.
\eeq
From the arguments of \cite{BGW,GLM99},
the T-moduli mass appears to be too light
even in the marginal case $b_C= -6$,
which gives the upper bound for $m_T$.
Taking the $b_C = -6$ limit for each of the
values of $\bGS$ (except $\bGS= -24$ which
corresponds to Pattern 1.1 discussed above---where
it seems $m_T \approx 0$ is unavoidable),
we find upper bounds of $m_T^{\stxt{max}}/m_{\tilde G}
\approx 4,3,2,1$ for $\bGS = -18,-15,-12,-9$ respectively.
Thus, the light T-moduli mass problem is a
general feature of the \bsa\ models.

Most of the 20 patterns contain $(3 + \bar 3,1)$
representations under $SU(3)_C \times SU(2)_L$.  
It is necessary to find a
vacuum solution which gives these fields
vector mass couplings at a high enough scale.
The greater the number of such pairs,
the more difficult this is to achieve,
since one must simultaneously
avoid high scale supersymmetry
breaking;  more and more
fields must be identified as Xiggses in order
to give all of the required effective
supersymmetric mass couplings.  As each new
Xiggs is introduced, it is harder to avoid
nonzero F-terms at the scale $\LamX$.
Similarly, large vector masses are generally
required for the many additional
$(1,2)$ and $(1,1)$ representations
present in all of the models.  The electroweak
hypercharges of these representations
depend on how the several {\uone}s
are broken in choosing a D-flat direction.
States with exotic electric charge (i.e.,
leptons with fractional charges and quarks
which may form fractionally charged color
singlet bound states) typically occur.
We will address constraints on the presence
of such matter in Section~\ref{app} below.

The distinction between observable and hidden
sectors is blurred by twisted states
in nontrivial representations of both
$G_O$ and $G_H$.  
Gauge interactions
communicating with both sectors
are a well-known effect in orbifold
models.  Communication via {\uone}s
was for example noted in Refs.~\cite{IMNQ88,CKM89,FINQ88a},
while the occurence of states in nontrivial
representations of both observable and
hidden nonabelian factors has been
noted in other orbifold constructions,
for example in a $Z_3 \times Z_3$ model
in Ref.~\cite{BL92}.
Cases 2 through 4 (cf. Table \ref{gcs}) have
at least one hidden $SU(2)$ factor (which
we denote $SU(2)'$),
and $(1,2,2)$ representations
under $SU(3)_C \times SU(2)_L \times SU(2)'$
occur in several of the patterns.  No 
$(\bar 3, 1, 2)$ representations occur, so
it is not possible to use $SU(2)'$
to construct a left-right symmetric model
in any of the 175 models studied here.
(Left-right symmetric models would place
the $u^c$- and $d^c$-type quarks
in $(\bar 3, 1, 2)$ representations.)
All 175 models
contain twisted states 
in nontrivial irreps of $SU(3)_C \times SU(2)_L$
charged under
{\uone}s contained in $G_H$.  

It is an
interesting question to what degree
these features might communicate
supersymmetry breaking to the observable
sector.  A similar scenario has
been considered by
Antoniadis and Benakli \cite{AB92}.
Specifically, they examined hidden
sector matter with supersymmetric
masses $M$ and a soft mass $\delta M$
splitting the matter scalars from fermions,
gauginos from vector gauge bosons,
with the assumption $\delta M \ll M$;
this ``hidden'' matter was also assumed
to be in nontrivial irreps of $\GSM$.
They found significant
contributions to the soft terms
which break supersymmetry in the
MSSM.  To evaluate the implications
of such gauge mediation of supersymmetry
breaking in the 175 models at hand
requires a significant extension of
their results, given the strong dynamics
of the hidden sector in a gaugino
condensation scenario; much of
the hidden sector matter now consists
of bound states of $G_C$ which
are $\GSM$ neutral (certainly
the case for those condensates which
acquire the supersymmetry
breaking nonvanishing vevs) yet contain
particles in nontrivial $\GSM$ irreps.
We leave these matters to future research.

The generic presence of an anomalous \ux\
has implications for low energy 
supersymmetric models which aim
to be ``string-inspired'' or ``string-derived.''
The effective theory in 
the low energy limit is obtained
by integrating out states which get
large masses due to the \ux\ FI term.
The surviving spectrum
of states will generally contain superpositions
of the original states, mixing
the various sectors.  Thus, assigning
each state in the MSSM to a {\it definite}
sector (i.e., the untwisted sector or one
of the 27 $(n_1,n_3,n_5)$ twisted sectors)
is in many cases inconsistent with 
the mixing which occurs in the presence
of a \ux, as was for instance remarked recently
in Ref.~\cite{Den01}.  Mixings of sectors
was considered for quarks, for example,
in the FIQS model and in the toy model
of Ref.~\cite{Gie01a}.
In addition to modified properties for
the spectrum, integrating out the
massive states will modify the interactions
of the light fields and create
threshold effects for running couplings.
These threshold effects can be
large due to the large number of
extra states, and need to be
considered in any analysis of gauge
coupling unification, for example.

\mysection{Hypercharge}
\label{hyc}
\vspace{-30pt}

\subsection{Normalization in GUTs}
An important feature of GUTs
is that the \uone\ generator corresponding to
electroweak hypercharge does not have arbitrary normalization.
This is because the hypercharge generator is embedded into
the Lie algebra of the GUT group.
That is, $\GGUT \supset \GSME$.
The {\it unified normalization}
is most clear when one identifies a Cartesian basis for the
GUT group generators $T^a$ for a given representation $R$:
\beq
\tra{R} T^a T^b = X(R) \; \delta^{ab}.
\eeq
The normalization prevalent in phenomenology has
$X(F)=1/2$ for an $SU(N)$
fundamental representation $F$.  Because of the GUT symmetry,
the interaction strength of a gauge particle with matter
is given by
\beq
g_U(\mu) \; T^a, \quad \forall \; a,
\label{4a}
\eeq
where $g_U(\mu)$ is the running coupling for the GUT
gauge group at the scale $\mu \geq \LamU$,
with $\LamU$ the unification scale.  One of the
$T^a$, say $T^1$, is then identified with the
electroweak hypercharge generator.  However,
to obtain the usual eigenvalues for MSSM particles
(e.g., $Y=1$ for $e^c$) we 
generally must rescale the generator:
\beq
Y \equiv \sqrt{k_Y} \; T^1 .
\label{3b}
\eeq
The reason for writing the rescaling
constant in this way will become clear
below.  Because of (\ref{4a},\ref{3b}), 
the hypercharge coupling $g_Y(\mu)$ will
be related to $g_U(\mu)$ at the boundary scale $\LamU$.  More
precisely,
\beq
g_U(\LamU) \; T^1 = g_Y(\LamU) \; Y 
=  \sqrt{k_Y} \; g_Y(\LamU) \; T^1,
\eeq
since the interaction strength should not depend on normalization
conventions for the generators.
We maintain the GUT normalization for the generators $T^a$
which correspond to the unbroken $SU(2)$ and $SU(3)$ groups,
so that there are no rescalings analogous to \myref{3b} for
these two groups; their running couplings are denoted by
$g_2(\mu)$ and $g_3(\mu)$ respectively.  Because of \myref{4a},
they too must be matched to the boundary value $g_U(\LamU)$
when $\mu=\LamU$; thus, we obtain the well-known
GUT boundary conditions
\beq
g_3(\LamU) = g_2(\LamU) = \sqrt{k_Y} \; g_Y(\LamU) 
= g_U(\LamU).
\label{4c}
\eeq

For example, consider an SU(5) GUT \cite{GG74}.
The $SU(5)$ embedding of hypercharge,
which we write as $T^1$, can
be determined from the requirement
that $\tr (T^1)^2 = 1/2$ for a
fundamental or antifundamental irrep.  For example,
\beq
T^1 = {1 \over \sqrt{60}} \diag ( -3,
-3, 2, 2, 2), \qquad \mtxt{for} \qquad \bar 5 =
{L \choose d^c} .
\label{gyc}
\eeq
Here, $L$ is a $(1,2)$ lepton,
and $d^c$ is a $(\bar 3,1)$ down-type quark,
where we denote \nasm\ quantum numbers.
On the other hand, the electroweak normalization
has by convention
\beq
Y = {1 \over 6} \diag ( -3,
-3, 2, 2, 2)
\eeq
for the same set of states.
Since $Y = \sqrt{5/3} \; T^1$,
we see from \myref{3b} that
\beq
k_Y = 5 / 3 .
\label{s5y}
\eeq
It is this value which, when assumed in \myref{4c},
yields the amazingly successful
gauge coupling unification in the MSSM,
detailed for example in Refs.~\cite{MSSMu,LP93}.

\subsection{Normalization in String Theory}
As in GUTs, the normalization of \uone\
generators in string-derived field theories
requires care.  Above,
we have alluded to the fact that
gauge coupling unification at the 
heterotic string scale $\LamH$ is
a prediction of the underlying theory \cite{Gin87}.
Just as
in GUTs, unification of the hypercharge coupling with the
couplings of other factors of the gauge symmetry
group $G$ corresponds to a particular normalization.
However, the unified normalization of hypercharge
is often different than the one which appears in $SU(5)$
or $SO(10)$ GUTs; in fact it is often difficult
or impossible
to obtain \myref{s5y}.  Examples
of this hypercharge normalization ``difficulty''
will be examined below.
We will show how the unified normalization can be identified
from very simple arguments.
In the process
we will make it very clear why, in the class of
orbifold models considered here, nonstandard hypercharge
normalization is generic and fractionally charged exotic matter
is abundant.

It was noted in Section \ref{mss} that the
basis \myref{2g} is larger by a factor
of two than the phenomenological normalization.
Thus, $\tr (T^a)^2 = 2$ for
an $SU(N)$ fundamental representation.
For instance, consider an untwisted $SU(2)_L$ doublet
with respect to $\alpha_{1,1}$ in \myref{2m} above,
for CMM 2 observable sector embeddings.
(The embedding label here corresponds to Table I of
Ref.~\cite{Gie01b}.)
The lowest and highest weight states are respectively
\beqa
K_1 & = & (0,1,0,0,0,1,0,0;0,\ldots,0), \nnn
K_2 & = & K_1 + \alpha_{1,1} = (1,0,0,0,0,1,0,0;0,\ldots,0).
\label{aad}
\eeqa
Using Eqs.~(\ref{2m},\ref{wtd}),
the corresponding weights are $\mp 1$; this gives
$\tr (H_1^1)^2 = 2$,
where $H_1^1 = T^3$,
the isospin operator of
$SU(2)_L$.  To get to the phenomenological
normalization, we should rescale generators
by $1/2$.  Thus, instead of \myref{2g},
we define our properly normalized Cartan generators
$\hat H_a^i$ according to
\beq
\hat H_a^i = \sum_{I=1}^{16} \hat h_a^{iI} H^I
\equiv \sum_{I=1}^{16} \half \alpha_{ai}^I H^I .
\label{4d}
\eeq
In this case, the sixteen-vectors $\hat h_a^i$
satisfy 
\beq
(\hat h_a^i)^2 = 1/2.
\label{4i}
\eeq
It is hardly surprising that the 
properly normalized generator $\hat Q_a$
of $U(1)_a$ must also satisfy 
$(\hat q_a)^2 = 1/2$,
where $\hat q_a$ is the sixteen-vector
appearing in \myref{Qdf},
but now with a special normalization.  After all,
the generator of $U(1)_a$ just corresponds
to a different linear combination of the
\eetee\ Cartan generators $H^I$,
and taking a linear combination of
the same norm is the logical choice.
If, on the other hand, we work with
a generator $Q_a = \sqrt{k_a} \hat Q_a$,
then it follows that $q_a^2 = k_a/2$.
This is one way
of motivating the ``affine level''
of a \uone\ factor:
\beq
k_a = 2 \sum_{I=1}^{16} (q_a^I)^2 .
\label{kad}
\eeq
(This relation also follows from a
consideration of the double-pole
Schwinger term which occurs in the
operator product of \uone\ currents 
in the underlying conformal field
theory \cite{FIQS90,DFMR96,KN97,Die97},
details which we have purposely
avoided here.)
The unified normalization,
where nonabelian Cartan generators $\hat H_b^i$
and \uone\ generators $\hat Q_a$ have in
common $(\hat h_b^i)^2 = \hat q_a^2 = 1/2$,
corresponds to $k_a = 1$.

\subsection{${\bf SU(5)}$ Hypercharge Embeddings}
Note that the generator
\beq
Y_1 = \sum_{I=1}^{16} y_1^I H^I, \qquad
y_1 = {1 \over 6} (-3,-3,2,2,2,0,0,0;0,\ldots,0),
\label{aac}
\eeq
satisfies $k_{Y_1} = 5/3$, is orthogonal to the
$SU(3)_C \times SU(2)_L$ roots in \myref{nz32}, and
has nonzero entries only in the subspace where
the $SU(3)_C \times SU(2)_L$ roots have nonzero
entries.  Furthermore, it gives 
$Y_1 = y_1 \cdot K_{1,2} = -1/2$ to
the doublet in \myref{aad}, corresponding to the
lepton doublets $L$ or the $H_d$ Higgs
doublet of the MSSM.
The \bsa\ models with observable sector
embedding CMM 2
also include $(\bar 3, 1)$ states
in the untwisted sector with weights
\beq
K_{3,4,5} = (0,0,\underline{1,0,0},-1,0,0;0,\ldots,0).
\label{aaf}
\eeq
These have $Y_1=y_1 \cdot K_{3,4,5}= 1/3$,
corresponding to the $d^c$ states.  Finally,
the untwisted sector contains $(3,2)$ states
with weights
\beq
K_{6,\ldots,11} =
(\underline{-1,0},\underline{-1,0,0},0,0,0;0,\ldots,0)
\label{4z}
\eeq
which have $Y_1=y_1 \cdot K_{6,\ldots,11}=1/6$, corresponding to the
quark doublets $Q$.
Thus, the untwisted sector
contains a $\bar 5$ and an incomplete $10$ under
the ``would-be'' $SU(5)$ into which we
wish to embed $SU(3)_C \times SU(2)_L \times U(1)_{Y_1}$,
taking \myref{aac} to be
the hypercharge generator.
The fact that the
$e^c$ and $u^c$ representations needed to
fill out the $10$ irrep are not present in
the untwisted sector is a troubling feature
which is generic to the 175 models studied
here.

In Table~\ref{unpat} (Appendix B) we display the
irreps present in the untwisted sector for
each of the twenty patterns.  In no case do
we have the required irreps to build
a $10$ of $SU(5)$.  In those cases where
one finds $(3,2) + (\bar 3,1)$, the states
which are singlets of the observable
$SU(3) \times SU(2)$ are in nontrivial
irreps of the hidden sector group.  One
could imagine breaking the hidden sector
group and using a singlet of the surviving
group to give the necessary $(1,1)$ irrep
to fill out a $10$.  For instance, in
Pattern 2.2, the $(1,1,1,2)$ irrep,
a $2$ of the hidden $SU(2)'$,
would give two singlets if we break
$SU(2)'$ with nonvanishing vevs
for a pair of twisted
sector $(1,1,1,2)$ irreps along a
D-flat direction.  (A pair is required
to have vanishing D-terms for $SU(2)'$.)
We would thereby obtain three
generations of two $(1,1,1)$ irreps
with respect to the surviving
nonabelian gauge symmetry
$SU(3) \times SU(2) \times SU(5)$,
where the $SU(5)$ shown here is the
hidden condensing gauge group.  However,
the untwisted $(1,1,1,2)$ irrep which
gives these states has an \eetee\ weight vector
$K$ of the form $K=(0;\beta), \; \beta \in \elat$,
since it is an untwisted state
charged under the hidden sector gauge
group.  Then it has vanishing charge
with respect to the generator $Y_1$
according to \myref{aac}, rather than
the required $Y_1=1$.
We could overcome this by modifying $y_1$
to have nonzero entries in the hidden
sector portion, represented by $0,\ldots,0$
in \myref{aac}.  However, according to
\myref{kad}, this would increase $k_Y$
over the value of $5/3$ which $y_1$
gives.  Moreover,
it can be seen that one never has enough
untwisted $(\bar 3,1)$ irreps to give three
generations of both $u^c$- and $d^c$-type
quarks, and that untwisted $(1,2)$ irreps always occur
when an untwisted $(\bar 3,1)$ is present.  Thus,
even if we break the hidden gauge
group, use a singlet to complete the
$10$, are willing to consider $k_Y > 5/3$,
and find the $(\bar 3,1)$ has $Y_1 = -2/3$
so that it fits into a $10$,
the $(1,2)$ would
stand for an incomplete $\bar 5$.
It is inevitable that we use states from
the twisted sectors to fill out the MSSM;
as we have already alluded to in Section~\ref{mss},
twisted states have unusual \uone\ charges 
(partly) because
the \eetee\ weights are shifted by the
embedding vectors $E(n_1,n_3,n_5)$.

Let us now examine the relationship
of \myref{aac} to $SU(5)$.
To begin with we
relabel the $SU(3) \times SU(2)$ simple roots in
(\ref{2m},\ref{2.39}) as
\beq
\alpha_1 \equiv \alpha_{1,1}, \qquad
\alpha_2 \equiv \alpha_{2,1}, \qquad
\alpha_3 \equiv \alpha_{2,2}.
\label{4e}
\eeq
These may be supplemented by a fourth \eetee\ root
\beq
\alpha_4 = (0, 1, -1, 0,0,0,0,0;0,\ldots,0)
\label{aaa}
\eeq
to give the correct Cartan matrix for $SU(5)$,
according to \myref{2c}.  In
this way we embed $SU(3) \times SU(2)$ into
a would-be $SU(5)$ subgroup of the observable $E_8$
factor of \eetee.
A (properly normalized)
basis $\hat H_1, \dots, \hat H_4$ for the Cartan
subalgebra of the would-be $SU(5)$
is given in terms of the \eetee\ Cartan generators
$H^I$ according to the methods described
in Section~\ref{mss}, supplemented by the
normalization considerations which led to
\myref{4d}.  That is, we take linear combinations
described by sixteen-vectors $\hat h^i = \alpha_i/2$,
so that
\beq
\hat H^i = \sum_{I=1}^{16} \half \alpha_i^I H^I.
\label{4g}
\eeq
However, when we decompose
$SU(5) \supset SU(3) \times SU(2) \times U(1)$
we want to take the \uone\ generator to be
orthogonal to the generators $\hat H^{1,2,3}$
associated with the
simple roots \myref{4e},
unlike $\hat H^4$.
(This is the analogue of \myref{2s}.)
We thus
make a change of basis, keeping $\hat h^i = \alpha_i/2$
for $i=1,2,3$ while taking the fourth vector
to be an orthogonal linear combination of the
four simple roots:
\beq
y = \sum_{i=1}^4 r_i \alpha_i, \qquad
\mtxt{where}
\qquad y \cdot \alpha_i = 0, \quad i=1,2,3.
\label{aab}
\eeq
The orthogonality constraint in \myref{aab}
and the fact that $\alpha_i^I=0$ for 
$I=6,\ldots,16$ requires
\beq
y = (a,a,b,b,b,0,0,0;0,\ldots,0),
\label{4m}
\eeq
while $\sum_I \alpha_i^I =0$ requires
$2a= -3b$.  From here it is easy to check
that with normalization $k_Y=5/3$, we have
$y=y_1$, Eq.~\myref{aac}.
Thus we see that $y_1$
corresponds to a natural completion of the $SU(3) \times SU(2)$
roots \myref{4e} into a would-be $SU(5)$ subgroup
of the observable $E_8$.  We note that
\myref{4m} has the form of a {\it minimal
embedding} of hypercharge, in the spirit
of the analysis carried out in \cite{DFMR96}.

\begin{figure}
\begin{center}
\unitlength=1mm
\begin{picture}(100,30)
\put(10,20){\makebox(0,0){$\alpha_1$}}
\put(20,20){\makebox(0,0){$\alpha_4$}}
\put(30,20){\makebox(0,0){$\alpha_2$}}
\put(40,20){\makebox(0,0){$\alpha_3$}}
\put(10,25){\circle*{1}}
\put(20,25){\circle*{1}}
\put(30,25){\circle*{1}}
\put(40,25){\circle*{1}}
\put(10,25){\line(1,0){10}}
\put(20,25){\line(1,0){10}}
\put(30,25){\line(1,0){10}}
\put(25,10){\makebox(0,0){Case 1}}
\put(60,20){\makebox(0,0){$\alpha_1$}}
\put(70,20){\makebox(0,0){$\alpha_4$}}
\put(80,20){\makebox(0,0){$\alpha_3$}}
\put(90,20){\makebox(0,0){$\alpha_2$}}
\put(60,25){\circle*{1}}
\put(70,25){\circle*{1}}
\put(80,25){\circle*{1}}
\put(90,25){\circle*{1}}
\put(60,25){\line(1,0){10}}
\put(70,25){\line(1,0){10}}
\put(80,25){\line(1,0){10}}
\put(75,10){\makebox(0,0){Case 2}}
\end{picture}
\end{center}
\caption{Would-be $SU(5)$ Dynkin diagrams.}
\label{wb5}
\end{figure}
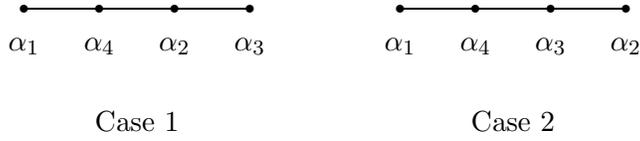

Now we come to the origin of the subscript in \myref{aac}.
It turns out that \myref{aaa} is not the unique $E_8$
root which may be appended to  $\alpha_1, \alpha_2, \alpha_3$
to obtain the simple roots of an $SU(5)$ subalgebra
of the observable $E_8$.  The two ways that a
supposed $\alpha_4$ could be related to the
roots $\alpha_1, \alpha_2, \alpha_3$ are
shown in the Dynkin diagrams of Figure~\ref{wb5}.
A line connecting $\alpha_i$ to $\alpha_j$
indicates $\alpha_i \cdot \alpha_j = -1$;
if not connected by a line, 
$\alpha_i \cdot \alpha_j = 0$.

We define $y$ as in \myref{aab}, except that
now we allow $\alpha_4$ to be any observable
$E_8$ root (i.e., $\alpha_4 = (\beta;0)$,
$\beta \in \elat$, $\beta^2=2$)
consistent with Figure~\ref{wb5}.
We simultaneously demand $2y^2=5/3$, corresponding
to $k_Y=5/3$ from \myref{kad}.  This gives
solutions:
\beqa
y & = & \pm {1 \over 6} (3 \alpha_1 + 4 \alpha_2
+ 2 \alpha_3 + 6 \alpha_4) \qquad \mtxt{Case 1}, 
\label{4y} \\
y & = & \pm {1 \over 6} (3 \alpha_1 + 2 \alpha_2
+ 4 \alpha_3 + 6 \alpha_4) \qquad \mtxt{\rm Case 2}.
\label{4x}
\eeqa

In each of the 175 models we consider here, 
the only $(3,2)$ representations
under the observable $SU(3) \times SU(2)$ are contained
in the untwisted sector, and they all take the form
\myref{4z}.  To accomodate the MSSM
we require that this representation have 
$Y = y \cdot K_{6,\ldots,11} =1/6$.
It suffices to demand this for any of the
six $K_i$ since by \myref{aab}
\beq
(K_i + \alpha_j)\cdot y = K_i \cdot y,
\qquad \forall \quad i=6,\ldots,11, \quad j=1,2,3.
\eeq
(Recall from the discussion in Section \ref{mss}
that the weights $K_{6,\ldots,11}$ are related
to each other by the addition
of $SU(3) \times SU(2)$ roots.)
We choose to employ 
\beq
K_6=(-1,0,-1,0,0,0,0,0;0,\ldots,0).
\eeq
It is easy to check that for Eq.~\myref{4y},
$K_6 \cdot y = 1/6$ imposes  
\beq
K_6 \cdot \alpha_4 = \left\{
\begin{array}{cc}
4/3 & (+), \\
1 & (-).
\end{array}
\right.
\eeq
Since $\alpha_4$ can only have integral or
half-integral entries, we must take the
negative sign in \myref{4y} 
and $K_6 \cdot \alpha_4=1$.
For Eq.~\myref{4x},
$K_6 \cdot y = 1/6$ imposes 
\beq
K_6 \cdot \alpha_4 = \left\{
\begin{array}{cc}
1 & (+), \\
2/3 & (-).
\end{array}
\right.
\eeq
Now we must take the
positive sign in \myref{4x}.
To summarize, imposing that the quark doublet have
$Y=1/6$ constrains $\alpha_4$ to satisfy
the additional constraint
\beq
K_6 \cdot \alpha_4 = 1
\eeq
and determines the signs in (\ref{4y},\ref{4x}):
\beqa
y & = & - {1 \over 6} (3 \alpha_1 + 4 \alpha_2
+ 2 \alpha_3 + 6 \alpha_4) \qquad \mtxt{Case 1}, 
\\
y & = & {1 \over 6} (3 \alpha_1 + 2 \alpha_2
+ 4 \alpha_3 + 6 \alpha_4) \qquad \mtxt{Case 2}.
\eeqa

As noted briefly in Section~\ref{mss}, the ordering
by which nonzero \eetee\ roots are determined to
be positive is arbitrary.  A particular 
{\it lexicographic ordering} for the first
$E_8$ can be specified by an eight-tuple
$(n_1,n_2,\ldots,n_8)$.  Here, $n_1$ tells us
which entry should be checked first, $n_2$
tells us which entry should be checked second,
etc.  For example, $(8,7,6,5,4,3,2,1)$ would
instruct us to determine positivity by
reading the entries of a given $E_8$ root
vector backwards, right to left.
It is easy to see that several
{\it lexicographic orderings} are consistent with
$\alpha_1, \alpha_2, \alpha_3$ being regarded
as positive; in fact, the number of such
orderings is 3360.  
Our final restriction on $\alpha_4$ is
that it for one of these 3360
orderings, $\alpha_4$ is also positive.
This is necessary if it is to be regarded
as a simple root of a would-be $SU(5)$.

When all of the conditions described above are taken
into account, the complete list
of observable $E_8$ roots $\alpha_4$ and the corresponding
vectors $y$ which result can be
determined by straightforward analysis
of the 240 nonzero $E_8$ roots.  The
results are given in Table \ref{su5e}.
We label the four additional
$y$ solutions according to:
\beqa
y_2 & = & {1 \over 6} (0,0,-1,-1,-1,-3,-3,-3;0,\ldots,0), \nnn
y_{3,4,5} & = & 
{1 \over 6} (0,0,-1,-1,-1,\underline{-3,3,3};0,\ldots,0).
\eeqa
In what follows, we refer to $Y_i$, $i=1,\ldots,5$, as the
five possible $SU(5)$ {\it embeddings} of the hypercharge
in the \bsa\ models.

\begin{table}[ht!]
\begin{center}
$$
\begin{array}{cc}
\alpha_4 & y \\ \hline
(0,1,-1,0,0,0,0,0;0,\ldots,0) & 
{1 \over 6} (-3,-3,2,2,2,0,0,0;0,\ldots,0) \\
(-\half,\half,-\half,\half,\half,\half,\half,\half;0,\ldots,0) &
{1 \over 6} (0,0,-1,-1,-1,-3,-3,-3;0,\ldots,0) \\
(-\half,\half,-\half,\half,\half,
	\underline{\half,-\half,-\half};0,\ldots,0) &
{1 \over 6} (0,0,-1,-1,-1,\underline{-3,3,3};0,\ldots,0) \\
(-1,0,0,0,1,0,0,0;0,\ldots,0) &
{1 \over 6} (-3,-3,2,2,2,0,0,0;0,\ldots,0) \\
(-\half,\half,-\half,-\half,\half,
	\underline{-\half,\half,\half};0,\ldots,0) &
{1 \over 6} (0,0,-1,-1,-1,\underline{-3,3,3};0,\ldots,0) 
\vspace{5pt} \\
\hline
\end{array}
$$
\end{center}
\caption{Observable $E_8$ roots which embed
$SU(3)_C \times SU(2)_L$ into a would-be
$SU(5)$. \label{su5e}}
\end{table}

A model must also have $Y$ non-anomalous
for it to survive unmixed with other \uone\ factors below
$\LamX$.  Many models have a trace anomaly for one or
more of the five $Y_i$.  This would not
occur if complete $SU(5)$ irreps
were present.
We have already seen that the untwisted
sector does not contain complete would-be
$SU(5)$ irreps for any of the 175 models
(cf. Table~\ref{unpat}).
Of course, whether or not $Y_1$ is
anomalous in those models also depends on the matter content
of the twisted sectors.  This in turn depends on the
hidden sector embedding through \myref{tps};
consequently, each of the 175 models must be
studied separately.

We have determined the charges of all matter
irreps with respect to $Y_i \; (i=1,\ldots,5)$
for all of models.
In those models where a given $Y_i$
is not anomalous, the MSSM particle spectrum is never
accomodated.  That is not to say that we do not have
enough $(3,2)$s, $(\bar 3,1)$s, $(1,2)$s and $(1,1)$s;
in fact, we typically have too many of the latter
three types, as can be seen from 
Tables~\ref{pt1}-\ref{pt4}.
The difficulty comes in their
hypercharge assignments when we take $Y$ to be one of
the five $Y_i$.  Although there are always a few
irreps with the right hypercharges, there are never enough.

As suggested by the discussion in Section~\ref{mss},
the origin of bizzare
hypercharges with respect to the $SU(5)$ embeddings
$Y_i$ is due to the fact that twisted states
generically have \eetee\ weights on a shifted
lattice, as is apparent in \myref{tps}.
To further understand these matters, 
we now discuss the decomposition
of the two lowest lying $E_8$ representations,
of dimension 248 and 3875 respectively.
The decomposition of these irreps under
$E_8 \supset SU(5)$ is tabulated,
for instance, in the review by Slansky~\cite{Sla81}.
We identify this $SU(5)$ as the
subgroup of $E_8$ in which
irreps of $\GSM$ are embedded.  The decompositions
are (numbers in parentheses denote $SU(5)$ irreps)
\beqa
248 & = & 24(1) + (24) + 10(5+\bbar{5}) + 5(10 + \bbar{10}), \nnn
3875 & = & 100(1) + 65(5 + \bbar{5}) + 50(10 + \bbar{10})
+ 5(15 + \bbar{15}) + 25(24) + 5(40+\bbar{40}) \nnn
& & \quad {} + 10(45 + \bbar{45}) + (75).
\label{edc}
\eeqa
Although these are real representations, a chiral
four-dimensional theory is obtained by compactification
on a quotient manifold (i.e., the $Z_3$ orbifold),
a mechanism pointed out some time ago \cite{CGS}.
Also from Slansky, we take the decomposition of
the $SU(5)$ irreps shown in \myref{edc} with
respect to $SU(5) \supset SU(3) \times SU(2) \times U(1)$,
with the standard electroweak normalization for the
\uone\ charge given in the last entry:
\beqa
1 & = & (1,1,0) \nnn
5 & = & (1,2,1/2) + (3,1,-1/3) \nnn
10 & = & (1,1,1) + (\bar 3,1,-2/3) + (3,2,1/6) \nnn
15 & = & (1,3,1) + (3,2,1/6) + (6,1,-2/3) \nnn
24 & = & (1,1,0) + (1,3,0) + (3,2,-5/6)
+ (\bar 3, 2, 5/6) + (8,1,0) \nnn
40 & = & (1,2,-3/2) + (3,2,1/6) + (\bar 3,1,-2/3)
+ (\bar 3,3,-2/3) + (8,1,1) + (\bar 6,2,1/6) \nnn
45 & = & (1,2,1/2) + (3,1,-1/3) + (3,3,-1/3)
+ (\bar 3,1,4/3) + (\bar 3,2,-7/6) \nnn & & \quad {} + (\bar 6,1,-1/3)
+ (8,2,1/2) \nnn
75 & = & (1,1,0) + (3,1,5/3) + (\bar 3,1,-5/3)
+ (3,2,-5/6) + (\bar 3,2,5/6)  \nnn & & \quad {} + (6,2,5/6)
+ (\bar 6,2,-5/6) + (8,1,0) + (8,3,0)
\label{4k}
\eeqa
While the higher dimensional $SU(5)$ irreps certainly
contain states with unusual hypercharge
(e.g., the $(1,2)$ irrep in the $40$ of $SU(5)$ with
$Y=-3/2$),
given the number of $5$, $\bar 5$ and $10$ representations
present in \myref{edc} it is perhaps
surprising that we do not obtain the
$SU(3) \times SU(2) \times U(1)$ irreps to fill out
the MSSM for any of the 175 models.

Beside the projections (\ref{2i},\ref{ums})
in the untwisted sector---which lead to incomplete
would-be $SU(5)$ irreps as discussed in
detail above---the problem, 
of course, is that in the twisted sectors
the \eetee\ weights do not correspond to the decomposition
of $E_8$ representations described by
(\ref{edc},\ref{4k}).  The weights
are of the form $\tK = K + E(n_1,n_3,n_5)$;
whereas $K \in \eelat$,
for any twisted sector with
solutions to \myref{tps}
the embedding vector is a 
strict fraction of a lattice vector:
\beq
3 E(n_1,n_3,n_5) \in \eelat,
\qquad
E(n_1,n_3,n_5) \not\in \eelat.
\eeq
Specializing \myref{2t},
the hypercharge for
any of the $Y_i$ is given by
\beq
Y_i(\tK;n_1,n_3,n_5)  =  
y_i \cdot K + \delta_{y_i}(n_1,n_3,n_5) , \qquad
\delta_{y_i}(n_1,n_3,n_5) =  y_i \cdot E(n_1,n_3,n_5).
\eeq
For a massless state, the value of $y_i \cdot K$
will take values corresponding 
to the decompositions \myref{4k};
$y_i \cdot K$ values from the 3875 of $E_8$
occur because $K^2 > 2$ is possible,
as discussed in Section~\ref{mss}.
The second term on the right-hand side
is the Wen-Witten defect,
briefly discussed above in Section~\ref{mss}.
Since each $y_i$ is nonzero
only in the first eight entries, the Wen-Witten defect
only depends on the observable sector embeddings
enumerated by CMM.  It is easy to 
check that
for each of the $y_i$ the defect in each twisted
sector is a multiple of $1/3$.  This is
consistent with general arguments 
\cite{AADF88,Ant90} which show that
fractionally charged color
singlet (bound) states in $Z_N$ orbifolds
have electric charges which are quantized
in units of $1/N$.

\subsection{Extended Hypercharge Embeddings}
Having failed to accomodate the MSSM with
any of the five $Y_i$, we envision
the most general hypercharge consistent
with leaving at least a hidden $SU(3)'$
unbroken to serve as the condensing
group $G_C$.  (Such a $Y$ is of the
{\it extended} hypercharge embedding
variety, studied for example in Ref.~\cite{CHL96}.)
That is, we include the
possibility that Cartan generators of
the nonabelian hidden sector group
mix into $Y$ under a Higgs effect,
perhaps induced by the FI term.  
(A well-known example of the mixing of 
a nonabelian Cartan generator into
a surviving \uone\ is the
electroweak symmetry breaking 
$SU(2)_L \times U(1)_Y
\to U(1)_E$.)  Thus, we assume
a hypercharge generator of the form
\beq
6 Y = \sum_{a \not= X} c_a Q_a + \sum_{a,i} c_a^i H_a^i.
\label{gym}
\eeq
A factor of six has
been included for later convenience.
The Cartan generators written here
are {\it not} those of \myref{2g}
or \myref{4g}.  Rather, we choose
a basis where the $H_a^i$ are mutually
orthogonal (i.e., $\tra{R} H_a^i H_a^j
= 0$ for $i \not= j$, any irrep $R$ of
$G_a$).

Nontrivial irreps of the hidden sector
gauge group $G_H$ may decompose
under the partial breaking of $G_H$ implied
by \myref{gym} to give some of
the $(1,2)$ and $(1,1)$ irreps of the MSSM.
For instance, if the pattern of 
gauge symmetry breaking
in an irrep Pattern 2.5 model is
\beq
SU(3)_C \times SU(2)_L \times SU(5) \times SU(2)' \times U(1)^8
\to SU(3)_C \times SU(2)_L \times SU(3)' \times U(1)_Y,
\eeq
then we have the following decompositions
of nontrivial irreps of $G_H$ onto the
surviving gauge symmetry group:
\beqa
(1,1,5,1) & \to & (1,1,3) + 2(1,1,1), \nnn
(1,1,10,1) & \to & 2(1,1,3) + (1,1,\bar 3) + (1,1,1), \nnn
(1,2,1,2) & \to & 2(1,2,1) .
\label{4f}
\eeqa
Thus, we get many candidates for $e^c$
as well as candidates for $L,H_d$ or $H_u$.
The Cartan generator of $SU(2)'$ is allowed
to mix into $Y$; this is also true of the
two Cartan generators of $SU(5)$ which commute
with all of the generators of the surviving
$G_C=SU(3)'$.  The weights of the 
$(1,2,1)$ and $(1,1,1)$ states in
\myref{4f} with respect to these generators
then contribute to the hypercharges of
these states.

Corresponding to \myref{gym} is an
assumption for the sixteen-vector $y$
which describes the linear combination
of \eetee\ Cartan generators $H^I$ which
give $Y$:
\beq
6 y = \sum_{a \not= X} c_a q_a + \sum_{a,i} c_a^i h_a^i.
\label{4h}
\eeq
To calculate $k_Y$, we use Eq.~\myref{kad}
and the orthogonality of the
sixteen-vectors appearing in \myref{4h}:
\beq
k_Y = {1 \over 36} \left(
\sum_{a \not= X} c_a^2 k_a 
+ \sum_{a,i} 2 (c_a^i h_a^i)^2 \right).
\eeq
We define, as above, $\hat H_a^i$ to be the
generator $H_a^i$ rescaled to the unified
normalization (e.g., $\tr (\hat H_a^i)^2 = 1/2$
for an $SU(N)$ fundamental irrep).
We express the rescaling by $H_a^i = \sqrt{k_a^i}
\hat H_a^i$.  Then in terms of the sixteen-vectors
associated with these generators, using
Eq.~\myref{4i},
\beq
2 (h_a^i)^2  = 2 k_a^i (\hat h_a^i)^2 = k_a^i.
\eeq
Thus, the hypercharge normalization
may be expressed as
\beq
k_Y = {1 \over 36} \left(
\sum_{a \not= X} c_a^2 k_a 
+ \sum_{a,i} (c_a^i)^2 k_a^i \right) .
\label{4j}
\eeq

Eq.~\myref{4j} gives $k_Y$ as a quadratic form
of the real coefficients $c_a$ and $c_a^i$,
a function which is easy to minimize subject
to the linear constraints imposed by demanding
that the seven types of chiral supermultiplets
in the MSSM ($Q,u^c,d^c,L,H_d,H_u,e^c$)
be accomodated, including hypercharges.
(For instance, we used standard
routines available on the math package Maple.)
We have performed an automated analysis
to determine the minimum $\dkY \equiv k_Y- 5/3$ 
values allowed by each model,
for each possible assignment of the
MSSM to the full pseudo-massless
spectrum.
Our results are shown in Table~\ref{mkytab}.

\begin{table}[ht!]
\begin{center}
\begin{tabular}{cccccccc}
Pattern & $\dkYm$ & Pattern & $\dkYm$
   & Pattern & $\dkYm$ & Pattern & $\dkYm$ \\
\hline
1.1 & 0       & 2.4 & 8/29    & 3.3 & -4/61  & 4.4 & 16/61 \\
1.2 & 1/5     & 2.5 & 11/73   & 3.4 & 16/59  & 4.5 & -1/31 \\
2.1 & 4/29    & 2.6 & 4/11    & 4.1 & -8/113  & 4.6 & 11/73 \\
2.2 & -8/167  & 3.1 & 1/7     & 4.2 & -8/113  & 4.7 & -1/31 \\
2.3 & 0       & 3.2 & -8/119  & 4.3 & 8/81  & 4.8 & 14/5 \\
\hline
\end{tabular}
\caption{Minimum values of $\dkY = k_Y - 5/3$. \label{mkytab}}
\end{center}
\end{table}

It can be seen from the table
that $k_Y=5/3$ is possible
in some patterns.  We remark, however,
that this value has lost most of
its motivation in the present context.
Whereas in a GUT the normalization
$k_Y=5/3$ came out
naturally, we now obtain this value
by artifice, choosing a ``just so''
linear combination of observable
and hidden sector generators.  Perhaps
this is to be expected, since \nasm\
was obtained from
the start at the string scale, without
ever being---properly speaking---embedded
into a GUT.

For some of the assignments of
$Q,u^c,d^c,L,H_d,H_u,e^c$ to the pseudo-massless
spectrum, other states in the spectrum
may have the right charges with
respect to $SU(3)_C \times SU(2)_L \times U(1)_Y$
to also be candidates for some of these MSSM states.
In this case, the MSSM states will generally
be a mixture of all the candidate states from
the pseudo-massless spectrum,
as described above in Section \ref{mds}.
An example of this will be seen in the
following section.  This, however,
does not alter our conclusions for the
coefficients $c_a$ and $c_a^i$, as
well as the hypercharge normalization $k_Y$.

\mysection{Example: ${\rm {\bf BSL}}_{\bf A}$ 6.5}
\label{app}
The model labeling here is the same as described in Section \ref{mds}:
the observable embedding is
CMM~6 from Table~I of Ref.~\cite{Gie01b}
and the hidden sector embedding is No.~5 from Table VI
of Ref.~\cite{Gie01b}.
Thus, the model has embedding
\beqa
3 V & = & (-1,-1,0,0,0,2,0,0;2,1,1,0,0,0,0,0), \nnn
3 a_1 & = & (1,1,-1,-1,-1,2,1,0;-1,0,0,1,0,0,0,0), \nnn
3 a_3 & = & (0,0,0,0,0,1,1,2;2,0,0,-1,1,0,0,0).
\label{5e}
\eeqa
(Recall that $a_5 \equiv 0$ in the class of models
studied here.)
Using the recipes of Section~\ref{mss}, it is easy to
determine the simple roots and to
check that the unbroken gauge group is
\beq
G = SU(3)_C \times SU(2)_L \times SU(5) \times SU(2)'
\times U(1)^8.
\label{5d}
\eeq
The untwisted sector pseudo-massless
matter states are also obtained by simple
calculations; the twisted sectors
are somewhat tedious because of the
large number of states involved.  The full
spectrum of pseudo-massless matter states
is given in Table~\ref{tb5} (Appendix B).
Each entry corresponds
to a {\it species} of chiral matter multiplets,
with three families to each species.  We have
assigned labels 1 through 51 to the species
for convenience of reference in the discussion
which follows.  The irrep of each species with
respect to the nonabelian factors of $G$
is given in the second column of Table~\ref{tb5},
with the order of entries corresponding to
the order of the nonabelian factors in \myref{5d}.
It is not hard to check that
the model falls into Pattern~2.6 of Table~\ref{pt2}.
This pattern has the attractive feature
that it contains only three
extra $(3 + \bar 3,1)$ representations.
Thus, we can expect less finagling with flat
directions to arrange masses
for these exotic isosinglet quarks.
The subscript on the Irrep column data denotes
the sector to which a species belongs:  ``U''
is for untwisted, while for the twisted
species, $n_1,n_3$ pairs of fixed
point labels are given.  The $n_5$
fixed point label now serves as a family
index, so that for each twisted species, it takes
on all three values $n_5=0,\pm 1$.
Twisted oscillator matter states do
not occur in the pseudo-massless spectrum
of this model.  The
remainder of the columns in Table~\ref{tb5}
provide information about \uone\ charges.

As discussed in Section~\ref{mss},
the eight \uone\ generators correspond to
sixteen-dimensional vectors $q_a$ which
are orthogonal to the simple roots and
to each other.  It is not hard to determine
a set of eight $q_a$s.  However, once
the pseudo-massless spectrum of matter
states has been calculated using the recipes
of Section~\ref{mss}, one finds that a naive choice
of the $q_a$s
does not isolate the trace anomaly to a single
\uone.  Using the redefinition technique
described in Section~\ref{mss}, we have isolated
the anomaly to the eighth generator, which
we denote $Q_X$.  Unfortunately,
the redefinitions required to do this, while
maintaining orthogonality of the $q_a$s,
lead to large entries for many of the $q_a$s
when the charges of states are kept integral.
We display our choice of $q_a$s
in Table~\ref{tb6}, along
with $k_a$ (determined by Eq.~\myref{kad}) and
$\tr Q_a$ (determined from the pseudo-massless
spectrum).
We note that $q_1/6=y_1$ of \myref{aac}.
States 27 and 42 would be
electrically neutral exotic isoscalar quarks
if we took $Q_1/6$ as hypercharge.
This provides an explicit example of
the effects of charge fractionalization;
in the low energy theory these states
would bind with ordinary quarks
to form fractionally charged color
singlet composite states.

\begin{table}[ht!]
\begin{center}
$$
\begin{array}{cccc}
a & q_a & \tr Q_a & k_a/4 \\ \hline
1 & (-3,-3,2,2,2,0,0,0;0,0,0,0,0,0,0,0) & 0 & 15 \\
2 & 3(-1,-1,-1,-1,-1,15,0,0;0,0,0,0,0,0,0,0) & 0 & 1035 \\
3 & 3(3,3,3,3,3,1,-46,0;0,0,0,0,0,0,0,0) & 0 & 9729 \\
4 & {3 \over 2}
	(-3,-3,-3,-3,-3,-1,-1,-47;0,0,0,0,0,0,0,0) & 0 & 2538 \\
5 & {3 \over 2}
	(-15,-15,-15,-15,-15,-5,-5,5;12,-12,-12,-48,-12,0,0,0)
   & 0 & 4590 \\
6 & \half (-15,-15,-15,-15,-15,-5,-5,5;-22,-12,-12,20,22,0,0,0)
   & 0 & 357 \\
7 & 3(0,0,0,0,0,0,0,0;1,0,0,0,1,0,0,0) & 0 & 9 \\
X & \half (-3,-3,-3,-3,-3,-1,-1,1;4,6,6,4,-4,0,0,0)
   & 504 & 21 \\ \hline
\end{array}
$$
\caption{Charge generators of \bsa\ 6.5 (cf.~\myref{Qdf}).
\label{tb6}}
\end{center}
\end{table}

For fields which are not $Q_X$ neutral,
we see from Table~\ref{tb5} that $|Q_X|$
has minimum value 1 and maximum value 6.
On the other hand, from Table~\ref{tb6}
we see that $k_X = 84$.  Then
the generator with unified normalization is
$\hat Q_X = Q_X / \sqrt{84}$ and
for fields which are not $\hat Q_X$ neutral,
$|\hat Q_X|$ has minimum value $1/\sqrt{84}
\approx 0.11$ and maximum value $6/\sqrt{84}
\approx 0.65$.  We appealed to this range
in Section~\ref{mds} above.

Finally, we note that the $SU(5)$ charged
states in the model consist of
\beq
3 \; [ \; (1,1,5,1) + 3(1,1,\bar 5,1)
+ (1,1,10,2) \; ] .
\label{4n}
\eeq
Using $C(SU(5))=5$, $X(5)=X(\bar 5)=1/2$,
and $X(10)=3/2$ (apparent from \myref{4k}
taking $\tr T^a T^a$ with respect to
a generator of an $SU(3)$ subgroup of $SU(5)$),
we find that 
\beq
\tbeta{5}= - 3 \cdot 5 + 3(4 \cdot 1/2 + 2 \cdot 3/2) = 0 .
\eeq
Thus, in order to have supersymmetry
broken by gaugino condensation in
the hidden sector,
it is necessary that vector masses be
given to some of the states in \myref{4n}.
If we can arrange to give large masses to
the $3(5 + \bar 5)$ vector pairs, then
the effective $\beta$ function coefficient
is only $b_5=-3$.  This gives a lower
$\LamC$ than the pure $G_C=SU(2)$ 
case ($b_C = -6$) which was regarded
as ``marginal'' in the 
Introduction.  Consequently, the
hidden $SU(5)$ must be broken to a
subgroup so that vevs can be given
to components of the $(\bar 5 \cdot \bar 5 \cdot 10)$
and $(5 \cdot 10 \cdot 10)$ invariants,
allowing more states to
get large masses.  (For the $SU(5)$ invariant
$(\bar 5 \cdot \bar 5 \cdot 10)$ to generate an
effective mass term, the hidden $SU(2)'$ would
also have to be broken since the 10s belong to
doublet representations of $SU(2)'$, as is
evident from Eq.~\myref{4n}.)

As an example, consider
breaking $SU(5) \to SU(4)$.  For many choices
of the hypercharge generator, some (but generally
not all) of the $5$ and $\bar 5$ irreps are hypercharge
neutral.  Decomposing these onto $SU(4)$
irreps, we have $5 = 4 + 1$ and $\bar 5 = \bar 4 + 1$.
The breaking can be achieved by
giving vevs to the $SU(4)$ singlets in
these decompositions, though one should
be careful to avoid generating
non-vanishing F- or D-terms in the process.
The $10$ of $SU(5)$ decomposes according
to $10 = 4 + 6$.  
The invariants mentioned above may generate
masses for many of the nontrivial $SU(4)$
irreps, since under the $SU(5) \supset SU(4)$ decomposition
\beq
(5 \cdot \bar 5) \ni (4 \cdot \bar 4), \qquad
(\bar 5 \cdot \bar 5 \cdot 10) \ni (1 \cdot \bar 4 \cdot 4), \qquad
(5 \cdot 10 \cdot 10) \ni (1 \cdot 6 \cdot 6).
\eeq
It is conceivable that {\it all} of the
$SU(4)$ charged matter may be given
$\ordnt{\LamX}$ masses in this way,
yielding $b_4 = -12$.  If some matter remains
light and $SU(4)$ is identified as the
condensing group $G_C$, values in the range $-12 < b_C \leq -6$
could be obtained.  To say whether or not these
arrangements can actually be made requires
an analysis of D- and F-flat directions which
is beyond the scope of the present work.

\subsection{Accomodating the MSSM}
\label{acm}
Inspection of Table~\ref{tb5} shows that
while appropriate \nasm\ charged multiplets are
present to accomodate the MSSM spectrum, the ``obvious''
choice for hypercharge, $Y_1=Q_1/6$,
does not provide for the
three $e^c$ supermultiplets nor does
it provide enough $(1,2)$ representations with
hypercharge $-1/2$ to accomodate three $L$s
and an $H_d$.
As discussed above, one problem is that
most of the twisted states have bizzare
$Y_1$ charges due to the Wen-Witten defect.
We also have the problem that
$\tK^2=4/3$ for twisted (non-oscillator)
states (versus $K^2=2$ for untwisted),
so that the \eetee\ weights are
``smaller'' and it is harder to obtain
the ``large'' $e^c$ hypercharge; this
explains why $k_Y > 5/3$ is generically
required.
Note that the $Y_1$ charges
are ordinary in the untwisted sector:
the hidden irreps $(1,1,10,2)$ and $(1,1,5,1)$
are $Y_1$ neutral while the observable irreps
$(3,2,1,1)$, $(1,2,1,1)$ and $(\bar 3,1,1,1)$
have $Y_1$ charges $1/6$, $1/2$ and $-2/3$
respectively.
Furthermore, if we subtract off the Wen-Witten
defect, we expect $Y_1$ charges which would
appear in the decompositions \myref{4k}
for twisted states.
With this in mind, we define {\it Z charge}
to be $Z=Y_1$ for untwisted states while
for twisted states
\beq
Z(n_1,n_3,n_5) \equiv Y_1 - y_1 \cdot E(n_1,n_3,n_5)
= {Q_1 \over 6} - \third + n_1 \twthird,
\eeq
where the last equality is easy to check using
the embedding vectors \myref{5e}.
The Z charges are given in
Table~\ref{tb5}.  To see that these
charges are ordinary,
one should compare to the decompositions
(\ref{edc},\ref{4k}).
Checking the Z charges and $SU(3) \times SU(2)$ irrep
labels from Table~\ref{tb5}, it can be seen
that all are in correspondence to some irrep
contained in a decomposition of the 248 and 3875
irreps of $E_8$.
An example of the role of the 3875 irrep
can be seen in state 11 of Table \ref{tb5},
which is a $(1,2)$ irrep of \nasm\
with Z charge $-3/2$; from
\myref{4k} we see that this occurs in
the 40 of $SU(5)$, which itself occurs in
the 3875 but not the 248 of $E_8$.
This shows how it is precisely
the peculiar role of higher dimensional
\eetee\ irreps and the shift $E(n_1,n_3,n_5)$ 
that is responsible for
the bizzare $Y_1$ charges
in the twisted sectors.

Thus, we are forced to assume hypercharge of
the more general form \myref{gym}, which in the present case
we write as
\beq
6 Y = c_1 Q_1 + \cdots + c_7 Q_7 + c_8 H_{(2')}
+ c_{9} H_{(5)}^1 + c_{10} H_{(5)}^2 .
\label{mhc}
\eeq
The generator $H_{(2')}$ is the Cartan
element for the hidden $SU(2)'$,
which we take to be
\beq
H_{(2')} = \diag (1,-1)
\label{5b}
\eeq
in the fundamental irrep.  The
generators $H_{(5)}^1, H_{(5)}^2$ are
the two Cartan elements for the hidden
$SU(5)$ which could combine into hypercharge
while still leaving unbroken a hidden
$SU(3)'$ for the condensing group $G_C$,
as explained in Section~\ref{hyc}.
We take them to be given by
\beq
H_{(5)}^1 = \diag (4,-1,-1,-1,-1),
\qquad
H_{(5)}^2 = \diag (0,3,-1,-1,-1),
\label{5c}
\eeq
for the fundamental representation.
We seek solutions
$c_1, \ldots,  c_{10}$ which
allow for the accomodation of the MSSM.
As mentioned in Section~\ref{hyc},
assigning the MSSM amounts to the
imposition of seven linear constraints
on the coefficients $c_i$,
one for each of the
species $Q,u^c,d^c,L,H_d,H_u,e^c$.
Because of the enormous number of species
to which $L,H_d,H_u$ and $e^c$ could be assigned,
a very large number of assignments
accomodate the MSSM.
However, it is also important to consider
the hypercharge normalization $k_Y$.
From the discussion given in
Section~\ref{hyc}, we know that
\beq
k_Y = {1 \over 36} (c_1^2 k_1 + \cdots + c_{10}^2 k_{10}),
\label{mky}
\eeq
with $k_1, \ldots, k_7$ given in Table~\ref{tb6},
and where $k_8, k_{9}, k_{10}$ depend on the normalization
of the hidden $SU(2)' \times SU(5)$
Cartan generators (\ref{5b},\ref{5c}).
It is easy to see that the generators
(\ref{5b},\ref{5c}) have been
rescaled from the unified normalization
according to
$$
H_{(2')} = \sqrt{k_8} \hat H_{(2')}, \qquad
H_{(5)}^1 = \sqrt{k_{9}} \hat H_{(5)}^1, \qquad
H_{(5)}^2 = \sqrt{k_{10}} \hat H_{(5)}^2,
$$
\beq
k_8 = 4, \qquad k_{9} = 40, \qquad k_{10} = 24.
\eeq

We have investigated the range of $k_Y$ that
is allowed in \bsa\ 6.5, consistent with assignment of
the MSSM spectrum to the model.  This is
not a difficult exercise.  We first obtain
seven linear constraint equations on the $c_i$s
from a given assignment of
the seven types of fields in the MSSM.
We use these constraint equations to rewrite
\myref{mky} in terms of a set of independent
$c_i$s.  The result is a quadratic form
$k_Y$ depending on the independent $c_i$s.
We minimize this quadratic form subject to
the constraint of real $c_i$ using
a standard algorithm provided with the
math package Maple.  We have verified
the automated results by hand in a few sample cases
and find agreement.  An exhaustive
analysis of all possible assignments of the MSSM to the
\bsa\ 6.5 spectrum shows that in every case $k_Y > 5/3$,
consistent with Table \ref{mkytab} (Pattern 2.6).
As above, it is convenient to define $\dkY = k_Y - 5/3$.
We find that constraining $\dkY \leq 2$
still gives 274 possible assignments.
A manageable set is obtained if we impose the limit
$\dkY \leq 1$.
The only possible assignments in this case are
given in Table~\ref{mkya}.  We also give
the minimum value $\dkYm$ for each of
the assignments.  For the cases where
$\dkYm=4/11$ or $\dkYm=1/2$, some of the
MSSM states have been assigned to
$(1,2,1,2)$ irreps, which are each effectively
two $(1,2,1)$ irreps when the hidden $SU(2)'$ is broken
to give an effective nonabelian gauge symmetry
group $SU(3) \times SU(2) \times SU(5)$.
None of the assignments in Table~\ref{mkya}
require breaking the hidden $SU(5)$ to provide
the $e^c$ species or $SU(5)$ Cartan generators
contributing to $Y$; that is, each of these
assignments has $c_{9}=c_{10}=0$ for the
minimum value $\dkYm$.  These two coefficients
are independent parameters for any of the
assignments in Table~\ref{mkya} and {\it could}
be made nonzero without affecting the $Y$
values of the MSSM spectrum; however, this
would alter the $Y$ charges of $SU(5)$ charged
states and would increase $\dkY$ above the
minimum value $\dkYm$.  In principle, $k_Y$
could be made arbitrarily large!
Subscripts on species labels in
Table~\ref{mkya} denote which of the two
$H_{(2')}$ eigenstates the MSSM state
has been assigned to.
For instance, in the $\dkYm=1/2$
assignments, $30_1$ and $30_2$ are states of opposite
$SU(2)'$ isospin.

\begin{table}[ht!]
\begin{center}
$$
\begin{array}{cccccc}
{\rm No.} & Q,u^c,d^c,\underline{L,H_d},H_u,e^c & \dkYm &
{\rm No.} & Q,u^c,d^c,\underline{L,H_d},H_u,e^c & \dkYm \\
\hline
1 & 1,3,10,11,30_1,2,48_2 & 4/11 & 10 & 1,3,42,30_1,30_2,2,29 & 1/2 \\
2 & 1,3,10,25,30_1,2,33_2 & 4/11 & 11 & 1,3,10,11,25,2,43 & 4/5 \\
3 & 1,3,10,30_1,31,2,28_2 & 4/11 & 12 & 1,3,10,11,31,2,49 & 4/5 \\
4 & 1,3,10,30_1,44,2,16_2 & 4/11 & 13 & 1,3,10,25,44,2,34 & 4/5 \\
5 & 1,3,42,11,30_1,2,48_2 & 4/11 & 14 & 1,3,10,31,44,2,23 & 4/5 \\
6 & 1,3,42,25,30_1,2,33_2 & 4/11 & 15 & 1,3,42,11,25,2,35 & 4/5 \\
7 & 1,3,42,30_1,31,2,28_2 & 4/11 & 16 & 1,3,42,11,31,2,24 & 4/5 \\
8 & 1,3,42,30_1,44,2,16_2 & 4/11 & 17 & 1,3,42,25,44,2,9  & 4/5 \\
9 & 1,3,10,30_1,30_2,2,29 & 1/2  & 18 & 1,3,42,31,44,2,17 & 4/5 \\
\hline
\end{array}
$$
\caption{Assignments satisfying $\dkY \leq 1$ 
in \bsa\ 6.5.  Underlining on $H_d$ and $L$
indicates that either permutation may be assigned to
the fourth and fifth entries.
Where applicable, the subscript on a
state label denotes which of the two $H_{(2')}$
eigenstates of a $(1,2,1,2)$ irrep is used in
an assignment.
\label{mkya}}
\end{center}
\end{table}

With these assignments and $\dkY$ set to its
minimum value $\dkYm$, the coefficients $c_i$ in
\myref{mhc} are uniquely determined for each
case; examples are:
\beq
\begin{array}{cccl}
{\rm Assign.~1:} &
(c_1,\ldots,c_{10})
& = & (1,3/253,1/11891,-4/517,0,0,2/11,-18/11,0,0), \\
{\rm Assign.~9:} &
(c_1,\ldots,c_{10})
& = & (2/5,1/10,0,0,1/68,-3/68,3/4,0,0,0), \\
{\rm Assign.~11:} &
(c_1,\ldots,c_{10})
& = & (1,-6/115,-2/5405,8/235,0,0,2/5,0,0,0).
\end{array}
\eeq
From these one can calculate the hypercharges
of the pseudo-massless spectrum, using
the $Q_a$ values and $SU(2)'$ irrep
data provided in Table~\ref{tb5}.
As an example, we have calculated
the hypercharges of the spectrum for
Assignment~11.  These are tabulated
in the last column of Table~\ref{tb5}.

For all of the $\dkYm=4/5$ cases, the \nasm\
charged exotic matter is
\beq
3 \; [ \; (3,1,1/15) + (\bar 3,1,-1/15)
+ 2(1,2,1/10) + 2(1,2,-1/10) \; ]
+ 2 \; [ \; (1,2,1/2)+(1,2,-1/2) \; ].
\eeq
The last number in each term 
gives the hypercharge of the corresponding
state.  We refer to the $SU(3)_C$ charged
states as {\it exoquarks} and to the $SU(2)_L$
charged states as {\it exoleptons.}
The last four exolepton states correspond to
the two extra families of $H_u$-like
and $H_d$-like states which are an artifact of the
three generation construction.
However, the other exoleptons have $Y = \pm 1/10$,
a rather bizzare value, and certainly not one
that appears in GUT scenarios,
as can be seen by comparison to \myref{4k}.
Here again we see the effect of
charge fractionalization.
Similar comments apply to the
exoquarks which have $Y= \pm 1/15$.

For all of the $\dkYm=1/2$ assignments, 
the \nasm\ charged exotic matter is
\beq
3 \; [ \; (3,1,-1/3) + (\bar 3,1,1/3) + 4(1,2,0) \; ]
+ 2 \; [ \; (1,2,1/2)+(1,2,-1/2) \; ].
\eeq
The exoquarks in these assignments have
SM charges of the colored Higgs fields
in an $SU(5)$ GUT.  Whether or not their
masses are similarly constrained 
by proton decay depends
on a detailed study of
the allowed effective superpotential couplings
along a given flat direction,
since we do not have the $SU(5)$ symmetry
to relate Yukawa couplings.
Since altogether we have six
$(\bar 3,1,1/3)$ representations, each of the
three $d^c$-type quarks and their
three exoquark relatives will generally be a mixture
of States 10 and 42, corresponding to
a cross between Assignments 9 and 10.
Such mixing was discussed above in Section~\ref{mds}.

For all of the $\dkYm=4/11$ assignments, 
the \nasm\ charged exotic matter is
\beqa
&& 3 \; [ \; (3,1,-2/33) + (\bar 3,1,2/33) 
+ (1,2,1/22) + 2(1,2,-3/22)
+ (1,2,5/22) \; ] \nnn
&& \hspace{1in} + 2 \; [ \; (1,2,1/2)+(1,2,-1/2) \; ].
\eeqa
Note that a portion of the exolepton spectrum is
chiral and would lead to a massless states
if the usual electroweak symmetry
breaking is assumed.  For this reason
the Assignments 1-8 are not viable.

\subsection{Gauge Coupling Unification}
\label{gcu}
Gauge coupling unification in semi-realistic
four-dimensional string models has been a
topic of intense research for several years.
The situation in the heterotic theory has
been reviewed by Dienes in Ref.~\cite{Die97},
which contains a thorough discussion
and extensive references to the original
articles.  We will only present a brief
overview; the interested reader is recommended
to Dienes' review for further details.

It has been known since the
earliest attempts \cite{SS74} to use closed string
theories as unified ``theories of everything'' that
\beq
g^2 \sim \kappa^2/\alpha',
\label{u01}
\eeq
where $g$ is the gauge coupling, $\kappa$ is the gravitational
coupling and $\alpha'$ is the {\it Regge slope,}
related to the string scale
by $\Lambda_{\stxt{string}} \approx 1/\sqrt{\alpha'}$.
In particular, this relation holds
for the heterotic string~\cite{GHMR85}.
However, $g$ and $\kappa$ in \myref{u01} are the 
ten-dimensional couplings.  By dimensional reduction 
of the ten-dimensional effective field theory
obtained from the ten-dimensional heterotic string
in the zero slope limit,
the relation \myref{u01} may be translated into
a constraint relating the heterotic 
string scale $\LamH$ to the 
four-dimensional Planck mass $m_P$.  
One finds, as expected on dimensional
grounds, $m_P \sim 1/\sqrt{\kappa}$,
where the coefficients which
have been supressed depend on the
size of the six compact dimensions;
similarly, the four-dimensional gauge coupling
satisfies $g_H \sim g$; for details
see Ref.~\cite{DS85}.
Then \myref{u01} gives
\beq
\LamH \sim g_H m_P.
\eeq
Kaplunovsky has made this relation more precise,
including one loop effects from
heavy string states \cite{Kap88}.
Subject to various conventions described
in \cite{Kap88}, including a choice
of the $\overline{\mbox{DR}}$ 
renormalization scheme in
the effective field theory, 
the result is:
\beq
\LamH \approx 0.216 \times g_H m_P
= g_H \times 5.27 \times 10^{17} \mtxt{GeV}.
\label{ssc}
\eeq

In \myref{ssc}, a single gauge coupling, $g_H$, is shown.
However, in the heterotic orbifolds under
consideration the gauge group $G$ 
has several factors, each of which will
have its own running gauge coupling.
One may ask how these running couplings are related to
$g_H$.  This question was studied by
Ginsparg \cite{Gin87}, with the result
that the running couplings unify to a common value $g_H$
at the string scale $\LamH$, up to string threshold effects
and affine levels (discussed below).
(In the case of {\uone}s, normalization conventions
must be accounted for, as we have described in
detail in Section~\ref{hyc}.)
Specifically, unification in four-dimensional string
models makes the following requirements
on the running gauge couplings $g_a(\mu)$:
\beq
k_a g_a^2(\LamH) = g_H^2, \quad \forall \; a.
\label{hbc}
\eeq
Here, $k_a$ for a nonabelian factor $G_a$ is
the {\it affine} or {\it Kac-Moody level} of
the current algebra in the underlying theory
which is responsible for the gauge symmetry in
the effective field theory.  It is unnecessary
for us to trouble ourselves with a detailed
explanation of this quantity or its string theoretic
origins, since $k_a = 1$ for any nonabelian
factor in the heterotic orbifolds we are considering.
For this reason, these heterotic orbifolds are referred
to as affine level one constructions.
In the case of $G_a$ a \uone\ factor, $k_a$
carries information about the normalization
of the corresponding current in the underlying theory, and
hence the normalization of the charge generator
in the effective field theory.  We saw explicit
examples of this 
in the previous section.

The important point, which has been emphasized many
times before, is that a gauge coupling
unification prediction is made by the underlying
string theory.  The SM gauge couplings
are known (to varying levels of accuracy), say,
at the Z scale (approximately 91 GeV).  Given the
particle content and mass spectrum 
of the theory between the Z scale
and the string scale, one can easily check at the one loop
level whether or not the unification prediction
is approximately consistent with the Z scale boundary values.
To go beyond one loop requires some knowledge of
the other couplings in the theory, and the analysis
becomes much more complicated.  However,
the one loop success is not typically spoiled by
two loop corrections, but rather requires a slight
adjustment of flexible parameters (such as superpartner
masses) which enter the one
loop analysis.

In what follows we briefly discuss
the one loop running of SM gauge couplings in \bsa\ 6.5,
Assignment 11 of Table~\ref{mkya}, estimating two loop
effects using previous studies of the MSSM.
Due to the presence of exotic matter,
we are able to achieve string scale unification.
This sort of unification scenario
has been studied
many times before, for example in 
Refs.~\cite{GX92,Far93,MR95,AK96}.
However, in contrast to the Refs.~\cite{GX92,MR95,AK96},
we have states which would not appear in
decompositions of standard GUT groups,
such as \myref{4k}.  Indeed, it was found
by Gaillard and Xiu in Ref.~\cite{GX92} that $(3 + \bar 3,2)$
representations with hypercharge
$Y=\pm 1/6$ were necessary to string
scale unification, while Faraggi
achieved string unification in Ref.~\cite{Far93}
in a model where
the only colored exotics were $(3 + \bar 3,1)$
states.  The resolution of this apparent
conflict is that the unification scenario of Faraggi
contains $(1,2)$ exoleptons with vanishing hypercharge
and $(3 + \bar 3,1)$ exoquarks with hypercharge
$Y=\pm 1/6$;
such states have exotic electric charge
and {\it do not} appear in \myref{4k}.
The appearance of these states is due to the
Wen-Witten defect in the free fermionic
construction used in the model of Ref.~\cite{Far93},
which has a $Z_2 \times Z_2$ orbifold underlying
it, leading to shifts in
hypercharge values by integer multiples of
$1/2$.  Because exotics with small hypercharge
values, much like the $(3 + \bar 3,2)$
representations used by Gaillard and Xiu,
appear in the model employed by Faraggi,
the $SU(3) \times SU(2)$ running can be
altered to unify at the string scale without
having an overwhelming modification on
the running of the $U(1)_Y$ coupling.

Similar to the unification scenario
of Faraggi, in the model studied here
exotic representations with small hypercharges
are present and allow us to unify
at the string scale without the presence
of exotic quark doublets.  However, we
also have nonstandard hypercharge normalization:
for Assignment~11 the minimum value is $k_Y=37/15>5/3$.
Nonstandard
hypercharge normalization has been
studied previously, for example in
Refs.~\cite{Iba93,DFMR96}.  In these analyses,
it was found that {\it lower} values
$k_Y < 5/3$ were preferred if only the
MSSM spectrum is present up to the unification
scale; the preferred values were between $1.4$
to $1.5$.  Unfortunately, we are faced with
the opposite effect---a larger than normal
$k_Y=37/15$.
This larger value requires a larger correction
to the running from the exotic states,
and has the effect of pushing down the
required mass scale of the exotics
from what was found in Faraggi's analysis---particularly
in the case of the exoquarks.  

Standard evolution of the gauge couplings from
the Z scale (i.e., the solution to \myref{bfdef}
for groups other than $G_C$),
together with the unification
prediction \myref{hbc}, 
leads to three constraint equations:
\beqa
4 \pi \icoup{H} & = & {1 \over k_Y} \left[
   4 \pi \icoup{Y}(m_Z) - b_Y \ln {\LamHsq \over m_Z^2}
   - \Delta_Y \right], \label{rgeY} \\
4 \pi \icoup{H} & = & 4 \pi \icoup{a}(m_Z)
   - b_a \ln {\LamHsq \over m_Z^2} - \Delta_a,
   \quad a=2,3. \label{rge23}
\eeqa
The notation is conventional, with 
$\alpha_a = g_a^2/ 4 \pi \; (a=H,Y,2,3)$.
Corrections are captured by the quantities $\Delta_a$,
and will be discussed below.
The quantities $b_a, \; a=Y,2,3$ are the
$\beta$ function coefficients
\beq
b_{a} = -3 C(G_a) + \sum_R X_a(R)
\label{5a}
\eeq
evaluated for the MSSM spectrum.  
Here, $C(SU(N))=N$ while $C(U(1))=0$.  For
a fundamental or antifundamental
representation of $SU(N)$ we have $X_a=1/2$
while for hypercharge $X_Y(R) = Y^2(R)$.
This gives
\beq
b_{Y} = 11, \qquad b_{2} = 1,
\qquad b_{3} = -3.
\label{bcm}
\eeq

Throughout, we use Z scale boundary
values from the Particle Data Group 2000
review \cite{PDG00}, which are given in the $\bbar{\rm MS}$
scheme.  For a supersymmetric running,
these boundary values should be converted 
to the $\bbar{\rm DR}$ scheme, so that the
supersymmetry
algebra is kept four-dimensional \cite{DRB,BAM}.
These scheme conversion effects are included
in the corrections $\Delta_a$.
Due to very small
errors (relative to other uncertainties in
the analysis), we take as precise
\beq
m_Z = 91.19 \mtxt{GeV}, \qquad \icoup{e}(m_Z) = 127.9 .
\eeq
For the other couplings we utilize global fits to
experimental data \cite{PDG00}:
\beq
\sin^2 \theta_W(m_Z) = 0.23117 \pm 0.00016, \qquad
\alpha_3(m_Z) = 0.1192 \pm 0.0028 .
\label{4b}
\eeq
Using
\beq
\icoup{2} = \icoup{e} \sin^2 \theta_W, \qquad
\icoup{Y} = \icoup{e} \cos^2 \theta_W,
\eeq
we obtain the boundary values
\beq
\icoup{Y}(m_Z) = 98.333 \pm 0.020 , \qquad
\icoup{2}(m_Z) = 29.567 \pm 0.020 , \qquad
\icoup{3}(m_Z) = 8.39 \pm 0.20 .
\label{Zin}
\eeq

We now discuss the various corrections contributing
to $\Delta_a \; (a=Y,2,3)$.  Each may be written as the
sum of six terms:
\beq
\Delta_a = \Delta_a^{\stxt{conv}} + \Delta_a^{\stxt{HL}} 
+ \Delta_a^{\stxt{string}} + \Delta_a^{\stxt{light}}
+ \Delta_a^{\stxt{exotic}} + \Delta_a^{\stxt{heavy}}.
\eeq

The quantities $\Delta_a^{\stxt{conv}}$ convert the
$\bbar{MS}$ renormalization scheme input values \myref{Zin}
to the $\bbar{DR}$ scheme \cite{BAM,LP93}.
They are given by:
\beq
\Delta_a^{\stxt{conv}} = \third C(G_a) \quad \Rightarrow \quad
\Delta_Y^{\stxt{conv}} = 0, \quad
\Delta_2^{\stxt{conv}} = 2/3, \quad
\Delta_3^{\stxt{conv}} = 1.
\eeq
As will be seen below, these corrections are negligible
in comparison to the other terms in $\Delta_a$,
and we could ignore them without changing our results
in a meaningful way.

The quantities $\Delta_a^{\stxt{HL}}$ represent corrections
from higher loop orders, which are sensitive to Yukawa
couplings for the MSSM spectrum and the exotic states.
If either the top or bottom Yukawa
coupling evolves to nonperturbative
values somewhere between Z scale and the string scale
(as can happen for small or very large
values of the ratio of MSSM Higgs vevs, $\tan \beta$),
the $\Delta_a^{\stxt{HL}}$ correction is out of control.  However,
if the Yukawa couplings arise from a weakly coupled heterotic
string theory, as we assume, then this does not occur;
$\Delta_a^{\stxt{HL}}$ will take more reasonable
values.  For example, Dienes, Faraggi and
March-Russell \cite{DFMR96} have studied the range of MSSM two loop
corrections with the Yukawa couplings taking
values $\lambda_t(m_Z) \approx 1.1$ and
$\lambda_b(m_Z) \approx 0.175$.  (Using $m_b(m_Z) \approx 3.0$ GeV
from Ref.~\cite{FK98} and $m_t(m_Z) \approx 174$ GeV from
\cite{PDG00}, these Yukawa couplings correspond to
$\tan \beta \approx 9.2$.)
These authors found that the two loop (TL) correction
terms took approximate values
\beq
\Delta_Y^{\stxt{TL}} \approx 11.6,
\qquad
\Delta_2^{\stxt{TL}} \approx 12.3,
\qquad
\Delta_3^{\stxt{TL}} \approx 6.0.
\label{5f}
\eeq
These should dominate $\Delta_a^{\stxt{HL}}$,
so we assume that to the same level of
approximation $\Delta_a^{\stxt{HL}}
\approx \Delta_a^{\stxt{TL}}, \; \forall \; a=Y,2,3.$
Relative to the boundary values
for $4 \pi \icoup{a}$, these
are 0.9\%, 3.3\% and 5.7\% corrections,
respectively.
By comparison, the largest experimental
error is 2.4\% for $\icoup{3}$.

The third type of correction is peculiar
to unified theories with large numbers of
gauge-charged states above or near the unification
scale.  These effects have been extensively
studied \cite{GUTth} in GUTs.  In attempts
to bring unification predictions into good
agreement with precision data
these corrections
play an important role \cite{LP93}.
When very large GUT group representations are
introduced near the unification scale, these
corrections can be considerable \cite{HY93}.
With the standard-like string constructions which
we study here, a GUT symmetry group and
heavy states which complete GUT multiplets
are not
restored at the unification scale.  Rather,
the chief concern is with threshold effects
due to the enormous towers of massive string states.
These may be computed from one loop diagrams
in the
underlying string theory, using background
field methods quite similar to those
exploited in ordinary field theory \cite{Kap88}.
As noted above,
in some four-dimensional heterotic theories,
string threshold corrections
exist which grow
in size as the T-moduli
vevs increase \cite{Nothr}.
This corresponds to the large volume limit for the compact
dimensions; the potentially large contribution
in this limit can also be understood from the
fact that the compactification scale drops below
the string scale and entire excited mass levels of
the string enter the running below the string scale.
In any event, such T-moduli dependent string
threshold effects are irrelevant for
the 175 models studied here, as they do not occur in $Z_3$
orbifold compactifications of the heterotic string
\cite{Nothr}.

However, threshold corrections which do not
increase with the vevs of T-moduli must
also be considered.  These threshold effects have
been calculated by Mayr, Nilles and Stieberger \cite{MNS93}
for an example model which is equivalent to one of the
175 studied here.  They find that the string threshold
effects are given by
\beq
\Delta^{\stxt{string}}_a = 0.079 \; \tbeta{a} + 4.41 \; k_a .
\label{stef}
\eeq
(Actually, Ref.~\cite{MNS93} states that \myref{stef}
is valid with $k_a=1$.  However, starting with
the hypercharge coupling in the
unified normalization $\icoup{1} = \icoup{Y} / k_Y$,
it can be seen from \myref{rgeY} that
by our conventions $\tbeta{1} = \tbeta{Y} / k_Y$
and $\Delta_1 = \Delta_Y / k_Y$.  Substituting these
expressions into \myref{stef} for $a=1$, and then
solving for $\Delta^{\stxt{string}}_Y$, one
finds that the formula is also valid for $a=Y$
where $k_Y \not= 1$.)
It is important to keep in mind that $\tbeta{a}$
is the $\beta$ function coefficient for $G_a$ with
the full spectrum of pseudo-massless states.
This includes
those states which get $\LamX \approx \LamH$
scale masses when the vacuum shifts
to cancel the FI term.  Because of the large
number of states with charge under a given
\uone\ factor, the hypercharge correction
$\Delta^{\stxt{string}}_Y$ is usually much larger
than $\Delta^{\stxt{string}}_2$
or $\Delta^{\stxt{string}}_3$.
The precise values of the coefficients in \myref{stef}
will vary from model to model; these must be worked
out by the numerical evaluation of a rather complicated
integral, as explained in \cite{MNS93}.  However,
Mayr, Nilles and Stieberger
analyzed a few other $Z_3$ orbifold
models, which do not fall into the class of models
considered here, and found that the threshold corrections
differed only slightly from \myref{stef}.  This
was found to be due to the fact that the leading
term in the integrand did not depend on the embedding.
From this we conclude that Eq.~\myref{stef}
gives a fair estimate of the string
threshold corrections in all 175 models
which we study here.

The hypercharge values of the 51 species
must be calculated in order to compute
$\tbeta{Y}$ for the example model.  
This of course depends on what linear
combination \myref{mhc} of generators
we take to be the hypercharge generator $Y$.
As an example we take Assignment 11
from Table~\ref{mkya}, which has
(for $\dkY=\dkYm$) hypercharge
normalization $k_Y=37/15$
and hypercharges $Y$ given in Table~\ref{tb5}.
It is easy to check that
\beq
\tbeta{Y} = \tr Y^2 = 171/5,
\qquad
\tbeta{2} = 9, \qquad \tbeta{3} = 0.
\eeq
Applying \myref{stef}, one finds
\beq
\Delta^{\stxt{string}}_Y \approx 13.6,
\qquad
\Delta^{\stxt{string}}_2 \approx 5.1,
\qquad
\Delta^{\stxt{string}}_3 \approx 4.4,
\eeq
which are comparable to the two loop corrections
in \myref{5f}.

Next we discuss one loop threshold corrections for
pseudo-massless states which have masses greater than the Z mass
but less than the string scale $\LamH$.
Heuristically, these corrections
may be understood as follows.  At a running
scale $\mu$, only states with masses less
than this scale contribute significantly to
the running of the gauge couplings.  Then the
more accurate one loop $\beta$ function coefficients
in this regime are calculated using the spectrum of
states with masses less than $\mu$.  If some of
the superpartner states are more massive than
$\mu$, the $\beta$ function coefficients will
not take the MSSM values given in \myref{bcm}.
Non-MSSM values for the coefficients will
also be obtained if exotic states with masses
less than $\mu$ are present.  In
(\ref{rgeY},\ref{rge23}) we assumed the MSSM
values for the $\beta$ function coefficients.
The threshold corrections we now discuss
account for the non-MSSM $\beta$ function coefficients
which ``should'' have been used over regimes
where the MSSM was not the spectrum of states
with masses less than $\mu$.  This simple
picture is valid in the $\bbar{DR}$ renormalization
scheme; in other schemes there are modifications
to the one loop threshold corrections
presented below, as has been recounted
for example in \cite{LP93}.

The first correction is
due to MSSM superpartners to the SM.
In the coefficients \myref{bcm}, we have implicitly
included these particles in the running all the
way from the Z scale; however, if they are more
massive than the Z scale, this is not quite right.
We introduce ``light'' threshold corrections
which subtract
out the running which should never have been there
in the first place:
\beq
\Delta_a^{\stxt{light}}
= - \sum_{m_i > m_Z} b_{a,i} \ln {m_i^2 \over m_Z^2} ,
\label{ltc}
\eeq
where $b_{a,i}$ is the contribution to the MSSM
$b_{a}$ coming from the state $i$ of mass $m_i$.
Properly speaking, the top quark and the light scalar Higgs doublet
threshold corrections should also be
included in $\Delta_a^{\stxt{light}}$.  The top mass is
near enough to the Z mass that the correction is
negligibly small for our purposes; we assume
that this is likewise true for the light scalar Higgs
doublet.  Following Langacker and Polonsky \cite{LP93},
one often defines effective thresholds $T_a \; (a=Y,2,3)$
which give the same $\Delta_a^{\stxt{light}}$
as \myref{ltc}:
\beq
\Delta_a^{\stxt{light}}
\equiv -(b_a - b_{a}^{\stxt{SM}}) \ln { T_a^2 \over m_Z^2}.
\label{5s}
\eeq
Here, $b_{a}^{\stxt{SM}}$ are the $\beta$
function coefficients in the SM (which we
take to include a light Higgs doublet and
the top quark):
\beq
b_{Y}^{\stxt{SM}} = 7, \quad b_{2}^{\stxt{SM}} = -3,
\quad b_{3}^{\stxt{SM}} = -7, \qquad
\Rightarrow \qquad b_a - b_{a}^{\stxt{SM}} = 4, \quad
a=Y,2,3,
\label{bsm}
\eeq
where we make use of \myref{bcm}.
Eq.~\myref{5s} has the interpretation that
it gives the equivalent threshold correction to
$\icoup{a}$ if all superpartners contributing
to $b_a$ had a uniform mass scale $T_a$.
One may study how the 
prediction for $\alpha_3(m_Z)$ in terms of
$\sin^2 \theta_W (m_Z)$ depends on $T_a$
and determine a combination of
the three effective thresholds which
would give the same effect as a uniform
superpartner mass threshold $\LamSB$ \cite{LP93}:
\beqa
\lefteqn{ \hspace{-15pt} (b_Y - b_3 k_Y)(b_2 - b_2^{\stxt{SM}}) \, \ln {T_2 \over m_Z}
- (b_2 - b_3)(b_Y - b_Y^{\stxt{SM}}) \, \ln {T_Y \over m_Z}
- (b_Y - b_2 k_Y)(b_3 - b_3^{\stxt{SM}}) \, \ln {T_3 \over m_Z}} 
& & \nnn
& {\hspace{-25pt} \equiv} & \hspace{-10pt} \[ (b_Y - b_3 k_Y)(b_2 - b_2^{\stxt{SM}})
- (b_2 - b_3)(b_Y - b_Y^{\stxt{SM}}) 
- (b_Y - b_2 k_Y)(b_3 - b_3^{\stxt{SM}}) \]
\, \ln {\LamSB \over m_Z}.
\label{5r}
\eeqa
From this one can define the
single effective threshold $\LamSB$ in terms
of a geometric average of superpartner masses \cite{CPW93}.
Because of terms of opposite sign
in \myref{5r}, it should be clear that $\LamSB$ can
be much lower than the typical superpartner
mass, which we denoted $\MSB$ in the
Introduction; $\LamSB \lappeq m_Z$ is not at all unreasonable,
even with the typical superpartner mass $\MSB$ several
hundred GeV.  Furthermore, it should be noted
that the formulae for $\LamSB$ given in Refs.~\cite{LP93,CPW93}
are modified in the present context due to the
nonstandard hypercharge normalization,
as has been accounted for in \myref{5r},
which holds generally.
(Our $b_a^{\stxt{SM}}$, as given in \myref{bsm},
also differ slightly due to the inclusion of a light scalar Higgs
doublet; however, Eq.~\myref{5r} has been written
such that it is valid in either case.)
Lastly, the effective threshold $\LamSB$
completely encodes the effects of split thresholds on the 
$\alpha_3(m_Z)$ versus $\sin^2 \theta_W (m_Z)$
prediction, but for other unification
predictions, such as the unified coupling and
scale of unification, a fixed value of
$\LamSB$ corresponds to many different
outcomes \cite{CPW93}; this is because other unification
predictions depend on
combinations of the $T_a$ other than
\myref{5r}.  In the present context,
simply using $\LamSB$ would
not cover the full range of
$g_H$, $\LamH$ and the predictions for
intermediate scales where exotic
matter thresholds alter the running.
An exhaustive analysis would require
scanning over the parameters $T_a \; (a=Y,2,3)$
independently, or subject to model constraints
on the generation of soft masses
by supersymmetry breaking.  Our
purpose here is simply to demonstrate the
possibility of string scale unification with
nonstandard hypercharge normalization and to
estimate the order of magnitude required
for the exotic scales.
For these purposes it is therefore
sufficient to take $\LamSB \approx T_a \; (a=Y,2,3)$.
Within this universal scale $\LamSB$ approximation,
\beq
\Delta_a^{\stxt{light}}
= - 4 \ln {\LamSB^2 \over m_Z^2}, \quad a = Y,2,3.
\eeq
If we limit $m_Z \lappeq \LamSB \lappeq 1$ TeV, then
\beq
0 \gappeq \Delta_a^{\stxt{light}} \gappeq -19.2,
\quad a=Y,2,3.
\eeq

The second set of mass threshold corrections
comes from exotic matter at intermediate
scales.  For the sake of
simplicity, we assume that exoleptons with mass
much less than the string scale enter the running
at a {\it single} scale $\Lambda_2$.  We assume that
{\it all} the exoquarks enter at a single scale $\Lambda_3$.
(Introducing only {\it some} of the exoquarks forces
$\Lambda_3$ to even lower values than we will
find below, which are already a bit of a problem
given the exotic hypercharges that these
exoquarks have.)
The exotic matter threshold corrections 
can be thought of as due to shifts
in the total $\beta$ function coefficients
between $\Lambda_{2,3}$ and the string scale.
Since we introduce $3(3 + \bar 3,1)$ chiral multiplets
$q_i$ and $q_i^c$ at $\Lambda_3$, we have
\beq
\Delta_3^{\stxt{exotic}}
= 3 \ln {\LamHsq \over \Lambda_3^2} .
\eeq
The shift in the $\beta$ function coefficient
for $SU(2)_L$ due to extra $(1,2)$ representations---the
exolepton chiral multiplets $\ell_i$ and
$\ell_i^c$ introduced at $\Lambda_2$---is given by
\beq
\dbtw = \sum_{\ell_i,\ell_i^c} \half .
\eeq
That is, $\dbtw$ is just the number of exolepton
pairs $\ell_i + \ell_i^c$.  The threshold corrections are
\beq
\Delta_2^{\stxt{exotic}}
= \dbtw \ln {\LamHsq \over \Lambda_2^2} .
\eeq
The exoquark and exolepton chiral multiplets
also carry hypercharge.  We denote the shifts in
the $\beta$ function coefficient for $U(1)_Y$ by
\beq
\dbY = \sum_{q_i,q_i^c} (Y_i)^2,
\qquad
\dbYp = \sum_{\ell_i,\ell_i^c} (Y_i)^2 .
\eeq
In this notation the threshold corrections are
\beq
\Delta_Y^{\stxt{exotic}}
= \dbY \ln {\LamHsq \over \Lambda_3^2} 
+ \dbYp \ln {\LamHsq \over \Lambda_2^2} .
\eeq

Let $m,n$ denote the numbers of
exolepton pairs entering the running at
$\Lambda_2$, where $m$ is the number of $Y=\pm 1/2$ exolepton
pairs and $n$ is the number of $Y=\pm 1/10$ exolepton
pairs.  We then have
\beq
\dbY = {2 \over 25}, \qquad \dbYp = m + {n \over 25},
\qquad \dbtw = m + n .
\eeq
For purposes of illustration below,
we will study only the case $(m,n)=(0,6)$, for which
\beq
\dbY = {2 \over 25}, \qquad \dbYp = {6 \over 25},
\qquad \dbtw = 6.
\label{5g}
\eeq
It is not difficult to generalize our results to
other $(m,n)$ values.

Finally, there is the spectrum of particles which
get masses of order $\LamX$ when the vacuum shifts
to cancel the FI term.  Since $\LamX < \LamH$
in \bsa\ 6.5 (cf.~Table~\ref{fir}, Pattern 2.6),
these can give an appreciable heavy threshold correction.
Corrections of this type have been noted
previously; for example, in Ref.~\cite{BL92}.
We assume that all pseudo-massless
states other than the MSSM spectrum
plus exotics associates with
$\Lambda_{2,3}$ enter the running at
$\LamX$, which is convenient because the ratio
\beq
\ln{\Lambda_H^2  \over \Lambda_X^2}
= 2 \ln {0.216 \times g_H m_P \over 0.170 \times g_H m_P} = 0.479
\eeq
is independent of $g_H$ (both $\LamX$ and $\LamH$
are proportional to $g_H$); here we use the value
for Pattern 2.6 from Table~\ref{fir}.  Taking into account the
exotic matter assumed at intermediate scales 
$\Lambda_{2,3}$ and
the total $\beta$ function coefficients mentioned
above, we have
\beq
\Delta_Y^{\stxt{heavy}} = (\tbeta{Y} - b_Y - \dbY - \dbYp)
\; \ln {\Lambda_H^2 \over \Lambda_X^2} = 10.3,
\eeq
\beq
\Delta_2^{\stxt{heavy}} = (\tbeta{2} - b_2 - \dbtw)
\; \ln {\Lambda_H^2 \over \Lambda_X^2} = 1.0,
\qquad
\Delta_3^{\stxt{heavy}} = 0 .
\eeq
The hypercharge threshold correction is
comparable to the larger corrections discussed
above.  On the other hand, we could just
as well ignore $\Delta_{2,3}^{\stxt{heavy}}$
at the level of approximation made here.

As we tune $\Lambda_{2,3}$ to satisfy
the unification constraints,
it is convenient to define the 
sum of all the corrections
{\it except} $\Delta_a^{\stxt{exotic}}$:
\beq
\Delta_a^0 \equiv
\Delta_a - \Delta_a^{\stxt{exotic}} = \Delta_a^{\stxt{conv}} + \Delta_a^{\stxt{HL}} 
+ \Delta_a^{\stxt{string}} + \Delta_a^{\stxt{light}}
+ \Delta_a^{\stxt{heavy}}.
\label{5u}
\eeq
Using the above estimates for each of the terms, we find
for the case of $\LamSB = m_Z$
\beq
\Delta_Y^0 \approx 35.5,
\qquad
\Delta_2^0 \approx 19.1,
\qquad
\Delta_3^0 \approx 11.4.
\label{5h}
\eeq
For the case of $\LamSB=1$ TeV, the estimate is
\beq
\Delta_Y^0 \approx 16.3,
\qquad
\Delta_2^0 \approx -0.1,
\qquad
\Delta_3^0 \approx -7.8.
\label{5i}
\eeq

We now proceed to study the unification
constraint in \bsa\ 6.5, Assignment 11, 
subject to the
assumptions described above.
For convenience, we define
\beq
a_H = 4 \pi \icoup{H}; \qquad
d_a = 4 \pi \icoup{a}(m_Z) - \Delta_a^0, \quad a=Y,2,3;
\eeq
\beq
t_2 = \ln { \Lambda_2^2 \over m_Z^2 }, \qquad
t_3 = \ln { \Lambda_3^2 \over m_Z^2 }.
\eeq
Because the string
scale $\LamH$ contains a dependence on $g_H$
through \myref{ssc}, it will prove convenient to write
\beq
\ln \left({\LamH \over m_Z}\right)^2 = t_P - \ln (4 \pi \icoup{H}),
\eeq
\beq
t_P \equiv 2 \ln \left( {4 \pi \LamH \over g_H m_Z} \right)
= 2 \ln \left( {4 \pi \times \zeta \times 5.27 \times 10^{17}
\over 91.19 } \right) = 77.6 + 2 \, \ln \zeta.
\label{5j}
\eeq
Here we introduce a coefficient $\zeta$
which expresses uncertainty in \myref{ssc}
described in \cite{Kap88};
we study 10\% deviations by
taking $0.9 \leq \zeta \leq 1.1$,
leading to $t_P = 77.6 \pm 0.2$.
Eqs.~(\ref{rgeY},\ref{rge23}) give the following equations
which must be simultaneously satisfied:
\beqa
a_H & = & d_3 + 3 t_3 \label{u22} \\
a_H & = & d_2 + \dbtw \, t_2
- (1 + \dbtw)(t_P- \ln a_H) \label{u23} \\
k_Y a_H & = & d_Y + \dbY \, t_3
+ \dbYp \, t_2
- (11 + \dbY + \dbYp)(t_P - \ln a_H) \label{u24}
\eeqa
The first equation shows the nice feature that since
the $SU(3)_C$ coupling becomes conformal above $\Lambda_3$,
the $\ln a_H$ dependence is gone and we can solve for
$a_H$ explicitly.  Since this equation does not depend
at all on $t_2$, we obtain $a_H = a_H(d_3,t_3)$.  Substituting
this into the second equation allows us to solve for
$t_2$ explicitly, yielding $t_2 = t_2(d_2,d_3,t_3)$.  Thus,
the last equation becomes the only nontrivial constraint,
which is transcendental and must be solved numerically.
Through it we can determine $t_3=t_3(d_Y,d_2,d_3)$ after having
substituted the expressions for $a_H$ and $t_2$ from
the first two equations.  Taking the values \myref{5g}
for the $(m,n)=(0,6)$ example, the implicit
equation for $t_3$ is
\beq
t_3 = {1 \over 540} \left[ 75 d_Y - 182 d_3 - 3 d_2 \right]
- {23 \over 15} \left[ t_P - \ln(d_3 + 3 t_3) \right],
\label{5q}
\eeq
which can easily be solved iteratively.  Once
$t_3$ is determined, $a_H$ is easily
obtained from \myref{u22} and
\beq
t_2 = {1 \over 6} \left[ a_H - d_2
+ 7 (t_P - \ln a_H) \right].
\eeq

Note that if $g_H$ and $\LamH$
were independent, as in the GUT case, we would have one
more degree of freedom and we could not uniquely determine
$t_2, t_3, g_H, \LamH$ in terms of $d_Y,d_2,d_3$.
Related to this is an alternative,
but equivalent, method of
solution to that employed above.  We could
treat $\LamH$ and $g_H$ as independent
and solve (\ref{rgeY},\ref{rge23}) keeping
$t_3$ as the extra free parameter.  Then
solutions to (\ref{rgeY},\ref{rge23})
would have $\LamH=\LamH(t_3)$ and
$g_H=g_H(t_3)$.  We could then determine
the range of $t_3$ which allow the fourth
constraint \myref{ssc} to be satisfied to within,
say, 10\%.  Instead we impose \myref{ssc}
from the start and
address uncertainty of $\pm$10\% with the
parameter $\xi$.  The results are
of course the same by either method.

In the case of $\LamSB=m_Z$, we find
\beqa
\Lambda_2 & =&  \( 2.25 \mp 0.07 \mp 0.006 \pm 0.09 \)
\times 10^{13} \; {\rm GeV}, \nnn
\Lambda_3 & = & \( 5 \mp 0.1 \mp 3 \mp 1 \)
\times 10^{6} \; {\rm GeV}, \nnn
g_H & = & 0.995 \pm 0.0004 \pm 0.0001 \pm 0.003, \nnn
\LamH & = & \( 5.1 \pm 0.002 \pm 0.0005 \pm 0.6 \)
\times 10^{17} \; {\rm GeV}.
\label{5k}
\eeqa
The first two uncertainties for each quantity
give the modified estimates if
$\swZ$ and $\icoup{3}(m_Z)$
are taken at the ends of
the $1\s$ ranges given in \myref{4b} and \myref{Zin}
respectively.  Upper signs in \myref{5k} correspond
to the upper limits of the $1\s$ ranges;
asymmetric uncertainties (due to logarithms)
have been rounded up to the larger of the two.
The third uncertainty gives
the modified estimates if the ``fudge parameter''
$\zeta$ in \myref{5j} is taken at the
ends of the range $0.9 \leq \zeta \leq 1.1$.
Again, the upper signs in \myref{5k}
correspond to the upper limit of the range
for $\zeta$.
Sensitivities are logical:
the exoquark scale $\Lambda_3$ is most
sensitive to $\icoup{3}(m_Z)$, while the
sensitivity to $\swZ$ is below significance.
Only the exolepton scale $\Lambda_2$ is
has significant sensitivity to $\swZ$;
$\Lambda_2, \; \LamH$ and $g_H$,
quantities more closely related
to the high scale physics, are sensitive
the high scale uncertainty $\zeta$.
For the case of $\LamSB=1$ TeV, we find
\beqa
\Lambda_2 & =&  \( 8.4 \mp 0.3 \mp 0.02 \pm 0.4 \)
\times 10^{12} \; {\rm GeV}, \nnn
\Lambda_3 & = & \( 7 \mp 0.1 \mp 4 \mp 1 \)
\times 10^{5} \; {\rm GeV}, \nnn
g_H & = & 0.972 \pm 0.0003 \pm 0.0001 \pm 0.003, \nnn
\LamH & = & \( 5.0 \pm 0.002 \pm 0.0004 \pm 0.5 \)
\times 10^{17} \; {\rm GeV}.
\label{5m}
\eeqa

\begin{figure}[p]
\begin{center}
\includegraphics[height=6.0in,width=5.0in,angle=90]{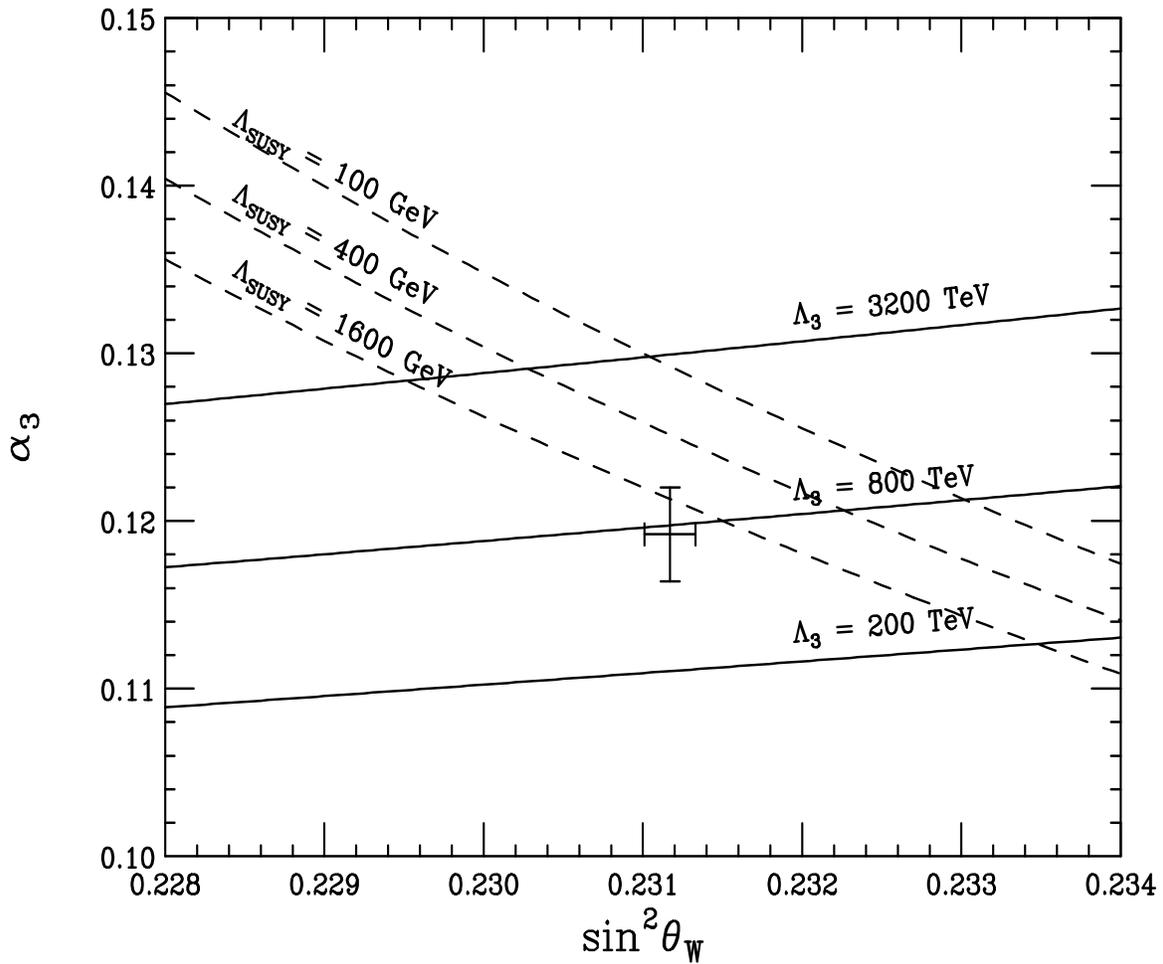}
\end{center}
\caption{Predicted Z scale values per the string
unification scenario (solid), for values of $\Lambda_3$
stepping by factors of four, with $\LamSB=1$ TeV.
For comparison, the MSSM unification prediction
is shown (dashed), with $\LamSB$ stepping by factors
of four.  Experimental values are show with error
bars.}
\label{fg1}
\end{figure}

We next address concerns over fine-tuning
in the unification scenario considered here.
Ghilencea and Ross have recently argued
that a realistic string model should not
disturb the ``significance of the prediction
for the gauge couplings'' which
occurs in the MSSM \cite{GR01}.
They note that for reasonable values of
$\LamSB$, the portion of the
$\alpha_3(m_Z)$ versus $\sin^2 \theta_W (m_Z)$
plane allowed by conventional MSSM unification
is a very small strip.  We can rewrite
Eq.~\myref{5q} as an implicit equation
$d_3=d_3(d_Y,d_2,t_3)$, so that for fixed
value of the exoquark scale, and thereby
$t_3$, we can predict $\alpha_3(m_Z)$ 
as a function of $\sin^2 \theta_W (m_Z)$.
In Figure~\ref{fg1} we show our results
for values of $\Lambda_3$ which step by
a factor of four; we assume $\LamSB=1$ TeV
for these (solid) curves.
For comparison we also show the MSSM
unification predictions (dashed) 
with $\LamSB$ stepping
by factors of four; in the
MSSM case we take $k_Y=5/3$
and assume threshold corrections
\beq
\Delta_a^{\stxt{MSSM}} \approx
\Delta_a^{\stxt{conv}} +
\Delta_a^{\stxt{HL}} +
\Delta_a^{\stxt{light}},
\qquad a=Y,2,3,
\eeq
where each of the quantities 
on the right-hand side are assumed
as above.  We also show with error bars
the experimental values \myref{4b}.
The experimental error bars define the
major and minor axes of an ``error ellipse.''
In any give direction, the distance from
the center of this ellipse to its edge
gives a measure which is independent
of how we scale the axes of the graph. 
We compare the widths of strips to those
of the MSSM in these units.
It can be seen
that the sensitivity to
$\Lambda_3$ is only a factor of
approximately three greater
than the sensitivity to
$\LamSB$ in the MSSM.  Roughly
speaking, the tuning is not much worse than
in the MSSM.  Another way to see that the
tuning is not ``fine''
is that deviations of up to roughly 60\%
in $\Lambda_3$
from the central value given in \myref{5m}
can be accomodated by the uncertainty in
$\icoup{3}(m_Z)$.
It is also interesting to note that
setting the scale $\Lambda_3$ is equivalent
to predicting $\icoup{3}(m_Z)$, since the
(solid) curves
in Figure~\ref{fg1}
are nearly horizontal; this is
reflected in that fact that uncertainty
in $\swZ$ had no
appreciable effects on the estimates
of $\Lambda_3$ in Eqs.~(\ref{5k},\ref{5m}).

\begin{figure}[p]
\begin{center}
\includegraphics[height=6.0in,width=5.0in,angle=90]{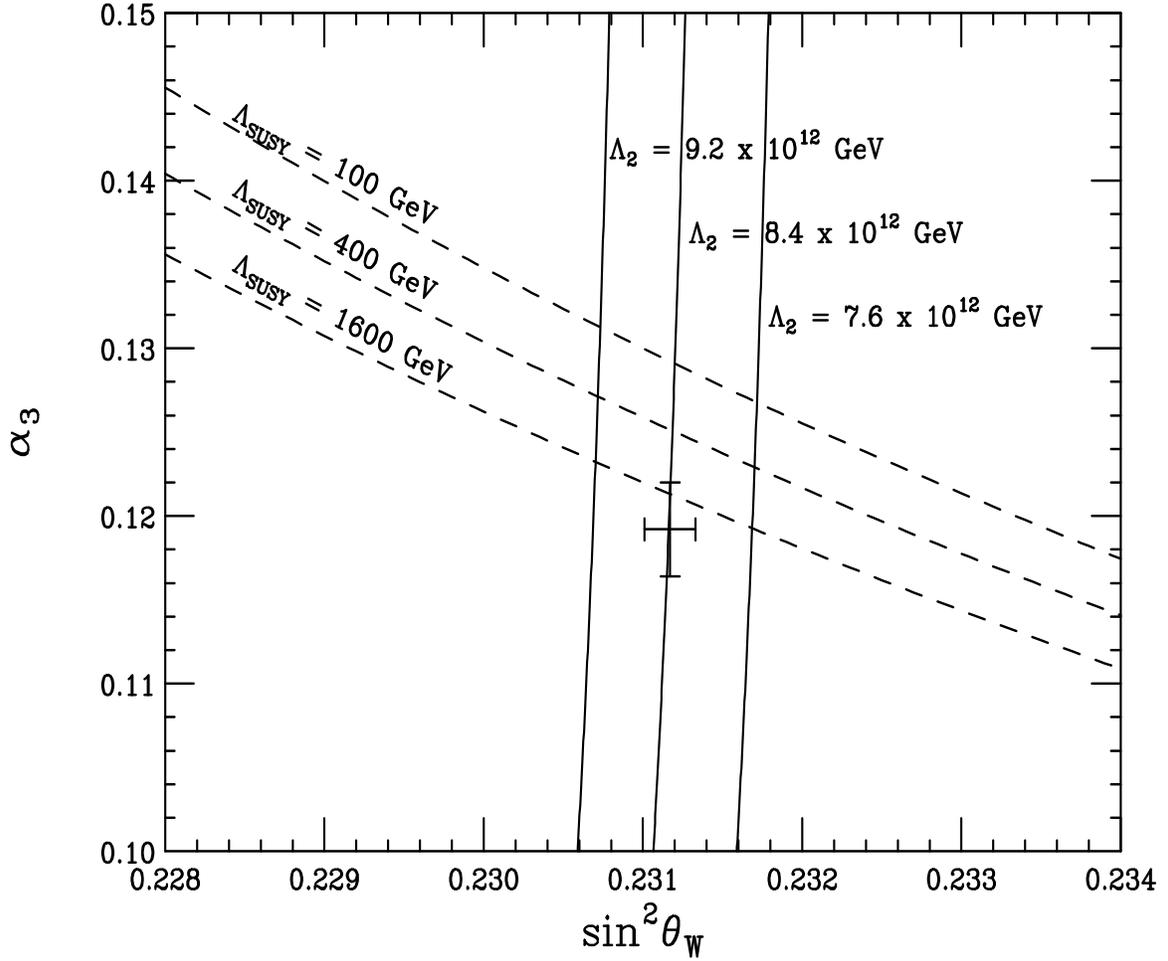}
\end{center}
\caption{Predicted Z scale values per the string
unification scenario (solid), for values of $\Lambda_2$
stepping by $\pm$10\% from the best fit value, with $\LamSB=1$ TeV.
For comparison, the MSSM unification prediction
is shown (dashed), with $\LamSB$ stepping by factors
of four.  Experimental values are show with error
bars.}
\label{fg2}
\end{figure}

In Figure~\ref{fg2} we present a similar analysis
for $\Lambda_2$, the exolepton scale.  
We fix $t_2$ and solve
Eqs.~(\ref{u22}-\ref{u24}) numerically 
eliminating $t_3$ and $a_H$ to obtain
$d_3=d_3(d_Y,d_2,t_2)$.  
For a given value of $t_2$ we
obtain a curve for $\alpha_3(m_Z)$ 
as a function of $\sin^2 \theta_W (m_Z)$;
we take $\LamSB=1$ TeV.
The sensitivity to the exolepton
scale is {\it much} higher,
so we only step by $\pm$10\% from
$\Lambda_2 = 8.4 \times 10^{12}$ GeV,
the approximate central value of \myref{5m}.
We compare
the widths of the strips to those of
the MSSM unification as describe
above.  It can be seen that
they are roughly three times wider, implying that a
10\% variation of $\Lambda_3$
in the string unification scenario
studied here is on a par with a 1200\% variation
of $\LamSB$ in the MSSM unification scenario.
That is, sensitivity to the exolepton scale is
roughly 120 times worse than the $\LamSB$
sensitivity of the MSSM.
From \myref{5m}
we note that deviations of up to
3.5\% for $\Lambda_2$ from the central
value can be accomodated by the uncertainty
in $\swZ$.  Although this tuning
is ``fine,'' it is not horrendous.
The vertical (solid) curves in Figure~\ref{fg2}
demonstrate that choosing $\Lambda_2$ is
essentially equivalent to predicting $\swZ$;
this is reflected in \myref{5m} by the fact
that $\Lambda_2$ has no significant
sensitivity to the uncertainty in $\icoup{3}(m_Z)$.

To summarize, relative to the tuning of superpartner
thresholds in the MSSM unification scenario, the
the exoquark scale is {\it not} finely-tuned, but the
exolepton scale {\it is} finely-tuned; however, the
fine-tuning of the exolepton scale is not astronomical
and is perhaps acceptable.  If one is prepared to
accept a tuning 120 times worse than the tuning of
$\LamSB$ in the MSSM, then one still must explain
{\it why} the exotic scales have the order of
magnitudes that they do.  Presumably, this would be
determined by a detailed study of the flat directions
which produce Xiggs vevs and the selection rules which
restrict couplings in the effective theory.  If the
leading couplings giving exoquarks mass were of
high enough dimension, a natural explanation of the low
exoquark scale may be possible; the exolepton scale
may be easier to explain because it is near the
condensation scale.

Using our results for the scales $\Lambda_{2,3}$, we can extract the
range of exotic thresholds corrections $\Delta_a^{\stxt{exotic}}$
which are required:
\beq
9 \lappeq \Delta_Y^{\stxt{exotic}} \lappeq 10, \qquad
120 \lappeq \Delta_2^{\stxt{exotic}} \lappeq 130, \qquad
150 \lappeq \Delta_3^{\stxt{exotic}} \lappeq 160.
\eeq
Comparing to (\ref{5h},\ref{5i}), it can be seen that the
exotic threshold corrections for $\icoup{2}$ and $\icoup{3}$ are
quite large compared to other effects; they represent
roughly 35\% and 150\% corrections to $4 \pi \icoup{2}$
and $4 \pi \icoup{3}$ respectively!  However, the
hypercharge correction is fairly modest (0.8\%).  
(To a good approximation, we could have neglected
the $\Delta^0_a$ of Eq.~\myref{5u} and solved for the
order of magnitude of the exotic
threshold corrections.)
This can
be traced to the fact that the exoquarks and exoleptons
which we have introduced at $\Lambda_3$ and $\Lambda_2$
have very small hypercharges.  This is precisely what
is needed to overcome the nonstandard hypercharge normalization.
It can be seen from \myref{rgeY} that as $k_Y$ is increased
above its standard value, the prediction for $\icoup{H}$
will tend to decrease, all other quantities being held constant
and ignoring the constraints (\ref{ssc},\ref{rge23}).
We can correct for this tendancy by making
$\Delta_2$ and $\Delta_3$ significantly larger
than what is typical in the MSSM, so long as we
do not greatly change $\Delta_Y$.  This is possible
because we have exoquarks and exoleptons with very
small hypercharge.

The bizzare hypercharges of the exotic particles
lead to fractionally charged particles;
the most problematic are the exoquarks, given
the rather low value of $\Lambda_3$.
Thermal production of exoquarks or exoleptons
at an early stage of the universe would violate
relic abundance bounds on fractionally charged
particles (FECs) by several orders of
magnitude, as discussed for example in
Refs.~\cite{AADF88,CCF96,Per01}.  
Thus, viabilty of this unification
scenario requires inflation, to dilute the abundances
of FECs, with a reheating temperature $T_R$ which
is sufficiently low that the FECs will not
be appreciably produced following inflation;
such scenarios have been examined
for example in free fermionic models \cite{CCF96}.
Chung, Kolb and Riotto \cite{CKR98} have
recently pointed out that the dilution
of heavy particle abundances
by inflation imposes
a much stronger limit than was initially
imagined:  to avoid thermal production of
heavy particles with $G_{SM}$ gauge quantum
numbers, the masses of these heavy particles
must be greater than $T_R$ by a factor
of roughly $10^3$.  Then to escape conflict with the
relic density data for fractionally charged
particles, we require inflation with
\beq
T_R \lappeq 10^{-3} \Lambda_3 \lappeq  5 \mtxt{TeV}.
\eeq
While inflationary scenarios with such low
reheating temperatures have certainly been
proposed (see for example Ref.~\cite{GRS00}),
it is not at all clear that
such scenarios can be achieved in the present
context.  We will not address this question
here, leaving it to further investigation.

\mysection{Conclusions}
\label{con}
In this work we have made a systematic
tabulation of detailed properties of
{\it all}\ orbifold models falling within
the \bsa\ class defined in the Introduction;
by so doing, we can legitimately say what
is ``typical'' in this class of models.
We have determined the hidden sector
gauge group $G_H$ and matter representations
charged under the nonabelian part of
$G_H$.  These details are key to predicting
the low energy phenomenology which arises
from supersymmetry breaking in a hidden
sector, such as in the effective theory of BGW.
We have listed all of the patterns of irreps 
under the nonabelian factors of $G$.  Using these
results, one can easily select a model from
the \bsa\ class which has the desired exotic matter.
The tables of irreps also suggest topics for
further study, such as gauge mediation of
hidden sector supersymmetry breaking from
mixed representations of the observable
and hidden sector gauge groups.  While such
communications may be suppressed by large masses,
they are likely competitive with gravity mediation,
which is suppressed by inverse powers of
the Planck mass.

For each model, we have given a number of
quantities which are useful for phenomenological
studies.  The FI terms in Table~\ref{fir} 
allow one to determine
the scale of initial gauge symmetry breaking.
An understanding of the details of how this occurs
is important to the construction of the low
energy effective theory.  Because many of the
low energy effective operators have coefficients
which at leading order depend on large powers of
the Xiggs vevs, $\ordnt{1}$ variations in the FI
term can be greatly enhanced.  For this reason,
an accurate determination of the FI term is
of practical interest.  Table~\ref{tbgs} gives
the Green-Schwarz coefficient $\bGS$ for
each model, which plays a prominent role
in formulae in the effective theory of 
BGW---for example, the T-moduli mass formula
\myref{mtm}.  In particular, we found that this
implies a problem of too light T-moduli masses
in the \bsa\ class.

The minimum hypercharge normalization $k_Y$
(consistent with accomodation of the MSSM and
at least $SU(3)'$ surviving in the hidden
sector to provide for gaugino condensation) was
determined for each model.  If one is determined
to obtain the standard normalization
$k_Y=5/3$, Table~\ref{mkytab}
spares effort on fruitless models where this
is not possible---over half of the 175 studied here.
We are able to conclude that ``extended'' hypercharge
embeddings allow for $k_Y < 5/3$ in some of
the models, similar to what was found for
free fermionic models in Ref.~\cite{CHL96}.
However, it is not possible
to obtain small enough $k_Y$, in the range of
$1.4$ to $1.5$, to achieve string scale
unification with only the
MSSM field content---a string
unification scenario studied in Refs.~\cite{Iba93,DFMR96}
and reviewed in \cite{Die97}.

All of the quantities tabulated here are necessary to
detailed model-building in the effective
supergravity theory and have implications for
soft terms in the MSSM and the unification
of running gauge couplings.  To our knowledge,
this is the first complete and systematic survey of
three generation standard-like bosonic
heterotic orbifold models performed {\it at
this level of detail.}
By organizing the models into twenty patterns
of irreps and enumerating various other
properties which are universal to models
within a given pattern, we allow the
phenomenologist to quickly select a subset
of the models within the \bsa\ class which
have the desired properties.  It is an interesting
result that so many of the features of the various
models within an irrep pattern are universal.
Cross-referencing with the
embeddings enumerated in \cite{Gie01a} using Table \ref{tb1},
one can employ the recipes provided in Section~2
to quickly generate the matter spectra
for a given model, without a detailed understanding
of the underlying theory; alternatively,
full tables of all 175 models are available
from the author upon request.  It is hoped
that through these efforts the \bsa\ class
of string-derived models has been rendered more
readily accessible for further study 
to a wider audience.

The unusual features of string-derived models,
charge fractionalization and nonstandard hypercharge
normalization, have been discussed in the simplest
of terms.  We have endeavored to make clear
as is possible how it is that states occur
which would not be discovered through
straightforward dimensional reduction
and irrep decompositions of the original
ten-dimensional \eetee\ theory.  We have
discussed at length the problems which
these features present for the construction
of a phenomenologically viable model.
We have described the size of Xiggs vevs in
general terms, and have found that large
T-moduli vevs would seem to spoil perturbativity
of the $\s$ model expansion of the effective
theory.

In an example model
where nonstandard hypercharge 
normalization cannot be avoided,
we have described the
lengths to which one must go in order to
achieve unification at the string scale.
Exotic matter states with very small
hypercharges were introduced at intermediate
scales to obtain agreement with Z scale
data for the gauge couplings.  The
exoquark scale was found to be rather
low.  The exotic hypercharges of the 
exotic matter in turn implied a low
reheating temperature to avoid problems
with FEC relic abundance constraints.
Fine-tuning of the intermediate scales
was examined and was shown to be,
in our opinion, rather mild.  However, we did not
study flat directions and superpotential couplings
in the example model,
and for this reason the intermediate scales
and intermediate field content remain to
be justified.

To defend the unification scenario
presented in Section~\ref{gcu}, 
one must be willing to take the position that
the apparent unification at roughly $2 \times
10^{16}$ GeV in the MSSM with $k_Y=5/3$ is
purely accidental; we find this
point of view difficult to accept.
On the other hand, the unification scenario
we have studied serves as
an illustration of how ugly things really
are when one attempts to refine many
of the models into a realistic theory.
Though we have studied only one example,
it can be seen from Table~\ref{mkytab}
that a good fraction of the \bsa\ class
models have $k_Y > 5/3$ and the unification
constraint in these models leads inevitably
to the contortions exhibited in our example.

In conclusion, the more promising models
will be those with $k_Y \leq 5/3$.  One
might invoke M-theory \cite{Mth}
to explain unification
at $2 \times 10^{16}$, as was done in Refs.~\cite{CFN99};
or, one might introduce {\it many} exotics
at a intermediate scale with a ``just so''
arrangement of irreps and charges in the hope that
with enough exotics the intermediate scale would
quite near the unification scale of the MSSM
and the apparent approximate unification
at $2 \times 10^{16}$ would not be an accident.
In either case, the classification performed
here, together with the identification of
equivalences performed in Ref.~\cite{CMM89,Gie01b},
has moved the effort further along 
for the \bsa\ class and has
narrowed down the number of ``attractive
models.''

\vspace{20pt}

\noindent
{\bf \large Acknowledgements}

\vspace{10pt}

\noindent
The author would like to thank Mary K.~Gaillard
for encouragement and useful discussions.  Other
individuals who provided helpful comments include
Robert Cahn, Amy Connolly, Dimitru Ghilencea, Hitoshi Murayama
and Brent Nelson.
This work was supported in part by the
Director, Office of Science, Office of High Energy and Nuclear
Physics, Division of High Energy Physics of the U.S. Department of
Energy under Contract DE-AC03-76SF00098 and in part by the National
Science Foundation under grant PHY-95-14797.

\vspace{20pt}

\myappendix

\mysection{Cancellation of the Modular Anomaly}
\label{cma}
For the $Z_3$ orbifold, $SU(3,3,\Zbf)$
reparameterizations of the nine T-moduli $T^{ij}$
are symmetries \cite{z3dual}
of the underlying perturbative string theory,
at least to one loop in string perturbation theory
\cite{TDTAF,Nothr}.
These are referred to as {\it target space modular transformations}
or {\it duality transformations} of the T-moduli.
Most commonly, projective $SL(2,\Zbf)$ subgroups acting
on the diagonal moduli are studied:
\beq
T^i \to {a^i T^i-i b^i \over ic^i T^i+d^i}, \qquad
a^i d^i - b^i c^i=1, \qquad \forall \; i=1,3,5,
\label{mtr}
\eeq
with $a^i,b^i,c^i,d^i$ all integers.  The indices on
these integers indicate that each of the three $T^i$
may transform with its own set.  In addition to
transformations on the T-moduli, accompanying T-dependent
reparameterizations of chiral matter superfields
must be made:
\beq
\Phi^A \to {\sum_B M^A_{\spc B} \Phi^B \over
\prod_{i=1,3,5} (ic^i T^i + d^i)^{q_i^A} } .
\label{6a}
\eeq
Here, $q_i^B$ is the modular weight of the field
$\Phi^B$; these quantities were
given in Section~\ref{mds}.
The matrix $M^A_{\spc B}$ is identity
for untwisted fields while it mixes 
subsets of twisted fields 
with the same modular weight \cite{TWMIX}
in a way which depends on the parameters
$a^i,b^i,c^i,d^i$.

Transformations (\ref{mtr},\ref{6a}) are symmetries of the
effective supergravity action at the classical 
level---isometries of the nonlinear $\s$ model.
However, at the quantum level there is a
$\s$ model anomaly \cite{NLSA} associated with the
duality tranformations, as originally
pointed in Refs.~\cite{DFKZ91,LCO92}.
To study this {\it modular anomaly}, 
one calculates the quantum corrections to the supergravity
lagrangian, in particular triangle diagrams involving
the composite $\s$ model connections of T-moduli to other
fields at one vertex and gauge boson currents at
the other two vertices.  Various calculations of the
modular anomaly have been performed.  Most often, supergravity
interactions have been studied at the component
level and then the anomaly written as a globally
supersymmetric
superspace integral, which is an approximation to
the true supergravity anomaly~\cite{DFKZ91,LCO92,KL945,LGT}.
The supergravity one loop effective lagrangian
and its transformation properties has been studied
in great detail by Gaillard and collaborators, using Pauli-Villars
regularization techniques \cite{PVREG}.  These calculations
were recently used to infer a
locally supersymmetric superspace
expression for the anomaly at one loop \cite{GNW99}.
Equivalent expressions have also been obtained
in Ref.~\cite{BMP00}.
Keeping only the leading term important to the
present analysis, the quantum part of the one loop effective
supergravity lagrangian transforms under \myref{mtr} as
\beq
\delta {\cal L}_Q
= \sum_{a,j} {\alpha_a^j \over 64 \pi^2}
\int d^4\theta {E \over R} \ln (ic^j T^j + d^j) 
\sum_i(\W^\alpha \W_\alpha)_a^i + \mtxt{h.c.}
\label{man}
\eeq
The expression on the right-hand side is a superspace
integral in the K\"ahler $U(1)$ formulation of supergravity
\cite{bgg}.  The quantity $E$ is the superdeterminant
of the {\it vielbein;} it generalizes the tensor density
$e = \sqrt{g}$ which appears in the Einstein-Hilbert action
to a superfield.  The superfield $R$ is
chiral and has as its lowest component the
scalar auxilliary field of supergravity.  The chiral
spinor superfield $\W_{\alpha,a}^i$ is the superfield-strength
corresponding to the generator $T_a^i$
of the factor $G_a$ of the gauge group $G$ and has
as its lowest component the gaugino $\lambda_{\alpha,a}^i$.
The coefficient $\alpha_a^j$ reflects particles 
in the triangle loop which
contribute to the anomalous transformation, 
and is given by \cite{LGT}
\beq
\alpha_a^j = - C(G_a) + \sum_A (1 - 2 q_j^A) X_a(R^A) ,
\label{A6}
\eeq
where the sum runs over matter
irreps $R^A$ of $G_a$ and $q_j^A$ is
the modular weight appearing in \myref{6a}.

Since the transformations (\ref{mtr},\ref{6a})
are known to be anomaly free in 
the underlying four-dimensional
string theory, we must add effective
terms to cancel the anomaly.
One possible cancellation is from the
shift in the T-moduli dependent threshold
corrections alluded to in Sections~\ref{mds} and~\ref{gcu}.
As mentioned there, however, such
threshold corrections are absent in $Z_3$ orbifold
compactifications \cite{Nothr}.
Thus, the entire modular anomaly given by \myref{man}
must be canceled by the Green-Schwarz mechanism.
That is, we include in the effective supergravity
lagrangian a term which will have an anomalous
transformation under (\ref{mtr},\ref{6a}),
just such as to cancel \myref{man}.
The overall coefficient 
$\bGS$ of the Green-Schwarz term is determined
by this matching.  

We now describe this
term in the BGW effective theory.
However, we note that
in expressions below,
we use a slightly different normalization
for the Green-Schwarz coefficient $\bGS$ than
BGW; rather, we adopt the more common
convention of Refs.~\cite{IL92,BIM}.
In the BGW notation, the Green-Schwarz
coefficient is written as $b$, which is related to
$\bGS$ by the equation $b = - \bGS / 24 \pi^2$.
In addition, in our formulae we do not use the BGW
conventions for the $\beta$ function coefficients of the
gauge groups.  The two
conventions are related by
$b_a^{\stxt{BGW}} = - b_a^{\stxt{here}}/ 24 \pi^2$.

In addition to the supergravity
multiplet, gauge multiplets, and matter multiplets,
string theory predicts the existence of other supermultiplets
of dynamic states.  One particularly important set of fields
is the following:  a real scalar field $\ell$ called
the ${\it dilaton}$, an antisymmetric tensor $B_{mn}$
whose field strength is dual to the universal
axion, and a Majorana spinor 
$\varphi$ which is referred to as the
${\it dilatino}$.  This is on-shell content of the superfield
$L$, which is a {\it linear} multiplet.  It satisfies
the modified linearity condition \cite{bgg}
\beq
(\Db^2 + 8 R) L = - \sum_{a,i}(\W^\alpha \W_\alpha)_a^i.
\label{6b}
\eeq
Following BGW,
we write the Green-Schwarz counterterm for the modular anomaly as
\beq
{\cal L}_{GS} = {\bGS \over 24 \pi^2} \int d^4\theta \;
E L \sum_j \ln (T^j + \bar T^j) .
\label{gst}
\eeq
Using \myref{mtr},
integration by parts in superspace \cite{Zumin79}, 
chirality of $T^j$
and the modified linearity condition \myref{6b}, 
\beqa
\delta {\cal L}_{GS} & = & {- \bGS \over 24 \pi^2}
\int d^4\theta E L \sum_j \ln (ic^j T^j + d^j) + \mtxt{h.c.} \nnn
& = & { \bGS \over 8 \cdot 24 \pi^2}
\int d^4\theta {E \over R}
(\Db^2 + 8 R)[ L \sum_j \ln (ic^j T^j + d^j)] + \mtxt{h.c.} \nnn
& = & { - \bGS \over 192 \pi^2} \sum_{j,a}
\int d^4\theta {E \over R}
\ln (ic^j T^j + d^j) \sum_i (\W^\alpha \W_\alpha)_a^i + \mtxt{h.c.}
\eeqa
Comparing to \myref{man},
it is easy to see that in the present context (i.e.,
in the absence of T-moduli dependent string threshold corrections),
\beq
\bGS = 3 \alpha_a^j \qquad \forall \; a,j.
\label{gsc2}
\eeq
A generic spectrum of massless states which is
free of chiral gauge anomalies will not satisfy
\myref{gsc2}, since it requires that we get the
same result, $\bGS$, for each factor $G_a$ in the gauge
group $G$.  Thus, \myref{gsc2} is a highly nontrivial
constraint on the matter spectrum.  This was
exploited by Ib\'a\~nez and L\"ust to draw a
number of phenomenological conclusions
for $Z_3$ orbifold models \cite {IL92}.

As discussed in Section~\ref{mss},
untwisted states come in families of three;
we make explicit the family index
$i=1,3,5$ by taking $A \to (\alpha,i)$ for
untwisted fields, so that $\alpha$ denotes the
species of untwisted field.  For the twisted
fields we take $A \to \rho$ to distinguish them,
but do not separate out the family label.
For nonabelian factors $G_a$ in the models considered
here, a nice simplification can be made.  
As mentioned in Section~\ref{mds}, none of
the pseudo-massless twisted 
fields which are in nontrivial representations
of $G_a$ are oscillator states.  Consequently,
it follows from the discussion of Section~\ref{mds} that
they all have modular weights $q_j^\rho = 2/3$.
Also from Section~\ref{mds}, we have for the
untwisted states $q^{(\alpha,i)}_j = \delta_j^i$.
With these
facts, it is easy to show that
Eqs.~(\ref{A6},\ref{gsc2}) can be rewritten
\beq
\bGS = - 3 C_a 
+ \sum_{(\alpha,i) \in \stxt{untw}} X_a(R^{(\alpha,i)})
- \sum_{\rho\in \stxt{tw}} X_a(R^\rho)
= \tbeta{a} - 2 \sum_{\rho\in \stxt{tw}} X_a(R^\rho),
\label{A2}
\eeq
where the last equality follows from 
\myref{5a},
only now it is the total $\beta$ function coefficient
which appears, since all pseudo-massless
states contribute.
In the absence of twisted states in nontrivial
irreps of $G_a$, the last term on the right-hand
side vanishes.  This occurs for $SO(10)$ in
Patterns 1.1 and 1.2.  But then for a $G_a$ with
only trivial irreps in the twisted sector $\bGS = \tbeta{a}$.
This is the source of the (approximately vanishing)
T-moduli mass problem
discussed in Section~\ref{mds} and Ref.~\cite{GG00}.

As an example of the surprising matching
of \myref{A2} for different $G_a$, 
we examine Pattern~1.1.
The $SO(10)$ factor of $G$ has no nontrivial matter representations,
as can be seen from Table~\ref{pt1}, which gives
\beq
\bGS = \tbeta{10} = -3 C(SO(10)) = -24 .
\label{A3}
\eeq
For the $SU(3)$ factor, we have $15(3 + \bar 3,1,1)$ beyond
the MSSM which gives 
$\dbth = \tbeta{3} - b_3 = 15$, and consequently
$\tbeta{3} = 12$.  Comparison
of Table~\ref{pt1} to Table~\ref{unpat} shows that
the twisted sector irreps are $15(3,1,1) + 21(\bar 3,1,1)$, which
gives
\beq
\bGS = \tbeta{3} - 2 \sum_{\rho \in \stxt{tw}} X_3(R^\rho)
= 12 - 36 = -24.
\label{A4}
\eeq
Finally, the $SU(2)$ factor has $40(1,2,1)$ beyond the
MSSM, so that $\dbtw = \tbeta{2} - b_2 = 20$ 
and $\tbeta{2}= 21$.
Again comparing Table~\ref{pt1} to Table~\ref{unpat},
we find that the $SU(2)$ charged twisted matter
is $45(1,2,1)$ and so
\beq
\bGS = \tbeta{2} - 2 \sum_{\rho \in \stxt{tw}} X_2(R^\rho)
= 21 - 45 = -24.
\label{A5}
\eeq
It is reassuring that each group $SO(10), SU(3)$ and $SU(2)$
gives the same answer for $\bGS$, as they must for universal
cancellation of the modular anomaly \cite{IL92}.
As a nontrivial check on
our results, we have verified that 
this matching holds among the nonabelian
factors in each of the twenty patterns.

\newpage

\mysection{Tables}
\label{tbs}

\setcounter{mytab}{\value{table}}

\begin{center}
\begin{tabular}{ll}
Pattern & \multicolumn{1}{c}{
   $SU(3) \times SU(2) \times SO(10)$ \hspace{3pt} Irreps}
   \\ \hline
1.1  &  $ 3[(3,2,1) + 5(3,1,1) + 7(\bar 3,1,1) + 15(1,2,1)
   + 48(1,1,1)_0 + 15(1,1,1)_1] $ \\
1.2  &  $ 3[(3,2,1) + 4(3,1,1) + 6(\bar 3,1,1) + 13(1,2,1)
   + (1,1,16) + 48(1,1,1)_0$ \\ & $+ 9(1,1,1)_1] $ \\
\hline
\end{tabular}
\mycap{pt1}{20pt}{Patterns of irreps in Case 1 models.}
\end{center}

\vspace{50pt}

\begin{center}
\begin{tabular}{ll}
Pattern & \multicolumn{1}{c}{ $SU(3) \times SU(2) \times SU(5)
   \times SU(2)$ \hspace{3pt} Irreps} \\ \hline
2.1 &  $ 3[(3,2,1,1) + 3(3,1,1,1) + 5(\bar 3,1,1,1) + 9(1,2,1,1) + (1,1,5,1) $ \\
   & $  + (1,1,\bar 5,1) + 6(1,1,1,2) + (1,2,1,2) + 34(1,1,1,1)_0 + 9(1,1,1,1)_1] $ \\
2.2 &  $ 3[(3,2,1,1) + 3(3,1,1,1) + 5(\bar 3,1,1,1) + 9(1,2,1,1) + (1,1,5,1) $ \\
   & $  + (1,1,\bar 5,1) + 6(1,1,1,2) + (1,2,1,2) + 37(1,1,1,1)_0 + 6(1,1,1,1)_1] $ \\
2.3 &  $ 3[(3,2,1,1) + 3(3,1,1,1) + 5(\bar 3,1,1,1) + 11(1,2,1,1) + (1,1,5,1) $ \\
   & $  + (1,1,\bar 5,1) + 8(1,1,1,2) + 33(1,1,1,1)_0 + 6(1,1,1,1)_1] $ \\
2.4 &  $ 3[(3,2,1,1) + 2(3,1,1,1) + 4(\bar 3,1,1,1) + 9(1,2,1,1) + (1,1,5,1) $ \\
   & $  + 2(1,1,\bar 5,1) + (1,1,10,1) + 6(1,1,1,2) + 32(1,1,1,1)_0 + 6(1,1,1,1)_1] $ \\
2.5 &  $ 3[(3,2,1,1) + 2(3,1,1,1) + 4(\bar 3,1,1,1) + 7(1,2,1,1) + (1,1,5,1) $ \\
   & $  + 2(1,1,\bar 5,1) + (1,1,10,1) + 4(1,1,1,2) + (1,2,1,2) + 36(1,1,1,1)_0 $ \\
   & $  + 6(1,1,1,1)_1] $ \\
2.6 &  $ 3[(3,2,1,1) + (3,1,1,1) + 3(\bar 3,1,1,1) + 5(1,2,1,1) + (1,1,5,1) $ \\
   & $  + 3(1,1,\bar 5,1) + (1,1,10,2) + 10(1,1,1,2) + (1,2,1,2) + 25(1,1,1,1)_0] $ \\
\hline
\end{tabular}
\mycap{pt2}{20pt}{Patterns of irreps in Case 2 models.}
\end{center}

\begin{center}
\begin{tabular}{ll}
Pattern & \multicolumn{1}{c}{ $SU(3) \times SU(2) \times SU(4)
   \times SU(2)^2$ \hspace{3pt} Irreps} \\ \hline
3.1 &  $ 3[(3,2,1,1,1) + 2(3,1,1,1,1) + 4(\bar 3,1,1,1,1) + 7(1,2,1,1,1) + 2(1,1,4,1,1) $ \\
   & $  + 2(1,1,\bar 4,1,1) + 6(1,1,1,2,1) + 4(1,1,1,1,2) + (1,2,1,1,2) + 27(1,1,1,1,1)_0 $ \\
   & $  + 6(1,1,1,1,1)_1] $ \\
3.2 &  $ 3[(3,2,1,1,1) + 2(3,1,1,1,1) + 4(\bar 3,1,1,1,1) + 7(1,2,1,1,1) + 2(1,1,\bar 4,1,1) $ \\
   & $  + 8(1,1,1,2,1) + 4(1,1,1,1,2) + (1,1,4,2,1) + (1,2,1,1,2) + 26(1,1,1,1,1)_0 $ \\
   & $  + 3(1,1,1,1,1)_1] $ \\
3.3 &  $ 3[(3,2,1,1,1) + 2(3,1,1,1,1) + 4(\bar 3,1,1,1,1) + 7(1,2,1,1,1) + 2(1,1,\bar 4,1,1) $ \\
   & $  + 6(1,1,1,2,1) + 6(1,1,1,1,2) + (1,1,4,2,1) + (1,2,1,2,1) + 26(1,1,1,1,1)_0 $ \\
   & $  + 3(1,1,1,1,1)_1] $ \\
3.4 &  $ 3[(3,2,1,1,1) + (3,1,1,1,1) + 3(\bar 3,1,1,1,1) + 5(1,2,1,1,1) + 2(1,1,4,1,1) $ \\
   & $  + 2(1,1,\bar 4,1,1) + 8(1,1,1,2,1) + 4(1,1,1,1,2) + (1,1,6,2,1) + (1,2,1,2,1) $ \\
   & $  + 24(1,1,1,1,1)_0 + 3(1,1,1,1,1)_1] $ \\
\hline
\end{tabular}
\mycap{pt3}{20pt}{Patterns of irreps in Case 3 models.}
\end{center}

\newpage

\begin{center}
\begin{tabular}{ll}
Pattern & \multicolumn{1}{c}{ $SU(3) \times SU(2) \times SU(3)
   \times SU(2)^2$ \hspace{3pt} Irreps} \\ \hline
4.1 &  $ 3[(3,2,1,1,1) + 2(3,1,1,1,1) + 4(\bar 3,1,1,1,1)
   + 9(1,2,1,1,1) + (1,1,3,1,1) $ \\
   & $  + (1,1,\bar 3,1,1) + 6(1,1,1,2,1) + 6(1,1,1,1,2)
   + 30(1,1,1,1,1)_0 + 3(1,1,1,1,1)_1] $ \\
4.2 &  $ 3[(3,2,1,1,1) + 2(3,1,1,1,1) + 4(\bar 3,1,1,1,1)
   + 7(1,2,1,1,1) + (1,1,3,1,1) $ \\
   & $  + (1,1,\bar 3,1,1) + 4(1,1,1,2,1) + 6(1,1,1,1,2)
   + (1,2,1,2,1) + 34(1,1,1,1,1)_0 $ \\
   & $  + 3(1,1,1,1,1)_1] $ \\
4.3 &  $ 3[(3,2,1,1,1) + (3,1,1,1,1) + 3(\bar 3,1,1,1,1)
   + 7(1,2,1,1,1) + 3(1,1,3,1,1) $ \\
   & $  + 3(1,1,\bar 3,1,1) + 4(1,1,1,2,1) + 4(1,1,1,1,2)
   + 36(1,1,1,1,1)_0 $ \\
   & $  + 3(1,1,1,1,1)_1] $ \\
4.4 &  $ 3[(3,2,1,1,1) + (3,1,1,1,1) + 3(\bar 3,1,1,1,1)
   + 7(1,2,1,1,1) + (1,1,3,1,1) $ \\
   & $  + 3(1,1,\bar 3,1,1) + 4(1,1,1,2,1) + 7(1,1,1,1,2)
   + (1,1,3,1,2) + 30(1,1,1,1,1)_0 $ \\
   & $  + 3(1,1,1,1,1)_1] $ \\
4.5 &  $ 3[(3,2,1,1,1) + (3,1,1,1,1) + 3(\bar 3,1,1,1,1)
   + 7(1,2,1,1,1) + (1,1,3,1,1) $ \\
   & $  + 3(1,1,\bar 3,1,1) + 4(1,1,1,2,1) + 7(1,1,1,1,2)
   + (1,1,3,1,2) + 33(1,1,1,1,1)_0] $ \\
4.6 &  $ 3[(3,2,1,1,1) + (3,1,1,1,1) + 3(\bar 3,1,1,1,1)
   + 5(1,2,1,1,1) + (1,1,3,1,1) $ \\
   & $  + 3(1,1,\bar 3,1,1) + 4(1,1,1,2,1) + 5(1,1,1,1,2)
   + (1,1,3,1,2) + (1,2,1,1,2) $ \\
   & $  + 34(1,1,1,1,1)_0 + 3(1,1,1,1,1)_1] $ \\
4.7 &  $ 3[(3,2,1,1,1) + (3,1,1,1,1) + 3(\bar 3,1,1,1,1)
   + 5(1,2,1,1,1) + 3(1,1,3,1,1) $ \\
   & $  + (1,1,\bar 3,1,1) + 4(1,1,1,2,1) + 5(1,1,1,1,2)
   + (1,2,1,1,2) + (1,1,\bar 3,1,2) $ \\
   & $  + 37(1,1,1,1,1)_0] $ \\
4.8 &  $ 3[(3,2,1,1,1) + 2(\bar 3,1,1,1,1) + 3(1,2,1,1,1)
   + (1,1,3,1,1) + 5(1,1,\bar 3,1,1) $ \\
   & $  + 8(1,1,1,2,1) + 6(1,1,1,1,2) + (1,2,1,1,2)
   + (1,1,3,2,2) + 25(1,1,1,1,1)_0] $ \\
\hline
\end{tabular}
\mycap{pt4}{20pt}{Patterns of irreps in Case 4 models.} 
\end{center}

\begin{center}
\begin{tabular}{cl}
Patterns & Untwisted Irreps \\
\hline
1.1 & $ 3[(3,2,1) + 3(1,1,1)_0] $ \\
1.2 & $ 3[(3,2,1) + (\bar 3,1,1) + (1,2,1) + (1,1,16)] $ \\
2.1 & $ 3[(3,2,1,1) + 3(1,1,1,1)_0] $ \\
2.2, 2.3 &
   $ 3[(3,2,1,1) + (\bar 3,1,1,1) + (1,2,1,1) + (1,1,5,1) + (1,1,1,2)] $ \\
2.4, 2.5 & $ 3[(3,2,1,1) + (1,1,10,1) + 2(1,1,1,1)_0] $ \\
2.6 &
   $ 3[(3,2,1,1) + (\bar 3,1,1,1) + (1,2,1,1) + (1,1,5,1) + (1,1,10,2)] $ \\
3.1 & $ 3[(3,2,1,1,1) + (1,1,4,1,1) + 2(1,1,1,1,1)_0] $ \\
3.2, 3.3 & $ 3[(3,2,1,1,1) + (\bar 3,1,1,1,1) + (1,2,1,1,1)
   + (1,1,1,1,2) + (1,1,4,2,1)] $ \\
3.4 & $ 3[(3,2,1,1,1) + (1,1,6,2,1) + 3(1,1,1,1,1)_0] $ \\
4.1, 4.2 & $ 3[(3,2,1,1,1) + (\bar 3,1,1,1,1) + (1,2,1,1,1)
   + (1,1,1,2,1) + (1,1,1,1,2)] $ \\
4.3 & 
   $ 3[(3,2,1,1,1) + (1,1,3,1,1) + (1,1,\bar 3,1,1) + 3(1,1,1,1,1)_0] $ \\
4.4, 4.6 & $ 3[(3,2,1,1,1) + (1,1,3,1,2) + 3(1,1,1,1,1)_0] $ \\
4.5, 4.7 & $ 3[(3,2,1,1,1) + (\bar 3,1,1,1,1) + (1,2,1,1,1)
   + (1,1,\bar 3,1,1) + (1,1,1,2,1) $ \\*
   & $  + (1,1,1,1,2) + (1,1,3,1,2)] $ \\
4.8 &
   $ 3[(3,2,1,1,1) + (1,1,3,1,1) + (1,1,3,2,2) + 3(1,1,1,1,1)_0] $ \\
\hline
\end{tabular}
\mycap{unpat}{20pt}{Irreps of the 
untwisted sectors for each pattern of
total irreps.}
\end{center}

\begin{center}
\begin{tabular}{ll}
Pattern & \multicolumn{1}{c}{Models} \\ \hline
1.1 & 1.1, 1.2, 1.3, 4.1, 4.2, 4.3, 8.1 \\
1.2 & 2.1, 2.2, 2.3, 6.1, 6.2, 6.3, 9.1 \\
2.1 & 1.4, 1.5, 1.11, 1.12, 4.4, 4.6, 4.9, 4.11, 8.2, 8.3 \\
2.2 & 2.4, 2.5, 2.6, 2.7, 6.4, 6.6, 6.9, 6.11, 9.2, 11.3 \\
2.3 & 2.9, 2.10, 2.12, 6.8, 6.10, 6.12, 9.3 \\
2.4 & 1.6, 1.8, 1.10, 4.5, 4.8, 4.10, 10.2 \\
2.5 & 1.7, 1.9, 4.7, 4.12, 10.1, 10.3 \\
2.6 & 2.8, 2.11, 6.5, 6.7, 11.1, 11.2 \\
3.1 & 1.14, 1.15, 1.16, 1.17, 4.13, 4.15,
   4.16, 4.18, 10.4, 10.5, 10.6, 10.7 \\
3.2 & 2.13, 2.14, 6.15, 6.17, 11.5 \\
3.3 & 2.15, 2.16, 2.17, 2.18, 6.13, 6.14, 6.16, 6.18, 9.4, 11.4 \\
3.4 & 1.13, 1.18, 4.14, 4.17, 8.4 \\
4.1 & 2.19, 2.20, 2.21, 6.22, 6.23, 6.29, 11.16 \\
4.2 &  2.22, 2.23, 2.27, 2.32, 6.24, 6.26, 6.30,
   6.31, 9.5, 11.9, 11.11, 11.14 \\
4.3 & 1.19, 1.32, 1.33, 4.20, 4.27, 4.31, 8.5 \\
4.4 & 1.21, 1.22, 1.23, 1.25, 1.28, 1.29, 4.21,
   4.24, 4.25, 4.28, 4.30, 4.33, 10.8, 10.11, \\
   & 10.14 \\
4.5 & 2.24, 2.25, 2.28, 2.29, 2.30, 2.31, 6.19,
   6.20, 6.21, 6.27, 6.28, 6.32, 11.6, 11.7, \\
   & 11.12, 11.13, 11.15 \\
4.6 & 1.20, 1.24, 1.27, 1.30, 4.19, 4.22, 4.26,
   4.29, 10.10, 10.12, 10.15, 10.16, 10.17 \\
4.7 & 2.26, 2.33, 6.25, 6.33, 11.8, 11.10 \\
4.8 & 1.26, 1.31, 4.23, 4.32, 10.9, 10.13 \\
\hline
\end{tabular}
\mycapw{tb1}{20pt}{Irrep patterns versus the models
enumerated in \cite{Gie01b}.  See explanation
of model labeling in Section~\ref{mds}.}
\end{center}

\newpage

\setcounter{table}{\value{mytab}}

\begin{longtable}{cccccccccccc}
\caption{\bsa\ 6.5 Pseudo-Massless Spectrum \label{tb5}} \\
No. & Irrep 
 & $Q_1$ & $Q_2$ & $Q_3$ & $Q_4$
 & $Q_5$ & $Q_6$
 & $Q_7$ & $Q_X$ & $Z$ & $Y$ \\ \hline \endfirsthead
\caption{\bsa\ 6.5 Pseudo-Massless Spectrum (Cont.)} \\
No. & Irrep 
 & $Q_1$ & $Q_2$ & $Q_3$ & $Q_4$
 & $Q_5$ & $Q_6$
 & $Q_7$ & $Q_X$ & $Z$ & $Y$ \\ \hline \endhead
1 & $(3, 2, 1, 1)_U$ & $1$ & $6$ & $-18$ & $9$ & $45$ & $15$ & $0$ & $3$ & $1/6$ & $1/6$ \\ 
2 & $(1, 2, 1, 1)_U$ & $3$ & $18$ & $-54$ & $27$ & $-45$ & $-15$ & $0$ & $-3$ & $1/2$ & $1/2$ \\
3 & $(\bar 3, 1, 1, 1)_U$ & $-4$ & $-24$ & $72$ & $-36$ & $0$ & $0$ & $0$ & $0$ & $-2/3$ & $-2/3$ \\
4 & $(1, 1, 10, 2)_U$ & $0$ & $0$ & $0$ & $0$ & $-18$ & $-6$ & $0$ & $3$ & $0$ & $0$ \\
5 & $(1, 1, 5, 1)_U$ & $0$ & $0$ & $0$ & $0$ & $36$ & $12$ & $0$ & $-6$ & $0$ & $0$ \\
6 & $(1, 1, 1, 1)_{-1,-1}$ & $0$ & $-20$ & $-32$ & $-31$ & $-35$ & $-23$ & $0$ & $1$ & $-1$ & $0$ \\
7 & $(1, 1, 1, 1)_{-1,-1}$ & $0$ & $-35$ & $13$ & $17$ & $25$ & $-3$ & $0$ & $5$ & $-1$ & $2/5$ \\
8 & $(1, 1, 1, 1)_{-1,-1}$ & $0$ & $10$ & $16$ & $-55$ & $25$ & $-3$ & $0$ & $5$ & $-1$ & $-2/5$ \\
9 & $(1, 1, 1, 1)_{-1,-1}$ & $0$ & $10$ & $-122$ & $14$ & $10$ & $-8$ & $0$ & $4$ & $-1$ & $0$ \\
10 & $(\bar 3, 1, 1, 1)_{-1,-1}$ & $2$ & $7$ & $25$ & $11$ & $-5$ & $-13$ & $0$ & $3$ & $-2/3$ & $1/3$ \\
11 & $(1, 2, 1, 1)_{-1,-1}$ & $-3$ & $7$ & $25$ & $11$ & $-5$ & $-13$ & $0$ & $3$ & $-3/2$ & $-1/2$ \\
12 & $(1, 1, 1, 2)_{-1,0}$ & $0$ & $-5$ & $61$ & $-7$ & $-5$ & $-13$ & $0$ & $-4$ & $-1$ & $0$ \\
13 & $(1, 1, 1, 1)_{-1,0}$ & $0$ & $-5$ & $61$ & $-7$ & $-95$ & $-9$ & $0$ & $1$ & $-1$ & $0$ \\
14 & $(1, 1, 1, 2)_{-1,0}$ & $0$ & $-20$ & $-32$ & $-31$ & $55$ & $7$ & $0$ & $0$ & $-1$ & $0$ \\
15 & $(1, 1, 1, 1)_{-1,0}$ & $0$ & $-20$ & $-32$ & $-31$ & $-35$ & $11$ & $0$ & $5$ & $-1$ & $0$ \\
16 & $(1, 1, 1, 2)_{-1,0}$ & $0$ & $25$ & $-29$ & $38$ & $40$ & $2$ & $0$ & $-1$ & $-1$ & $0$ \\
17 & $(1, 1, 1, 1)_{-1,0}$ & $0$ & $25$ & $-29$ & $38$ & $-50$ & $6$ & $0$ & $4$ & $-1$ & $0$ \\
18 & $(1, 1, 1, 2)_{-1,1}$ & $0$ & $-5$ & $61$ & $-7$ & $-5$ & $21$ & $0$ & $0$ & $-1$ & $0$ \\
19 & $(1, 1, \bar 5, 1)_{-1,1}$ & $0$ & $-5$ & $61$ & $-7$ & $49$ & $5$ & $0$ & $1$ & $-1$ & $0$ \\
20 & $(1, 1, 1, 1)_{-1,1}$ & $0$ & $-5$ & $61$ & $-7$ & $-5$ & $4$ & $-3$ & $5$ & $-1$ & $-1/5$ \\
21 & $(1, 1, 1, 1)_{-1,1}$ & $0$ & $-5$ & $61$ & $-7$ & $-5$ & $4$ & $3$ & $5$ & $-1$ & $1/5$ \\
22 & $(1, 1, 1, 1)_{0,-1}$ & $2$ & $-28$ & $38$ & $-19$ & $-5$ & $4$ & $-1$ & $5$ & $0$ & $2/5$ \\
23 & $(1, 1, 1, 1)_{0,-1}$ & $2$ & $17$ & $41$ & $50$ & $-20$ & $-1$ & $-1$ & $4$ & $0$ & $2/5$ \\
24 & $(1, 1, 1, 1)_{0,-1}$ & $2$ & $17$ & $-97$ & $-22$ & $-20$ & $-1$ & $-1$ & $4$ & $0$ & $0$ \\
25 & $(1, 2, 1, 1)_{0,-1}$ & $-1$ & $14$ & $50$ & $-25$ & $-35$ & $-6$ & $-1$ & $3$ & $-1/2$ & $-1/2$ \\
26 & $(1, 1, 1, 1)_{0,-1}$ & $-4$ & $-4$ & $-34$ & $17$ & $-5$ & $4$ & $-1$ & $5$ & $-1$ & $-3/5$ \\
27 & $(3, 1, 1, 1)_{0,-1}$ & $0$ & $-10$ & $-16$ & $8$ & $-50$ & $-11$ & $-1$ & $2$ & $-1/3$ & $1/15$ \\
28 & $(1, 1, 1, 2)_{0,0}$ & $2$ & $32$ & $-4$ & $2$ & $10$ & $9$ & $-1$ & $-1$ & $0$ & $0$ \\
29 & $(1, 1, 1, 1)_{0,0}$ & $2$ & $32$ & $-4$ & $2$ & $10$ & $-8$ & $2$ & $4$ & $0$ & $1/5$ \\
30 & $(1, 2, 1, 2)_{0,0}$ & $-1$ & $-16$ & $2$ & $-1$ & $-5$ & $4$ & $-1$ & $-2$ & $-1/2$ & $-1/10$ \\
31 & $(1, 2, 1, 1)_{0,0}$ & $-1$ & $-16$ & $2$ & $-1$ & $-5$ & $-13$ & $2$ & $3$ & $-1/2$ & $1/10$ \\
32 & $(1, 1, \bar 5, 1)_{0,1}$ & $2$ & $2$ & $-52$ & $26$ & $4$ & $7$ & $-1$ & $0$ & $0$ & $2/5$ \\
33 & $(1, 1, 1, 2)_{0,1}$ & $2$ & $2$ & $-52$ & $26$ & $40$ & $2$ & $2$ & $-1$ & $0$ & $3/5$ \\
34 & $(1, 1, 1, 1)_{0,1}$ & $2$ & $2$ & $-52$ & $26$ & $40$ & $-15$ & $-1$ & $4$ & $0$ & $2/5$ \\
35 & $(1, 1, 1, 1)_{0,1}$ & $2$ & $2$ & $-52$ & $26$ & $-50$ & $6$ & $2$ & $4$ & $0$ & $3/5$ \\
36 & $(1, 1, 1, 2)_{1,-1}$ & $-2$ & $3$ & $-9$ & $-19$ & $55$ & $7$ & $-2$ & $0$ & $0$ & $-3/5$ \\
37 & $(1, 1, \bar 5, 1)_{1,-1}$ & $-2$ & $3$ & $-9$ & $-19$ & $19$ & $12$ & $1$ & $1$ & $0$ & $-2/5$ \\
38 & $(1, 1, 1, 1)_{1,-1}$ & $-2$ & $3$ & $-9$ & $-19$ & $-35$ & $11$ & $-2$ & $5$ & $0$ & $-3/5$ \\
39 & $(1, 1, 1, 1)_{1,-1}$ & $-2$ & $3$ & $-9$ & $-19$ & $55$ & $-10$ & $1$ & $5$ & $0$ & $-2/5$ \\
40 & $(1, 1, 1, 1)_{1,0}$ & $-2$ & $-27$ & $-57$ & $5$ & $-5$ & $4$ & $1$ & $5$ & $0$ & $0$ \\
41 & $(1, 1, 1, 1)_{1,0}$ & $-2$ & $18$ & $84$ & $5$ & $-5$ & $4$ & $1$ & $5$ & $0$ & $-2/5$ \\
42 & $(\bar 3, 1, 1, 1)_{1,0}$ & $0$ & $15$ & $-45$ & $-1$ & $-35$ & $-6$ & $1$ & $3$ & $1/3$ & $-1/15$ \\
43 & $(1, 1, 1, 1)_{1,0}$ & $4$ & $-6$ & $18$ & $38$ & $-20$ & $-1$ & $1$ & $4$ & $1$ & $1$ \\
44 & $(1, 2, 1, 1)_{1,0}$ & $1$ & $-9$ & $27$ & $-37$ & $-35$ & $-6$ & $1$ & $3$ & $1/2$ & $1/10$ \\
45 & $(1, 1, 1, 1)_{1,0}$ & $-2$ & $-12$ & $36$ & $29$ & $-65$ & $-16$ & $1$ & $1$ & $0$ & $0$ \\
46 & $(1, 1, 1, 2)_{1,1}$ & $-2$ & $-12$ & $36$ & $29$ & $25$ & $14$ & $1$ & $0$ & $0$ & $0$ \\
47 & $(1, 1, 1, 1)_{1,1}$ & $-2$ & $-12$ & $36$ & $29$ & $25$ & $-3$ & $-2$ & $5$ & $0$ & $-1/5$ \\
48 & $(1, 1, 1, 2)_{1,1}$ & $4$ & $9$ & $-27$ & $-10$ & $10$ & $9$ & $1$ & $-1$ & $1$ & $3/5$ \\
49 & $(1, 1, 1, 1)_{1,1}$ & $4$ & $9$ & $-27$ & $-10$ & $10$ & $-8$ & $-2$ & $4$ & $1$ & $2/5$ \\
50 & $(1, 1, 1, 2)_{1,1}$ & $-2$ & $3$ & $-9$ & $-19$ & $-35$ & $-6$ & $1$ & $-4$ & $0$ & $-2/5$ \\
51 & $(1, 1, 1, 1)_{1,1}$ & $-2$ & $3$ & $-9$ & $-19$ & $-35$ & $-23$ & $-2$ & $1$ & $0$ & $-3/5$ \\
\hline
\end{longtable}

\end{document}